\documentclass[lettersize,journal, 10pt]{IEEEtran}
\IEEEoverridecommandlockouts

\usepackage{cite}
\usepackage{amsmath,amssymb,amsfonts,bm}
\usepackage{algorithmic}
\usepackage{graphicx}
\usepackage{textcomp}
\usepackage{xcolor}
\usepackage{subfigure}
\usepackage{tikz}
\usepackage{xspace}
\usepackage[belowskip=-16pt,aboveskip=-0pt,font=footnotesize]{caption}
\def\BibTeX{{\rm B\kern-.05em{\sc i\kern-.025em b}\kern-.08em
    T\kern-.1667em\lower.7ex\hbox{E}\kern-.125emX}}
\begin{document}

\title{\textbf{RadYOLOLet}: 
\textbf{Rad}ar Detection and Parameter Estimation Using \textbf{YOLO} and Wave\textbf{Let}
}
\author{
Shamik Sarkar,~\IEEEmembership{Member,~IEEE,}
Dongning Guo,~\IEEEmembership{Fellow,~IEEE,}
and Danijela Cabric,~\IEEEmembership{Fellow,~IEEE}
\thanks{This work was supported by SpectrumX, which is an NSF Spectrum Innovation Center funded via Award 2132700.}
\thanks{
Shamik Sarkar is with the Electronics and Communications Engineering Department, Indraprastha Institute of Information Technology Delhi, 110020, India. This work was done when Shamik was with UCLA (e-mail: shamik@iiitd.ac.in).
}
\thanks{
Danijela Cabric is with the Electrical and Computer Engineering Department,
University of California at Los Angeles, Los Angeles, CA 90095 USA (e-mail: danijela@ee.ucla.edu).
}
\thanks{
Dongning Guo is with the Electrical and Computer Engineering Department,
Northwestern University, Evanston, IL 60208 USA (e-mail: dGuo@northwestern.edu).
}
}

\newcommand{\sys}{RadYOLOLet\xspace}
\newcommand{\DL}{DL\xspace}
\newcommand{\NN}{neural network\xspace}
\newcommand{\YOLOCNN}{RadYOLO\xspace}
\newcommand{\WaveletCNN}{Wavelet-CNN\xspace}

\maketitle
\begin{abstract}
Detection of radar signals without assistance from the radar transmitter is a crucial requirement for emerging and future shared-spectrum wireless networks like Citizens Broadband Radio Service (CBRS). In this paper, we propose a supervised deep learning-based spectrum sensing approach called \sys that can detect low-power radar signals in the presence of interference and estimate the radar signal parameters. 
The core of \sys is two different convolutional neural networks (CNN), \YOLOCNN and \WaveletCNN, that are trained independently.
\YOLOCNN operates on spectrograms and provides most of the capabilities of \sys. However, it suffers from low radar detection accuracy in the low signal-to-noise ratio (SNR) regime. We develop \WaveletCNN specifically to deal with this limitation of \YOLOCNN. \WaveletCNN operates on continuous Wavelet transform of the captured signals, and we use it only when \YOLOCNN fails to detect any radar signal. We thoroughly evaluate \sys using different experiments corresponding to different types of interference signals. Based on our evaluations, we find that \sys can achieve 100\% radar detection accuracy for our considered radar types up to 16 dB SNR, which cannot be guaranteed by other comparable methods. \sys can also function accurately under interference up to 16 dB SINR.
\end{abstract}

\begin{IEEEkeywords}
Spectrum Sensing, Radar Detection, Deep Learning, YOLO, Wavelet Transform, Spectrum Sharing.
\end{IEEEkeywords}

\section{Introduction} \label{section:intro}
\textit{Motivation:} Radar bands are increasingly being shared by mobile broadband systems for better radio spectrum utilization via dynamic spectrum access~\cite{clegg2022radar}. One such well-known spectrum-sharing paradigm in the United States is CBRS~\cite{cbrs}. Hence, robust spectrum sensing methods for detecting radar signals are of prime importance.
In such spectrum sensing problems, the sensor does not have a priori knowledge of the radar transmitters' signal parameters, transmission activities, and location. 
Under these restrictions, prior works have shown that machine learning (ML) based spectrum sensing methods can detect radar with high accuracy when the peak radar signal to average interference and noise ratio (SINR)\footnote{For brevity, throughout the paper, we will use `SINR (and SNR)' to imply peak-to-average SINR (and SNR) per MHz.} per MHz at the sensor is above 20 dB~\cite{sarkar2021deepradar}. 

\textit{Goals}: Our goal is to push the minimum required radar SINR limit to below 20 dB, at which the radar signals can be detected by the spectrum-sensing sensor  (henceforth sensor) with high accuracy. Specifically, we investigate three fundamental aspects of spectrum sensing for radar signals. First, we aim to develop a method to detect low SNR radar signals. Second, we aim to have the capability of detecting radar signals in the presence of interference. Third, while aiming for the above goals, we also want to estimate the parameters of radar and interference signals, e.g., bandwidth, pulse width, pulse interval, etc. These capabilities will be instrumental in designing intelligent and efficient radar-communication spectrum-sharing systems in the future.

For our investigation, we rely on the CBRS framework. CBRS is a complex spectrum-sharing ecosystem with many details~\cite{WINNF}. Hence, we enumerate the main features of CBRS that are relevant to our problem. i) We consider five types of radar signals relevant to CBRS. ii) The sensor, known as environmental sensing capability (ESC) in CBRS, must detect the radar signals without assistance from the radar transmitter. iii) The interference at the sensor originates from a cellular network that shares the spectrum with the radar signals. The interference source has no coordination with the radar transmitter. More details about these features are discussed later in Section~\ref{section:system_model}.
In the current CBRS rules, the sensor must detect the radar signals with high accuracy when the SINR is above 20 dB. However, as mentioned before, we want to go below the 20 dB radar SINR requirement in CBRS without compromising the radar detection accuracy.

\textit{Challenges}: To achieve our ambitious goals, we must address several important challenges. 
\begin{itemize}
    \item First, our considered radar signals have a low duty cycle, which is the ratio of ON time to OFF time.
    Thus, it is difficult to detect radar signals using their ON times, which are much smaller than their OFF times. This challenge becomes more critical for low SNR radar. 
    \item Second, the sensor must detect different types of radar signals that are dissimilar from one another and have unknown signal parameters within a range. Hence, it is challenging to have a method that can achieve high detection accuracy on all relevant types of radar signals. 
    \item Third, due to the dissimilarity of the different radar types, their parameters belong to very wide ranges. 
    For example, some radar signals have narrow bandwidth, while some have narrow pulses. Hence, it is difficult to estimate the radar signal parameters accurately. 
   \item Fourth, the interference signals from communication systems do not have a low duty cycle. Consequently, their presence can significantly degrade the sensor's ability to detect ephemeral radar signals, especially when the interference-to-noise ratio (INR) is high.
   \item Fifth, if the interference signals have certain transmission activity patterns, then the sensor must be capable of distinguishing such patterns from those of the radar signals. Otherwise, there will be spurious false alarms, which can hinder the overall goal of spectrum sharing.
\end{itemize}

\textit{Approach}: To address the above-mentioned challenges, we propose a supervised deep-learning-based spectrum sensing method called \sys. To deal with the first challenge, \sys uses two different CNNs. While we develop the first CNN, which we call \YOLOCNN, to have different necessary capabilities, the second one is built specifically to deal with low SNR radar signals. \YOLOCNN operates on spectrograms and simultaneously detects the radar signals and estimates their parameters using the YOLO framework~\cite{redmon2016you}. We carefully design the formation of the spectrograms to assist \YOLOCNN in fulfilling its objectives. By virtue of being a data-driven supervised deep learning method, \YOLOCNN has the capability of detecting different types of radar signals using the same neural network and, thus, overcomes the second challenge. In \YOLOCNN, we take the ambitious step of treating each radar pulse as a different object, detecting, and localizing them. This enables us to deal with the third challenge. This approach also helps us counter the fourth challenge as it provides robustness 
against interference signals that switch between the ON and OFF phases. We develop several strategies to tackle the small-sized radar pulse objects in \YOLOCNN. \YOLOCNN treats radar and interference signals as different classes and learns to distinguish between their patterns, which is crucial for tackling the fifth challenge. Finally, \YOLOCNN can also extract the parameters of the interference signals. 

However, \YOLOCNN is not robust in detecting radar in the low SNR regime. Hence, we use the second CNN, which operates on images generated by Wavelet transform of the captured signals. We call this CNN \WaveletCNN. Here our intuition is to leverage Wavelet transform, which has been used as a robust method for detecting low SNR radar echoes in traditional radar signal processing~\cite{ball2008low}, where the receiver is aware of the radar signal parameters. However, in our case, the radar signal detection problem is more complex as the sensor is unaware of the transmitted radar signals.
Hence, we use a neural network for the detection task. Instead of directly using a Wavelet transformed signal as input to the CNN, we carefully design a preprocessing step before the Wavelet transforms that improves our chances of detecting radar signals. \WaveletCNN acts as a binary classifier that distinguishes between radar and non-radar signals. Thus, \WaveletCNN lacks the diversity (multi-class classification and signal parameter estimation) of \YOLOCNN. For this reason, we use \WaveletCNN only when \YOLOCNN does not detect any radar. While \WaveletCNN provides robustness to low SNR radar, it does not provide robustness to interference inherently. Hence, we develop several strategies in \WaveletCNN and its associated preprocessing such that it does not miss-classify interference signals as radar.

\begin{table*}[htp] 
\caption{Comparison of YOLO-based radar detection methods.} 
\label{tab:yolo_comparison}
    \centering
    {
    \begin{tabular}
    {|c|c|c|c|c|c|c|c|c|c|}
        \hline
         & \multicolumn{9}{c|}{\textbf{Capabilities}} \\
        \hline
         & Radar &  All CBRS & Interference & Radar  & Radar pulse  & Interference & Tolerance to  & Tolerance to  & Tolerance to \\
         & detection & radar types  & detection &  bandwidth  & parameters & parameters & low radar &  high power & different type \\
        \textbf{Methods}  &   &   &   & estimation  & estimation & estimation & power &  interference  & of interference \\
        \hline
        \sys & YES & YES & YES & YES & YES  &  YES & YES & YES & YES\\
        \hline
        DeepRadar~\cite{sarkar2021deepradar} & YES & YES & NO & YES    & NO  &  NO  & NO & NO & NO\\
        \hline
        Waldo~\cite{soltani2022finding} & YES & NO  & YES & YES & NO   & NO & NO & YES & NO\\
        \hline
    \end{tabular}
    }
\vspace{-5mm}
\end{table*}
\subsection{Related Work} \label{section:related_work}
In traditional monostatic radar, the radar transmitter emits pulses that are reflected by objects, received by the radar receiver, and processed for detecting the objects. An essential signal processing tool in this scheme is matched filtering of the received signal for improving the SNR of the reflected pulses~\cite{richards2010principles}. However, matched filtering-based techniques are not applicable in scenarios where the spectrum sensing sensor is unaware of the transmitted signal parameters.

To deal with interference, MIMO radars use beamforming methods like sampled matrix inversion (SMI) based minimum variance distortionless response (MVDR)~\cite{vorobyov2013principles} or ML-based MVDR for suppressing interference~\cite{li2022fast}. However, we cannot directly apply such beamforming techniques to our problem as the radar and interference signals arrive at the sensor antenna from opposite directions~\cite{WINNFWhisper}\footnote{The source of interference at the sensors in CBRS is unique due to the deployment factors, and it is different from the traditional interference models in radar signal processing. More details can be found in~\cite{WINNFWhisper}.}. Additionally, having an antenna array on the sensor can impact the location privacy (via the direction of arrival estimate) of navy radar transceivers~\cite{WINNF}. 

Electronic support and electronic intelligence (ES/ELINT) is a broad area where the task is to detect radar signals that have a low probability of intercept (LPI)~\cite{kong2018automatic}. In such problems, the radar signals are designed to be challenging to detect. The radar detection problem in CBRS differs from ES/ELINT as the radar transmitter is not trying to hide its signals from the sensor. 
However, LPI radar detection techniques can be leveraged in our problem. 

Since the inception of the idea of ESC in CBRS, there have been several works on detecting radar signals~\cite{selim2017spectrum,caromi2018detection,lees2019deep,caromi2019detection,sarkar2021deepradar,soltani2022finding}. Most of these works have relied on ML-based spectrum sensing.
These methods generally frame the radar detection problem as a classification task. Several feature representations and learning methods have been proposed for this problem in the literature.
For example, a combination of signal amplitude and phase difference can be used as input to a CNN for predicting the presence of radar signals~\cite{selim2017spectrum}. 
Instead of signal amplitude and phase, the classification task can be performed using spectrograms~\cite{lees2019deep}.
Computer vision-inspired objection detection methods can be applied to spectrograms for detecting radar signals and estimating their bandwidth~\cite{sarkar2021deepradar} or detecting non-radar signals that might be present on the spectrograms~\cite{soltani2022finding}.
Instead of deep learning, support vector machines (SVM) based classifiers can also be used for the classification task using features like higher-order and peak statistics \cite{caromi2019detection}.
As mentioned before, matched filtering is not directly applicable to our problem. However, if only one type of radar signal is considered and the radar pulse shape is assumed to be known at the sensor, then matched filtering can be applied~\cite{caromi2018detection}. 
Unlike \sys, none of these works aim at detecting low-power radar signals, estimating their parameters, and tolerating higher interference.

An important component of \sys is YOLO-based radar detection. Similar ideas have been explored in a couple of prior works~\cite{sarkar2021deepradar,soltani2022finding}. Hence, we point out the differences between \sys and other YOLO-based radar detection works in Table~\ref{tab:yolo_comparison}. We see that our proposed approach has several capabilities that are absent in other YOLO-based radar detection works.

Another vital component of our work is using Wavelet transform for radar detection. Hence, we briefly review the relevant works in this domain. Continuous Wavelet transform (CWT) can be used for low SNR radar target detection and can have better processing gain (input to output SNR ratio) than matched filtering~\cite{ball2008low}. While this work serves as an important motivation, it cannot be directly applied to our problem as it assumes the knowledge of the transmitted radar signal, does not consider different radar types and interference. Wavelet transform also has the capability of reducing noise, and that can be used as used to denoise the radar returns~\cite{wang2022electronic}.
Multi-scale product of discrete Wavelet transform can be applied to the power spectral densities (PSD) of the captured signals for noise reduction~\cite{zheleva2018airview}. However, these approaches differ from ours as we use Wavelet transform to mimic the operation of matched filtering without the cognizance of radar signal parameters.
Our idea of using a CNN on images generated via Wavelet transform has similarities with the work in~\cite{walenczykowska2022radar}. 
However, our approach and objectives differ from those in ~\cite{walenczykowska2022radar}. Specifically, the work in~\cite{walenczykowska2022radar} aims at modulation recognition and does not consider the presence of interference. In contrast, our focus is on signal detection rather than modulation classification. Additionally, an important focus of our approach is to deal with interference.

\subsection{Contributions}
Our main contributions to this paper are the following.
\begin{enumerate}
    \item We develop a deep learning-based object detection method, \YOLOCNN, that simultaneously detects radar and interference signals and estimates their parameters. For radar, \YOLOCNN estimates the center frequency, bandwidth, number of pulses, pulse width, and pulse interval. For interference, \YOLOCNN estimates center frequency, bandwidth, and ON times.    
    \item We develop a deep learning-based binary classifier, \WaveletCNN, that distinguishes between radar and non-radar signals, especially in the low radar SNR regime. \WaveletCNN  uses a CNN as its core, operating on the Wavelet transform of the captured signals.
    \item Our overall design, \sys, is a tight integration of \YOLOCNN and \WaveletCNN. \WaveletCNN strives to succeed when \YOLOCNN fails. At the same time, \WaveletCNN relies on \YOLOCNN for signal parameter estimation as \WaveletCNN cannot do that. Importantly, for both the CNNs, we design several preprocessing and postprocessing of the inputs and outputs, respectively, for achieving robustness to noise and interference.
    \item We thoroughly evaluate \sys using a diverse set of experiments involving different radar SNR and interference INR scenarios. Our evaluations show that, when interference signals are not present, \sys can achieve 100\% radar detection accuracy for all five radar types up to 16 dB SNR. 
    In the presence of different types of interference signals, \sys can detect radar signals with 100\% accuracy up to 16 dB SINR. 
\end{enumerate}

\subsection{Organization}
In Section~\ref{section:system_model}, we present our system model, the detail about the radar signals relevant to our problem, and the problem statement. Then, in Section~\ref{section:methods}, we describe our methodology in \sys. In Section~\ref{section:evaluations}, we explain our evaluation setup and present our results. Finally, Section~\ref{section:conclusions} provides conclusions and future work.


\begin{table}[b] 
\caption{Radar signal parameters \cite{sanders2017procedures}} 
\label{tab:radar_types}
\resizebox{3.5in}{!}{
    {
    \begin{tabular}
    {|c|c|c|c|c|c|}
        \hline
        Radar & Pulse &  Inter-pulse & Number of & Burst  & Band \\
        type & width &  interval & pulses  & duration & width   \\
        & ($\mu$sec) &  (msec) & per burst  & (msec)   & (MHz) \\
        \hline
        1 & 0.5 - 2.5 & 0.9 - 1.1 & 15 - 40 & 13 - 44  &  1\\
        \hline
        2 & 13 - 52 & 0.3 - 3.3 & 5 - 20  & 1 - 66  &  1 \\
        \hline
        3 & 3 - 5 & 0.3 - 3.3  & 8 - 24 & 2 - 80   & 50 - 100 \\
        \hline
        4 & 10 - 30 & 0.3 - 3.3  & 2 - 8   & 0.6 - 26 & 1 - 10 \\
        \hline
        5 & 50 - 100 & 0.3 - 3.3  & 8 - 24  & 2 - 80  &  50 - 100\\
        \hline
    \end{tabular}
    }
    }  
\end{table}

\begin{figure*}[t]
    \centering
    \hspace{40mm}
    \includegraphics[scale=0.51]{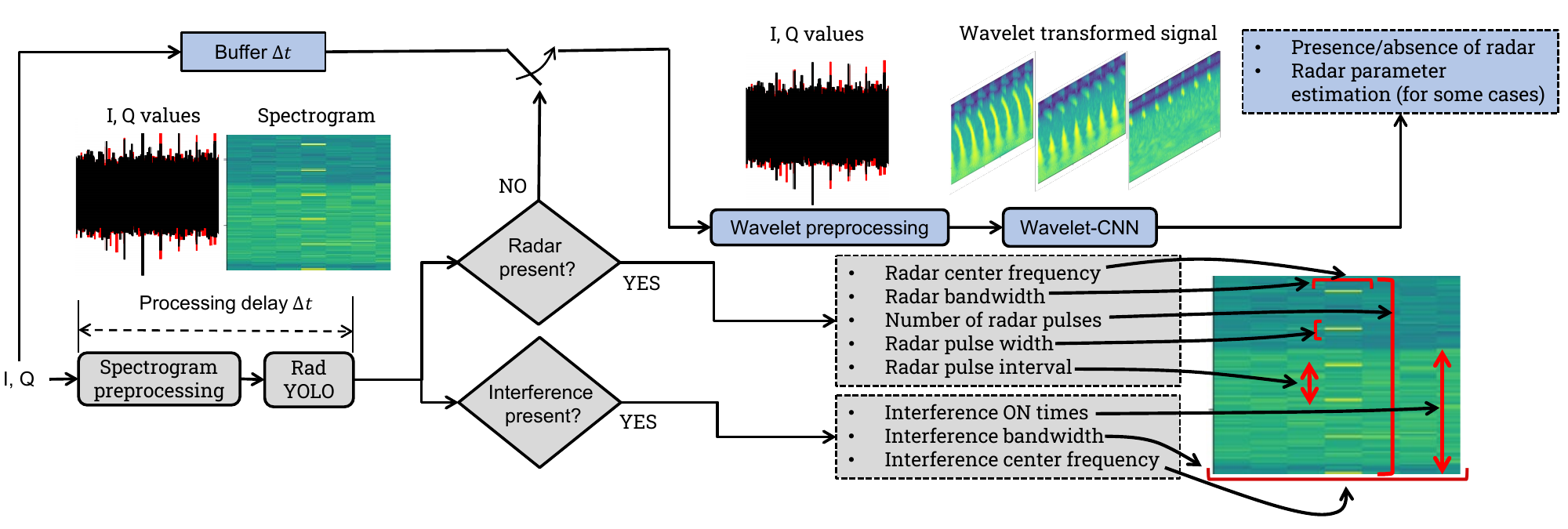}
    \caption{Overall flow diagram of \sys. The switch is closed when the branch labeled `NO' is triggered. For each block, we show on top the form of the signal on which the blocks operate.} 
    \label{fig:flow_diagran}
\end{figure*}
\section{System Model} 
\label{section:system_model}
\textit{Radar characteristics}: As in CBRS, we consider five radar types whose characteristics are shown in Table~\ref{tab:radar_types}. Radar types 1 and 2 are pulse-modulated, and the remaining ones are frequency-chirping. Thus, the bandwidth of radar types 3-5 in Table~\ref{tab:radar_types} are their chirp width. 

\textit{Sensor details:} We consider a sensor whose task is to detect the above-described radar signals with high accuracy and estimate their parameters. For all radar types, the parameters are in a range that is known to the sensor. The radar pulse parameters do not change within a burst (a set of pulses). However, at a particular time, the exact values of these parameters, along with the radar type, are unknown to the sensor. The sensor's task is to detect, at most, one radar signal at a time. I.e., we assume multiple radar signals
do not appear simultaneously. 
Due to the reasons described in Section~\ref{section:related_work}, we consider the sensor equipped with a single antenna. 
The instantaneous bandwidth of the sensor is $B$ MHz, and the sampling rate is also $B$ MS/sec. As the radar signals can appear anywhere on a 100 MHz portion of the CBRS band (3550-3650 MHz), ideally, we should have $B \geq 100$. Such sensors have been considered in~\cite{sarkar2021deepradar,soltani2022finding}. 
An alternative approach is to use $B=10$, which is what we consider in the work. In this approach, the 100 MHz band can be broken into 10 different non-overlapping 10 MHz sub-bands, and the spectrum sensing algorithm can be applied to each sub-band individually. Then, the decisions for all sub-band can be combined into a single decision for the whole 100 MHz band. Accordingly, without loss of generality, we focus on monitoring a $B=10$ MHz band for our sensor. Dividing the 100 MHz band into smaller sub-bands and making independent decisions is computationally more expensive than looking at the whole 100 MHz band altogether and making a single decision. We choose to use the computationally expensive approach as $B = 10$ suites \sys better in achieving high radar detection accuracy, as explained later in Section~\ref{subsection:wavelet_peprocess}. 
There is another implication of using $B=10$. For radar types 3 and 5 (refer to Table~\ref{tab:radar_types}), the signal bandwidth can be larger than 10 MHz. Hence, for these two radar types, we consider only the portion of the radar band that overlaps the sensor's monitoring bandwidth as the radar bandwidth.

\textit{Interference model}: The sensor 
should be able to operate in the presence of interference. However, the interference signals are not adversarial. As in CBRS, the source of interference is a cellular network that shares the spectrum with the radar signals. In our system model, we consider that the source of interference is a cellular base station (BS) whose downlink signals appear at the sensor as interference. Following the most popular channel occupancy of 10 MHz by the cellular operators in CBRS~\cite{NTIACBRS2023}, we assume the interference signal occupies the whole 10 MHz band on which the sensor operates. 
As part of our research in this work, we examine different types of interference signals and their impact on detecting radar signals. Considering the LTE sub-frame duration of 1 msec~\cite{matlabTDDFDD}, for all the interference signals considered in this paper, we assume that their ON time is at least 1 msec.

\textit{Decision granularity}: The sensor must continuously monitor the $B$ MHz band to detect radar signals. The sensor uses the sampled I, Q values from the RF frontend as its observations and makes decisions based on them. We discretize the decisions in contiguous, non-overlapping time windows of duration $T$ msec.  I.e., for every batch of $N = (T \times 10^{-3} \times B \times 10^{6})$ samples, the sensor makes a decision. The decisions of multiple time windows can be combined to make a single decision over a time duration longer than $T$ msec.
However, without loss of generality, we focus on the sensor's performance on $T$ msec time windows and do not consider combining decisions of multiple time windows. Finally, the sensor has enough computing capabilities to make decisions at a rate that is faster than the sampling rate.

\textit{Problem Statement}: Our problem is to develop a method that has the following attributes:
\begin{itemize}
    \item Capability to make accurate radar (possibly low SNR) detection decisions for each time window of duration $T$ msec, both in the presence and absence of interference.
    \item When a radar signal is detected, the capability to estimate radar center frequency, bandwidth, pulse width, pulse interval, and the number of pulses (within a time window of $T$ msec).
    \item Capability of detecting interference signals, both in the presence and absence of radar.
    \item When interference is detected, the capability to estimate its center frequency, bandwidth, and ON times.
\end{itemize}
While the radar center frequency and bandwidth estimation is an essential in CBRS, the remaining capabilities can serve as general tools of importance in spectrum sensing.  

\section{Description of \sys} \label{section:methods}
In this section, first, we present the overall flow diagram of \sys and then explain its components.

\textit{Overview of \sys}: Fig.~\ref{fig:flow_diagran} shows the flow diagram of \sys. The input to \sys is a set of I, Q values corresponding to a time window of $T$ msec. We denote the I, Q values as a complex vector $\mathbf{s}$
of size $N$. There are two interdependent flows in Fig.~\ref{fig:flow_diagran}.

In the first flow, the I, Q values go through the spectrogram preprocessing block, which generates a spectrogram using the Short-Term Fourier transform (STFT) of the I, Q values. The details of generating the spectrogram are described in Section~\ref{subsection:spectro}. The generated spectrogram is used as the input to the \YOLOCNN block. The \YOLOCNN block is essentially a CNN inspired by the YOLO framework. \YOLOCNN takes the spectrogram as input and produces as output its detection decisions, i.e., whether radar and/or interference signals are present or not. If it detects a signal, it also estimates the signal parameters as shown in Fig.~\ref{fig:flow_diagran}. The details of the \YOLOCNN are presented in Section~\ref{subsection:yolo_cnn}. When \YOLOCNN detects a radar signal, the second flow comprising of Wavelet preprocessing and Wavelet-CNN blocks in Fig.~\ref{fig:flow_diagran} are not used.

If \YOLOCNN predicts the absence of radar, the switch in Fig.~\ref{fig:flow_diagran} is closed, and the input I, Q values go through the second flow. When the second flow is used, its output overrides the output of the first flow.
The buffer block in Fig.~\ref{fig:flow_diagran} shows that both the flows operate on the same set of I, Q values, but the second flow is triggered only after the first flow makes its decision. 
In our second flow, the Wavelet preprocessing block performs frequency domain filtering on the input signal and performs continuous Wavelet transform on each filtered signal. The output of this block gives us three different images, which we stack along the depth dimension and form a 3-D tensor that is fed to the \WaveletCNN block. The details of the Wavelet preprocessing block are presented in Section~\ref{subsection:wavelet_peprocess}. The \WaveletCNN block, which is a CNN, takes the 3-D tensor as input, acts as a binary classifier, and produces its detection decision regarding the presence of radar signals as output. The details of the \WaveletCNN block are presented in Section~\ref{subsection:wavelet_cnn}.

\subsection{Spectrogram Preprocessing} \label{subsection:spectro}
\begin{figure}[t]
    \centering
    \includegraphics[scale=0.33]{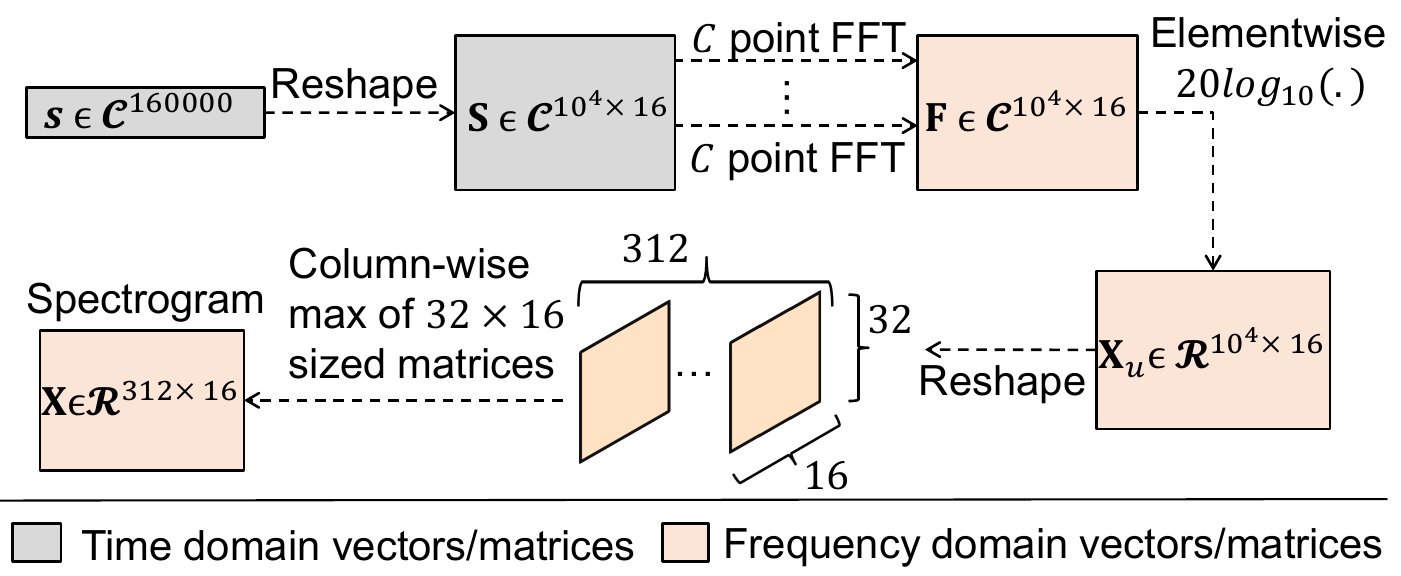}
    \caption{Operations inside the spectrogram preprocessing block.} 
    \label{fig:spectro_preprocess}
\end{figure}

The processing in this block is shown in Fig.~\ref{fig:spectro_preprocess}. 

\textit{Selection of $T$}: First, we form the complex vector $\mathbf{s}$ of size $N$, which, as defined earlier, is the set of I, Q values corresponding to a time window of $T$ msec. However, we must decide the value of $T$, which is the time unit at which \sys repeats its operations. Based on the duration of radar signals (burst duration in Table~\ref{tab:radar_types}), we choose $T = 16$ msec. Accordingly, $N=16 \times 10^4$.
$T = 16$ msec is a reasonable choice as it is not too high compared to the shorter radar signals and large enough to capture a significant portion of longer radar signals. 

\textit{Formation of spectrogram}: Next, we perform STFT on the I, Q samples corresponding to a time window. For the STFT, we reshape $\mathbf{s}$ to a matrix, $\mathbf{S}$, of size $R \times C$.
Then, we perform a $C$ point Fast Fourier Transform (FFT) on each of the rows of $\mathbf{S}$ and obtain a new complex matrix $\mathbf{F}$, which has the same dimension as $\mathbf{S}$. Finally, we take the logarithm of each element of $\mathbf{F}$ and multiply with 20 to obtain the spectrogram, $\mathbf{X}_u$. 
The columns of $\mathbf{X}_u$ represent the different frequency bins, and the rows correspond to the different short terms (we will use the phrase `time slots' to imply the short terms) in our STFT. There is no temporal overlap of the different time slots in our STFT, and we use rectangular windowing in our FFTs.

\textit{Dimensions of spectrogram}: Since we have fixed the value of $T$ (in turn, $N$), finding the dimensions of $\mathbf{S}$ requires fixing either $R$ or $C$. Based on the guidelines for selecting $C$ in~\cite{sarkar2021deepradar}, we use $C$ such that each of the rows in $\mathbf{S}$ corresponds to a time duration of 1.624 $\mu$sec. The intuitive reasoning behind using a very small time duration for the rows is that the radar pulses are of very short duration, whereas noise or interference is not. Hence, a longer time duration for a row would not increase the amount of radar energy, whereas the amount of energy from non-radar signals would increase significantly and hamper our chances of detecting radar. Selecting the time duration of a row to be 1.624 $\mu$sec implies $C = 1.624 \times 10^{-6} \times B = 1.624 \times 10^{-6} \times 10 \times 10^6 = 16.24$. Since $C$ must be an integer, we use $C = 16$. Consequently, $R$ becomes $\frac{N}{16} = 10^4$. This implies the number of rows in $\mathbf{X}_u$ is huge and much higher than the number of columns in $\mathbf{X}_u$. 

\begin{figure*}[t]
    \centering
    \begin{subfigure}[Radar type 1]
    {\includegraphics[scale=0.14]{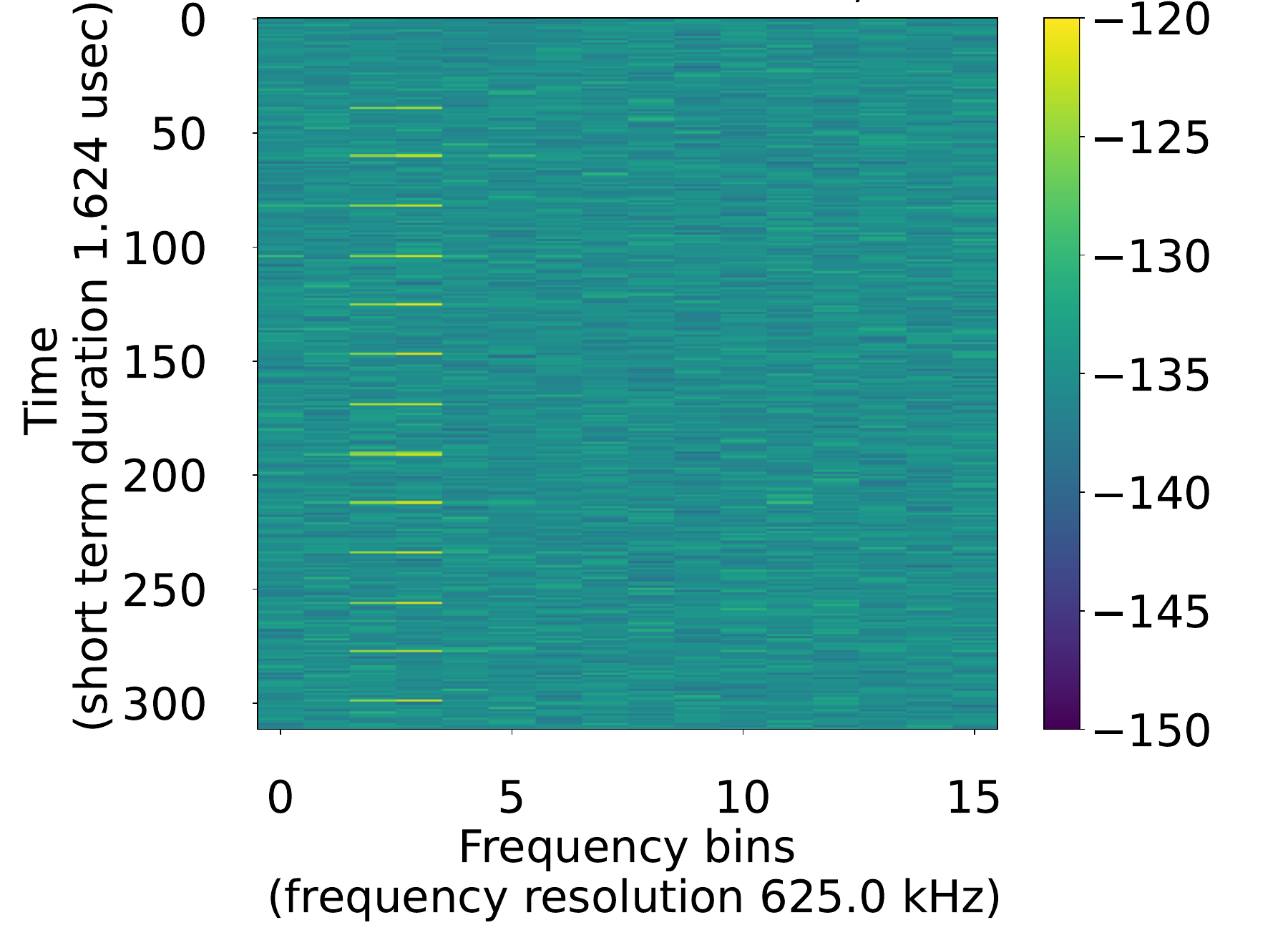}}
    \end{subfigure}
    \hspace{-13mm}
    \begin{subfigure}[Radar type 2]
    {\includegraphics[scale=0.14]{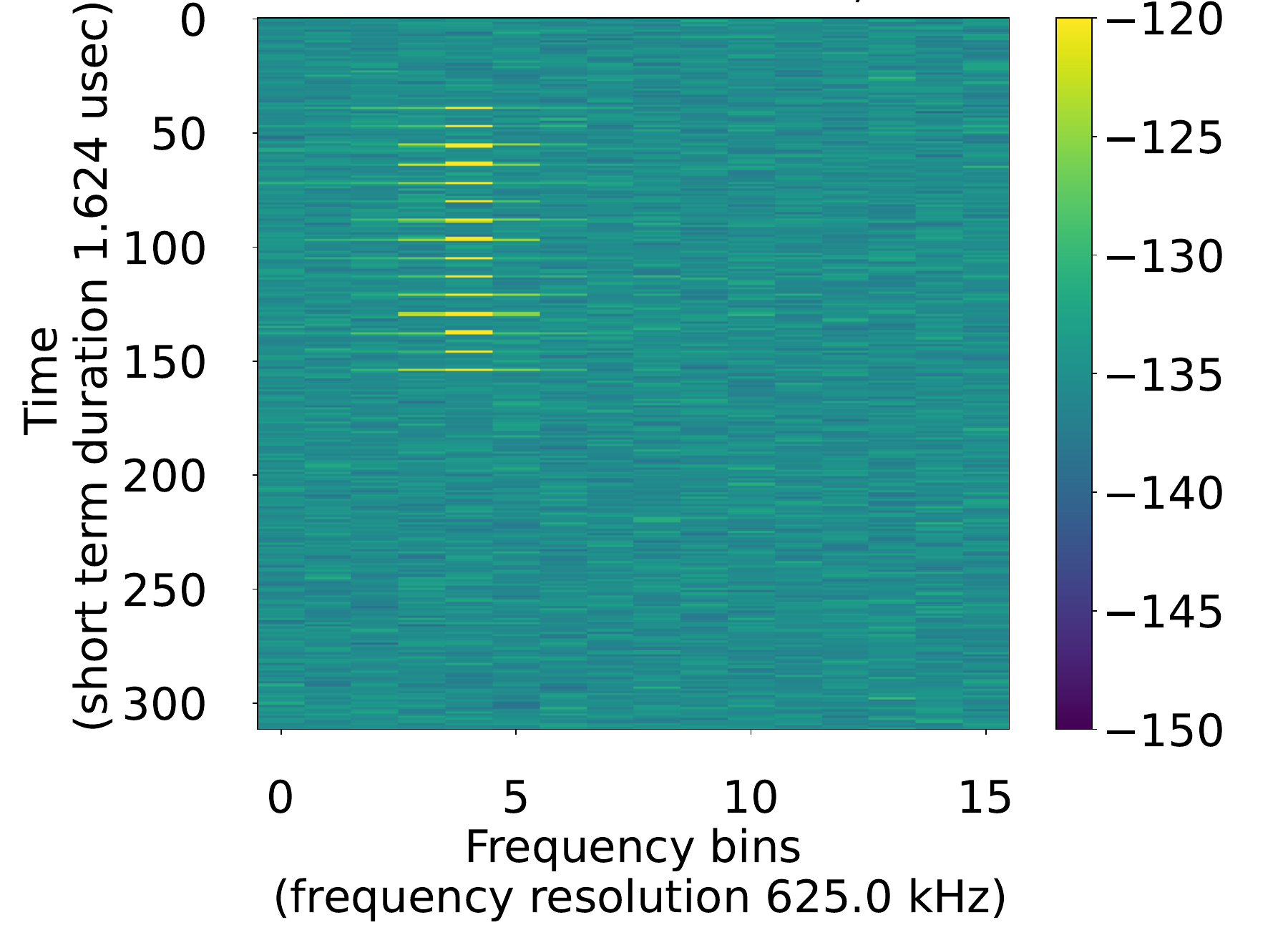}}
    \end{subfigure}
    \hspace{-13mm}
    \begin{subfigure}[Radar type 3]
    {\includegraphics[scale=0.14]{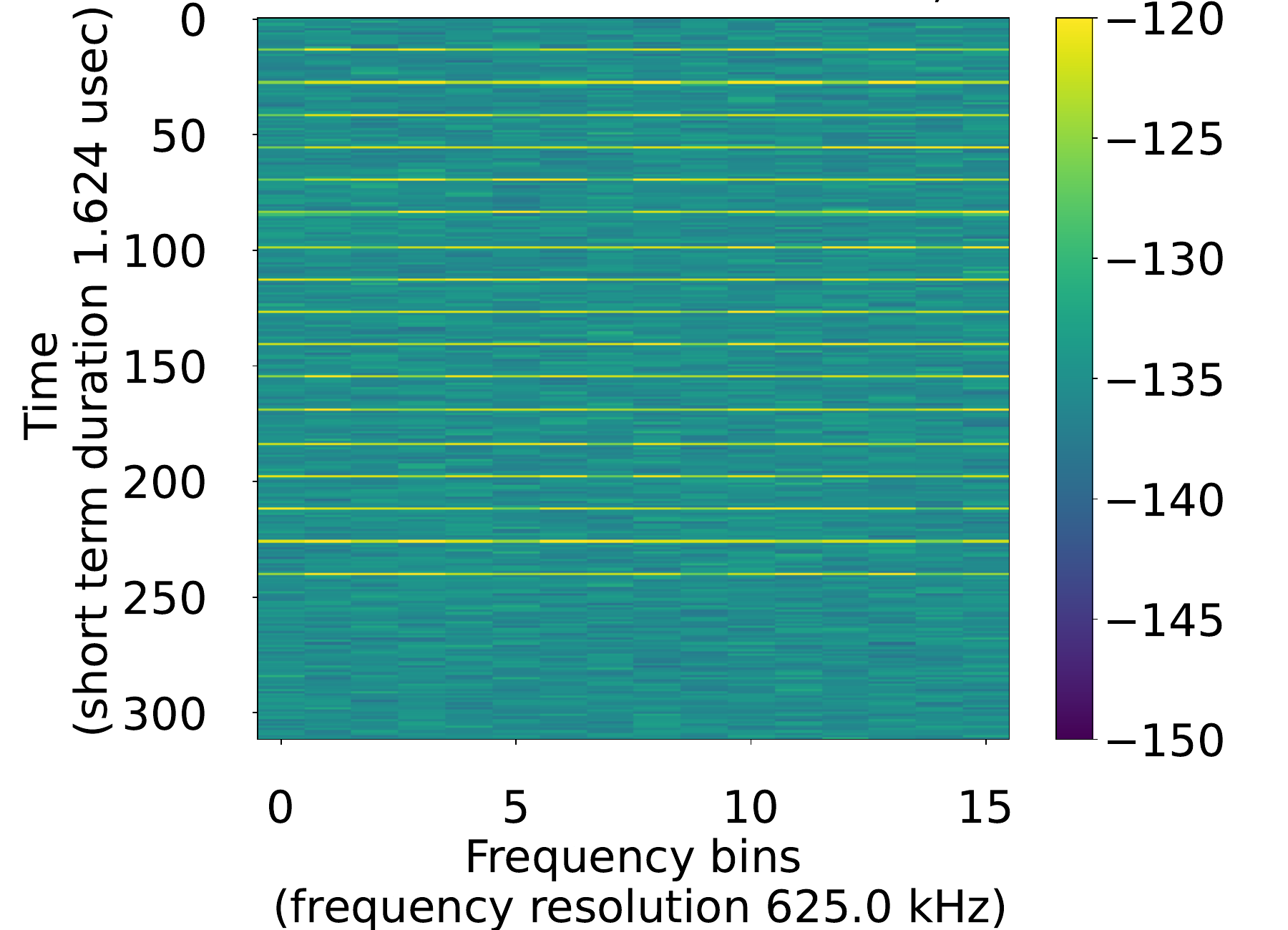}}
    \end{subfigure}
    \hspace{-13mm}
    \begin{subfigure}[Radar type 4]
    {\includegraphics[scale=0.14]{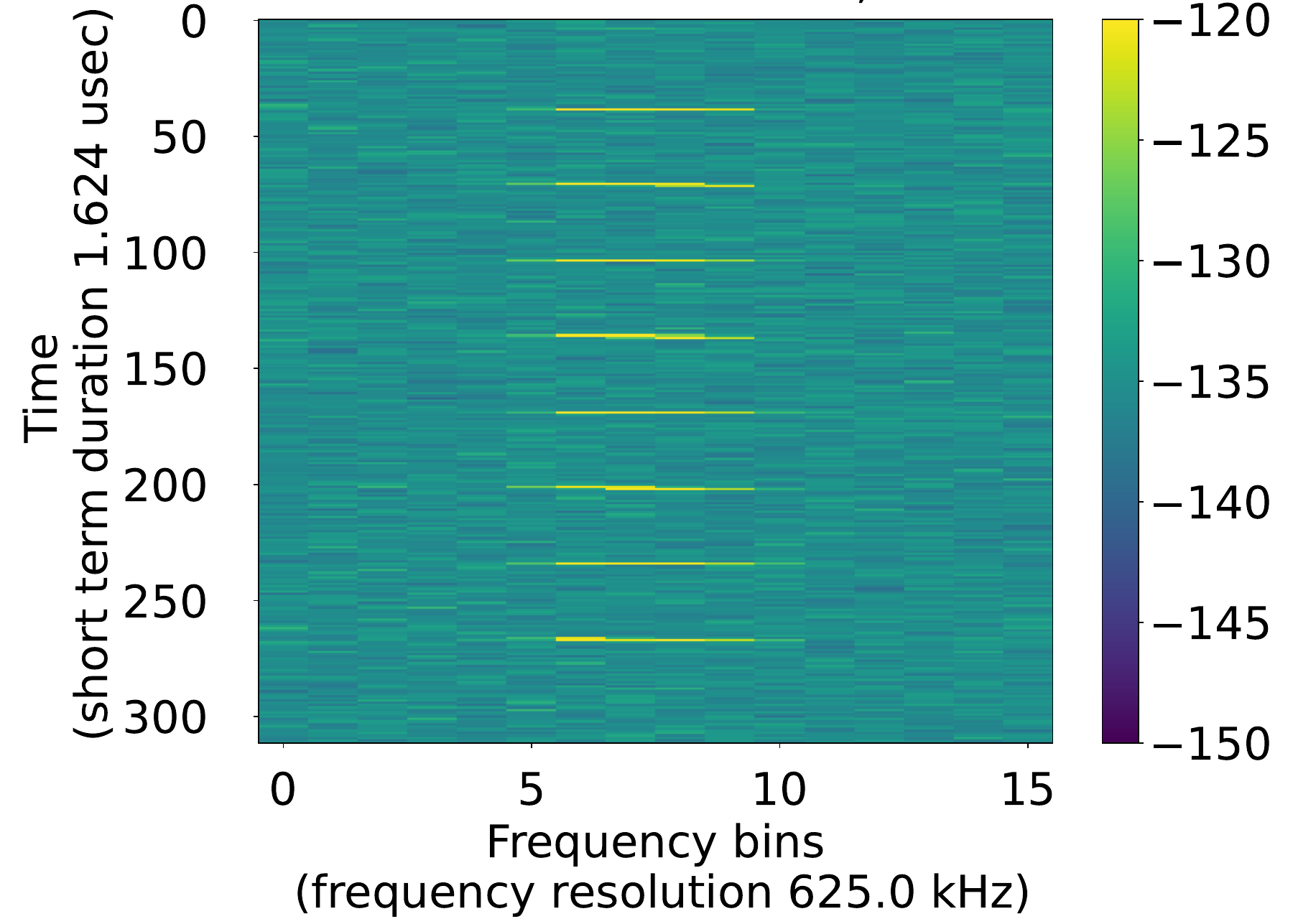}}
    \end{subfigure}
    \hspace{-13mm}
    \begin{subfigure}[Radar type 5]
    {\includegraphics[scale=0.14]{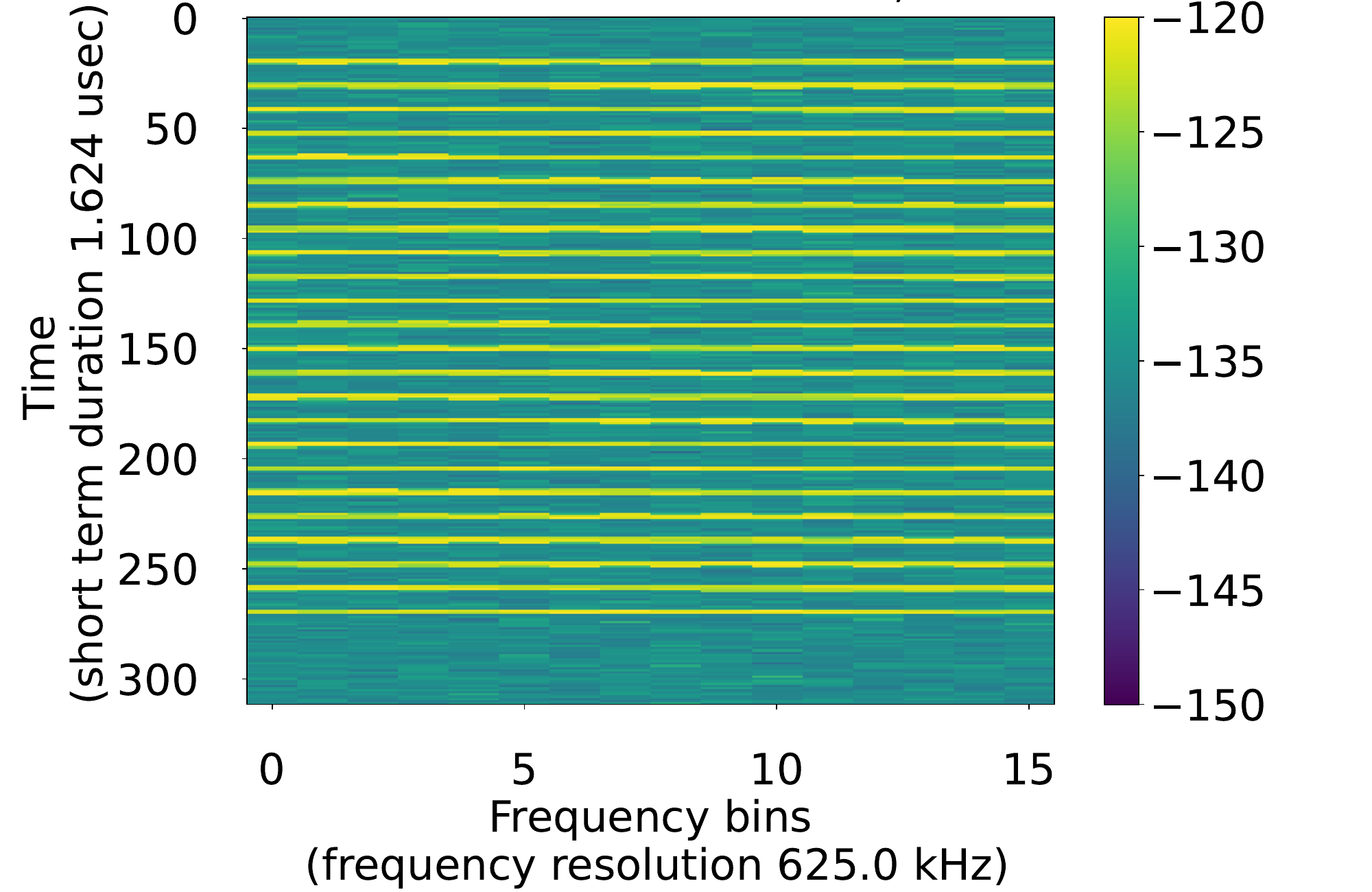}}
    \end{subfigure}
    \caption{Example spectrograms of five different radar (20 dB SNR) types after spectrogram preprocessing. No interference signal is present in these examples. 
    }
    \label{fig:spectro_examples}
\end{figure*}

\textit{Compression of spectrogram}: 
To deal with the large number of rows in $\mathbf{X}_u$, we use a compression method. We reshape $\mathbf{X}_u$ of size $(10^4 \times 16)$ to $(312 \times 32 \times 16)$, as shown in Fig.~\ref{fig:spectro_preprocess}. $\mathbf{X}_u$ can be seen as 312 matrices, each of size $(32 \times 16)$. Next, we compress each of these $(32 \times 16)$ sized matrices to vectors of size 16 by selecting the column-wise maximum for each column. This way, we obtain a new matrix $\mathbf{X}$ of size $(312 \times 16)$. 
Fig.~\ref{fig:spectro_examples} shows $\mathbf{X}$ for a set of realizations of the five different types of radar signals listed in Table~\ref{tab:radar_types}. 
Compression of the spectrogram helps the \YOLOCNN block in several ways, as explained in next section.
In spectrogram compression, we collapse 32 consecutive rows, which correspond to $32 \times 1.624 \approx 52 \mu$secs, to a single row. Referring to the inter-pulse interval column in Table~\ref{tab:radar_types}, we see the inter-pulse interval of all the radar types is much higher than 52 $\mu$sec. Hence, the ON-OFF patterns created by radar on the spectrograms are not lost by the compression. 
Additionally, since we assume that the ON time of interference signals is at least 1 msec, our compression technique would not make the interference signal patterns look like radar signal patterns and impact the detectability of radar.

\subsection{\YOLOCNN} \label{subsection:yolo_cnn}
The input to this block is the spectrogram, $\mathbf{X}$, obtained after spectrogram preprocessing. The core idea of this block is to pass $\mathbf{X}$ through a CNN and apply an object detection algorithm based on YOLO~\cite{redmon2016you} for jointly making radar and interference detection decisions and estimating their parameters. However, using the YOLO framework for signal parameter estimation is challenging. We develop several strategies to deal with these challenges, as described throughout this section.


 \begin{figure}[t]
    \centering
    \hspace{-5mm}
    \begin{subfigure}[Radar pulse objects. 
    ]
    {\label{fig:yololet_radar_obj}
    \includegraphics[scale=0.31]{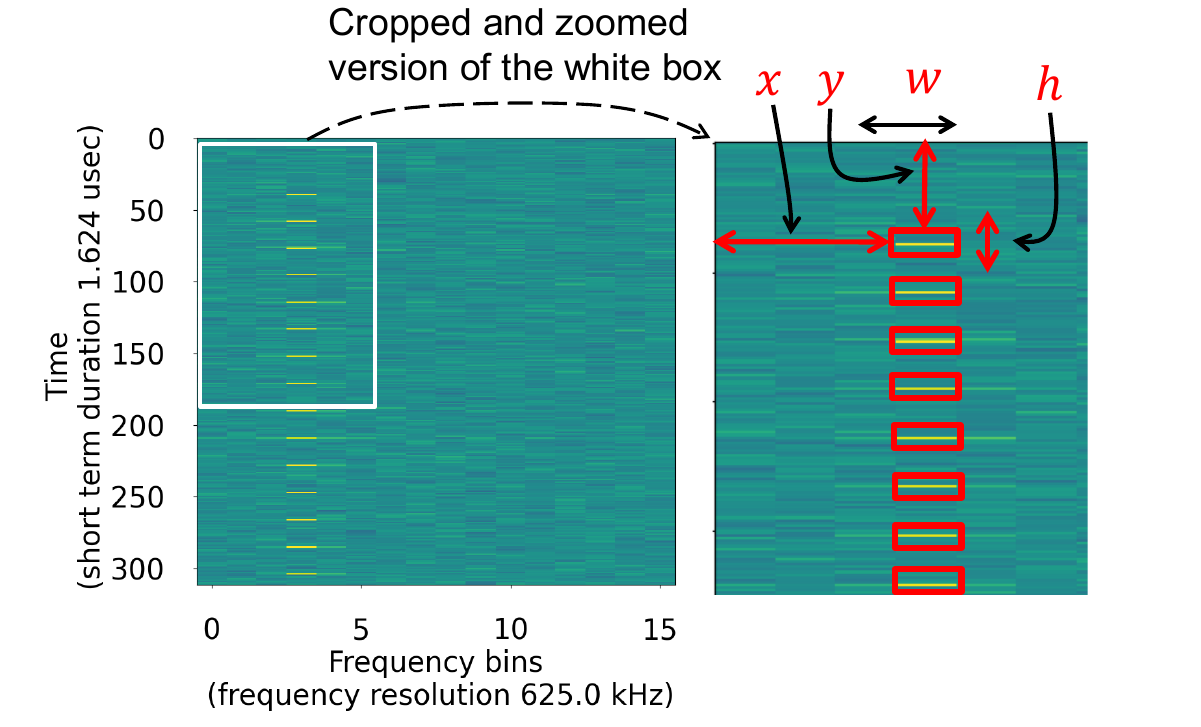}}
    \end{subfigure}
    \hspace{-9mm}
    \begin{subfigure}[Interference objects.]
    {\label{fig:yololet_if_obj}
    \includegraphics[scale=0.31]{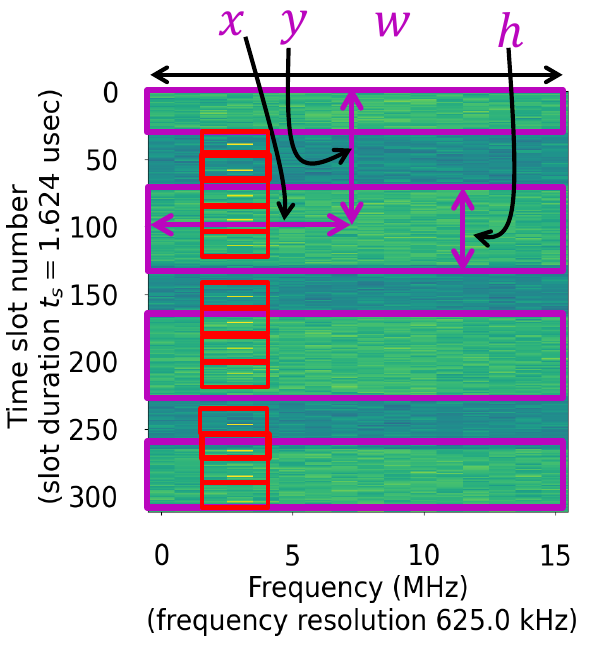}}
    \end{subfigure} 
    \caption{
    Definition of radar (red), interference (magenta) objects in \YOLOCNN. 
    }
    \label{fig:object_definition}
\end{figure}

\textit{Definition of objects}: We formulate the object detection task as shown in Fig.~\ref{fig:object_definition}. We take the ambitious step of treating each radar pulse as an object. This way, the number of detected objects can estimate the number of radar pulses, and their separation can estimate the radar pulse interval. Additionally, the height and width of the detected objects can estimate the radar pulse width and bandwidth, respectively. Finally, the $x$ parameter (see Fig.~\ref{fig:yololet_radar_obj}) of the detected objects can estimate the radar center frequency. However, as we consider the presence of interference, we must not confuse the interference signals as radar objects. Hence, we treat radar and interference as objects belonging to different classes, as shown in Fig.~\ref{fig:yololet_if_obj}. This enables us to detect radar and interference signals simultaneously. Similar to the radar pulses, we define each interference ON duration as an objects (see Fig.~\ref{fig:yololet_if_obj}). Using the detected interference objects, we can estimate their ON times, 
center frequency, and bandwidth, similarly as described for radar objects.

\textit{Object detection framework}: Now we explain the object detection framework in \YOLOCNN.
We divide each spectrogram in a grid of size $K_C \times 1$ (can be visualized as a set of $K_C$ grid cells stacked vertically). The number of grid cells along the frequency axis is 1 because, based on our system model, we do not anticipate multiple signals with nonoverlapping bands on the same spectrogram. Then for each grid cell, we define a bounding box, $(x_i, y_i, w_i, h_i); i = 1, ..., K_C$, that specifies the location of the object within that grid cell. An object is associated with a grid cell if that object's center lies within the grid cell. The size of an object can be bigger than the size of a grid cell. However, not every grid cell contains an object. Hence, for every bounding box, we define confidence, $c_i; i = 1, ..., K_C$ (as we use one bounding box per grid cell), which defines the confidence that an object is present in that box. Finally, we must associate the objects with classes. For each grid cell, we use two probabilities, $(p_i^R, p_i^I); i = 1, ..., K_C$, that are the probabilities that the object in grid cell $i$ belongs to the radar class or interference class, respectively. These probabilities are conditioned on the presence of an object in grid cell $i$. Note that, with our formulation, we can have only one object per grid cell. Thus, at most, one element of $(p_i^R, p_i^I)$ is non-zero for grid cell $i$. However, that does not deter us from multi-class classification (detecting both radar and interference on the same spectrogram) as long as all the radar and interference objects do not share the same grid cells. To avoid such undesired situations, we must choose the value of $K_C$ carefully. The choice of $K_C$ is also impacted by the fact that we want different radar pulses/objects to fall in different grid cells so that we can detect them individually. Based on the above factors, we choose $K_C$ to be 32. This choice of $K_C$ implies each of the grid cells corresponds to a duration of $T/K_C = 16/32 = 0.5$ msec, which is comparable to the lowest inter-pulse interval of the different radar types (refer to the Inter-pulse interval column in Table~\ref{tab:radar_types}). Hence, different radar objects will fall in different grid cells with a high probability. Additionally, since the interference signals' ON time is at least 1 msec, the choice of $K_C = 32$ implies that different interference objects will be in different grid cells, and, with high probability, there will be some grid cells that contain only radar objects. This will improve our chances of detecting radar signals even in interference. Finally, since the number of interference objects on a spectrogram is, in general, smaller than the number of radar objects (see Fig.~\ref{fig:yololet_if_obj}), we associate a grid cell with interference if both radar and interference objects share the grid cell. This will reduce the chances of missing the interference signal altogether.

\textit{Training procedure}: Based on the above description, we must train the CNN in \YOLOCNN so that it can predict $(p_i^R, p_i^I, x_i, y_i, w_i, h_i, c_i)$ for each of the $K_C$ grid cells. Thus, for each input spectrogram, the output of the CNN is of size $K_C \times 7 = 32 \times 7$. During training, we minimize the following loss function using the Adam optimizer~\cite{kingma2014adam}. 
\begin{equation} \label{eq:yolo_cnn_loss}
\begin{aligned}
    \mathcal{L}_Y = 
    & \lambda_{coord} \sum_{b \in \mathcal{B}} \sum_{i=1}^{K_C} \mathbf{1}_{b,i}^{obj} \big[ (x_{b,i} - \hat{x}_{b,i})^2 + (y_{b,i} - \hat{y}_{b,i})^2 \big] \\
    & + \lambda_{coord} \sum_{b \in \mathcal{B}} \sum_{i=1}^{K_C}  \mathbf{1}_{b,i}^{obj} \bigg[ \bigg(\sqrt{w_{b,i}} - \sqrt{\hat{w}_{b,i}}\bigg)^2  \\
    & \hspace{40mm} + \bigg(\sqrt{h_{b,i}} - \sqrt{\hat{h}_{b,i}}\bigg)^2 \bigg] \\
    & + \lambda_{obj} \sum_{b \in \mathcal{B}} \sum_{i=1}^{K_C}  \mathbf{1}_{b,i}^{obj} (c_{b,i} \times \text{IOU}_{b,i} - \hat{c}_{b,i}) ^ 2 \\
    & + \lambda_{nobj}  \sum_{b \in \mathcal{B}} \sum_{i=1}^{K_C}  \mathbf{1}_{b,i}^{nobj} (c_{b,i} \times \text{IOU}_{b,i} - \hat{c}_{b,i})^2  \\
    & + \lambda_{class} \sum_{b \in \mathcal{B}}  \sum_{i=1}^{K_C} \sum_{j \in \{R, I\}} \mathbf{1}_{b,i}^{obj} \big(p_{b,i}^{j} - \hat{p}_{b,i}^{j}\big)^2
\end{aligned}
\end{equation}
where $b$ denotes an example belonging to a batch $\mathcal{B}$, $\mathbf{1}$ denotes an indicator function, and $\lambda_{coord}$, $\lambda_{obj}$, $\lambda_{nobj}$, and $\lambda_{class}$ are hyperparameters. $\text{IOU}_{b,i}$ is the intersection over union (IOU) of the predicted bounding box and ground truth bounding box for grid cell $i$ of training example $b$. IOU is defined as the ratio of intersection and union of the predicted bounding box and true bounding box, respectively. IOU represents the quality of localization of an object on the spectrogram.

An important thing to note in~\eqref{eq:yolo_cnn_loss} is that the ground truth confidence score, $c_{b,i}$, is multiplied by $\text{IOU}_{b,i}$ before it is compared to the predicted confidence $\hat{c}_{b,i}$. This way, the CNN is trained to output high confidence in predicting the presence of an object only when the IOU of the predicted bounding box is high. However, this creates a challenging problem in our object detection formulation. Since, in our formulation, individual radar pulses are treated as an object, the radar objects are very small with respect to the spectrogram. Consequently, a small localization error for a radar object can result in a very low IOU. If the IOU of the predicted bounding boxes for the radar objects is always low, the network will not be incentivized to predict high confidence, $\hat{c}_{i,b}$, for the radar objects (see~\eqref{eq:yolo_cnn_loss}). In such cases, it would be difficult for the trained CNN to differentiate between radar objects and background. 

To deal with this challenge, we use three strategies. The first strategy is the spectrogram compression, described in Section~\ref{subsection:spectro}, which increases the radar objects' size compared to the spectrogram's size. Second, we penalize localization errors (first two terms in~\eqref{eq:yolo_cnn_loss}) higher than the other terms such that the IOU of the predicted bounding boxes improves. This is done via choosing $\lambda_{coord}$ to be higher than other hyperparameters.
At the same time, we also use a higher value of $\lambda_{nobj}$ such that the predicted confidence is further reduced when no object is present. This way, we aim to have high confidence for radar objects and low confidence for background and, in turn, better distinguishability between radar objects and background. However, using a high value of $\lambda_{coord}$ causes overfitting. I.e., the localization error is low on the training dataset but not on the validation set. Hence, we carefully choose the values of $\lambda_{coord}$ and $\lambda_{nobj}$ via cross-validation such that overfitting does not happen during training. Our third strategy is slightly modifying the loss function in~\eqref{eq:yolo_cnn_loss}. Specifically, we modify the definition of IOU as the following:
\begin{equation} \label{eq:iou_modification}
\text{IOU}_{b,i}= 
    \begin{cases}
    \text{IOU}_{b,i} + 0.5 \text{ if } h_{b,i} < 2\% \text{ of spectrogram height}  \\
    \text{IOU}_{b,i} \text{ o.w.}
    \end{cases}
\end{equation}
From Table~\ref{tab:radar_types}, we can see that the maximum possible radar pulse width is 100 $\mu$sec, which is less than 1\% of $T = 16$ msec; the duration (height) of the spectrograms. Hence, whenever the true height of an object is less than 2\% of the spectrogram height, we provide a boost of 0.5 to the IOU. The value of the boost parameter is chosen to be 0.5 via cross-validation. It is important to note that we must use the second and third strategies simultaneously. If we only use the second strategy while avoiding overfitting, we will not have sufficient confidence in detecting the radar objects. On the other hand, if we only use the third strategy, the network will not learn to perform accurate localization of radar objects. Finally, the interference objects are not affected by the challenge of small objects as they are much larger than the radar objects. 
Hence, our IOU modification does not affect interference objects as they do not fulfill the criteria in~\eqref{eq:iou_modification}. 

After training, we extract the following statistical parameters used in the prediction phase.

$c_{R,O}^{max}$: We pass all training examples through the trained CNN. For a training example $b$, we note the predicted radar confidence score $\hat{c}_{b,i}^{R} = \hat{c}_{b,i} \times \hat{p}_{i}^R$ for each of the cells, $i$, where radar objects are present (based on ground truth). Then, for that training example, we compute $\hat{c}_{b,max}^{R} = \max_{i} \hat{c}_{b,i}^{R}$.
Next, we form a set, $\mathcal{C}_{R,O}^{max}$, that contains $\hat{c}_{b,max}^{R}$, for all the training examples where radar was present. Finally, we compute $c_{R,O}^{max}$ as the $10^{th}$ 
percentile of the set of values in  $\mathcal{C}_{R,O}^{max}$. Essentially, $c_{R,O}^{max}$ indicates the confidence of the trained model in detecting radar objects.
    
$c_{R,O}^{min}$: For computing $c_{R,O}^{min}$, we use the same procedure as $c_{R,O}^{max}$, but for each training example we compute $\hat{c}_{b,min}^{R} = \min_{i} \hat{c}_{b,i}^{R}$, instead of $\hat{c}_{b,max}^{R} = \max_{i} \hat{c}_{b,i}^{R}$.
    
$c_{I,O}^{max}$: We compute this using the same procedure as $c_{R,O}^{max}$, but only for interference objects. $c_{I,O}^{max}$ indicates the confidence of the trained model in detecting interference objects.
    
$c_{I,O}^{min}$: We compute this using the same procedure as $c_{R,O}^{min}$, but only consider interference objects.
    
$c_{B,NO}^{max}$: For a training example $b$, we note the predicted confidence in background $\hat{c}_{b,i}^{B} = \hat{c}_{b,i} \times [1 - (\hat{p}_{i}^R + \hat{p}_{i}^I)]$ for each of the cells, $i$, where no object is present. Then, for that training example, we compute $\hat{c}_{b,max}^{B} = \max_{i} \hat{c}_{b,i}^{B}$.
Next, we form a set, $\mathcal{C}_{B,NO}^{max}$, that contains $\hat{c}_{b,max}^{B}$, for all the training examples where at least one grid cell is present with no object. Finally, we compute $c_{B,NO}^{max}$ as the 
$95^{th}$ percentile of the set of values in $\mathcal{C}_{B,NO}^{max}$. $c_{B,NO}^{max}$ indicates the false object detection confidence of the trained model when no object is present.

\begin{figure}[t]
    \centering
    \includegraphics[scale=0.27]{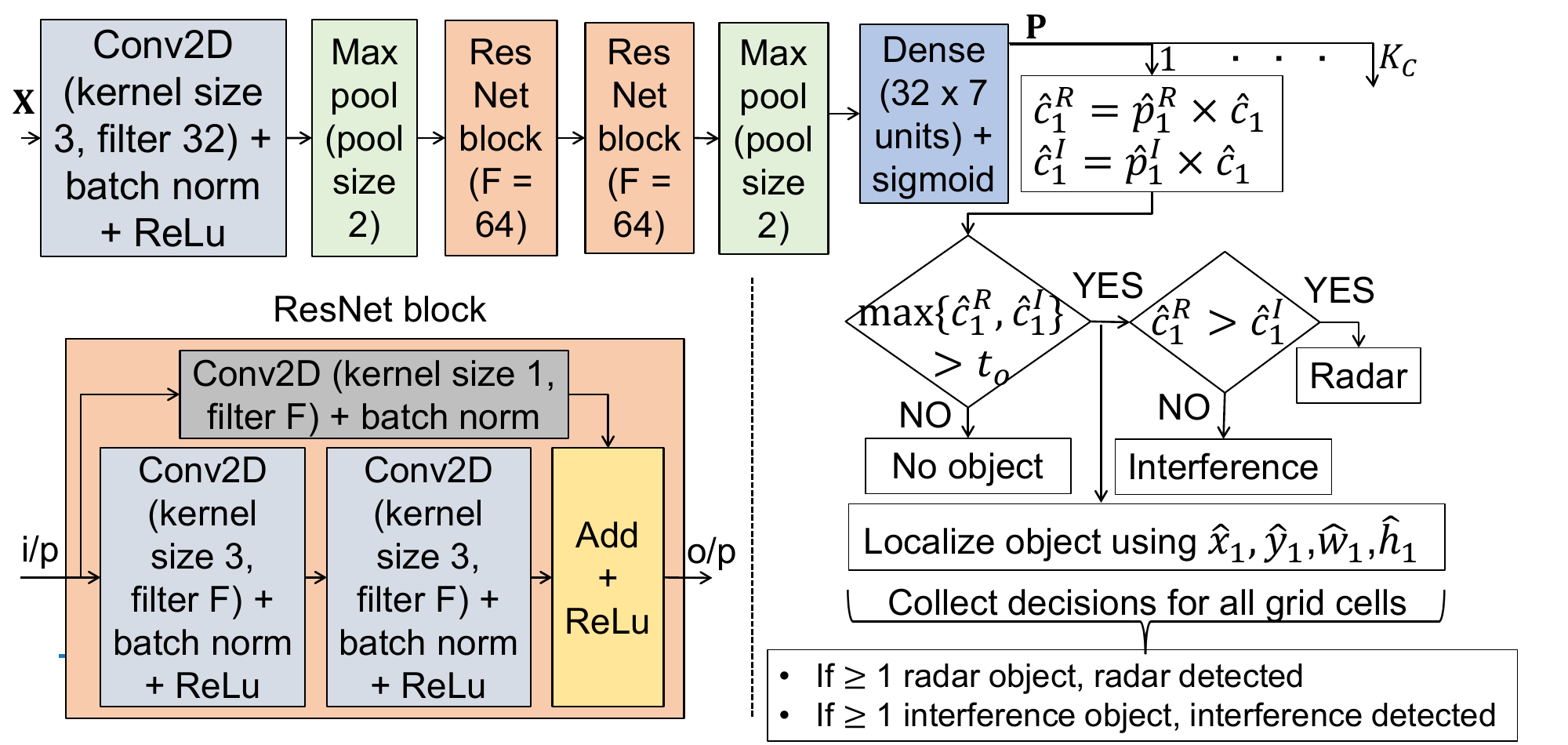}
    \caption{CNN used in \YOLOCNN, along with the prediction procedure (shown for grid cell 1, but used for all $K_C$ grid cells.)} 
    \label{fig:yolo_cnn}
\end{figure}
\textit{Prediction procedure}: The predictions in \YOLOCNN are made as shown in Fig.~\ref{fig:yolo_cnn}, which also shows the architecture of our CNN in \YOLOCNN. We select this architecture based on experimentation. Importantly, the compression method in Section~\ref{subsection:spectro} simplifies the CNN architecture by reducing the size of the input spectrogram. 
The CNN takes $\mathbf{X}$ as the input and produces as output $\mathbf{P} \in \mathcal{R}^{K_C \times 7}$ which consists of $(\hat{p}_i^R, \hat{p}_i^I, \hat{x}_i, \hat{y}_i, \hat{w}_i, \hat{h}_i, \hat{c}_i)$ for each of the $K_C = 32$ cells. First, we multiply $\hat{c}_i$ with $\hat{p}_i^R$ and $\hat{p}_i^I$ to get the class-specific confidence scores, $\hat{c}_i^R = \hat{p}_i^R \times \hat{c}_i$ and $\hat{c}_i^I = \hat{p}_i^I \times \hat{c}_i$, for each of the grid cells. As mentioned before, $\hat{p}_i^R$ represents $\text{Pr}[\text{radar} | \text{object is present}]$ for grid cell $i$, and $\hat{c}_i$ represents the confidence that an object is present in grid cell $i$. Hence, $\hat{c}_i^R$ represents the probability of a radar object's presence in grid cell $i$. Similarly, $\hat{c}_i^I$ represents the probability of an interference object's presence in grid cell $i$. Next, we find the maximum of $\hat{c}_i^R$ and $\hat{c}_i^I$ and compare it with a threshold, $t_o$. If $\max \{\hat{c}_i^R, \hat{c}_i^I\} \geq t_o$, we predict the presence of an object in grid cell $i$. If the presence of an object is predicted, we associate that object with the radar class if $\hat{c}_i^R >  \hat{c}_i^I$; otherwise, we associate the object with the interference class. If an object is detected for grid cell $i$, we estimate its location using $\hat{x}_i, \hat{y}_i, \hat{w}_i, \hat{h}_i$.
For a test spectrogram, $b$, if a radar object is detected in any grid cell, we decide the presence of radar signal in example $b$. In such cases, we estimate the radar signal parameters using all the grid cells where radar objects have been detected. Let us denote those grid cells as the set $\mathcal{K}_C^{R}$. We estimate the number of radar pulses as $|\mathcal{K}_C^{R}|$, the center frequency as the mean of $\hat{x}_i; i \in \mathcal{K}_C^{R}$, the bandwidth of the radar signal as $\cup_i \hat{w}_i; i \in \mathcal{K}_C^{R}$, the pulse width as the minimum of $\hat{h}_i; i \in \mathcal{K}_C^{R}$, the pulse interval as the minimum difference between any pair of $\hat{y}_i; i \in \mathcal{K}_C^{R}$. Similarly, if an interference object is detected for any of the cells in test example $b$, we decide the presence of interference. The interference signal parameters can be estimated using the same procedure as described above for radar signals. For interference signals, we care about center frequency (mean of $\hat{x}_i$), bandwidth ($\cup_i \hat{w}_i$), and ON times ($\cup_i \hat{y}_i$ and the associated $\hat{h}_i$),
where $i$ runs over the grid cells where an interference object has been detected.

\textit{Selection of threshold, $t_o$}: In the prediction procedure, the threshold $t_o$ plays a vital role in deciding whether an object is present. We carefully choose $t_o$ to be $\max\{c_{B,NO}^{max},  \min\{c_{R,O}^{max}, c_{I,O}^{max}\}\}$ and our choice is justified below. As defined earlier, $c_{R,O}^{max}$ and $c_{I,O}^{max}$ are the confidence of the trained model in detecting radar and interference objects, respectively. Thus by choosing $t_o$ to be $\min\{c_{R,O}^{max}, c_{I,O}^{max}\}$, we declare the presence of an object only when the predicted confidence of the trained model is high enough for our target objects.
However, we must also ensure that $t_o$ is higher than $c_{B,NO}^{max}$ (false object detection confidence of the trained model when no object is present) to minimize the number of false detection of objects. Hence, instead of using $t_o = \min\{c_{R,O}^{max}, c_{I,O}^{max}\}$, we use $t_o = \max\{c_{B,NO}^{max},  \min\{c_{R,O}^{max}, c_{I,O}^{max}\}\}$.

\begin{figure}[t]
    \centering
    \includegraphics[scale=0.14]{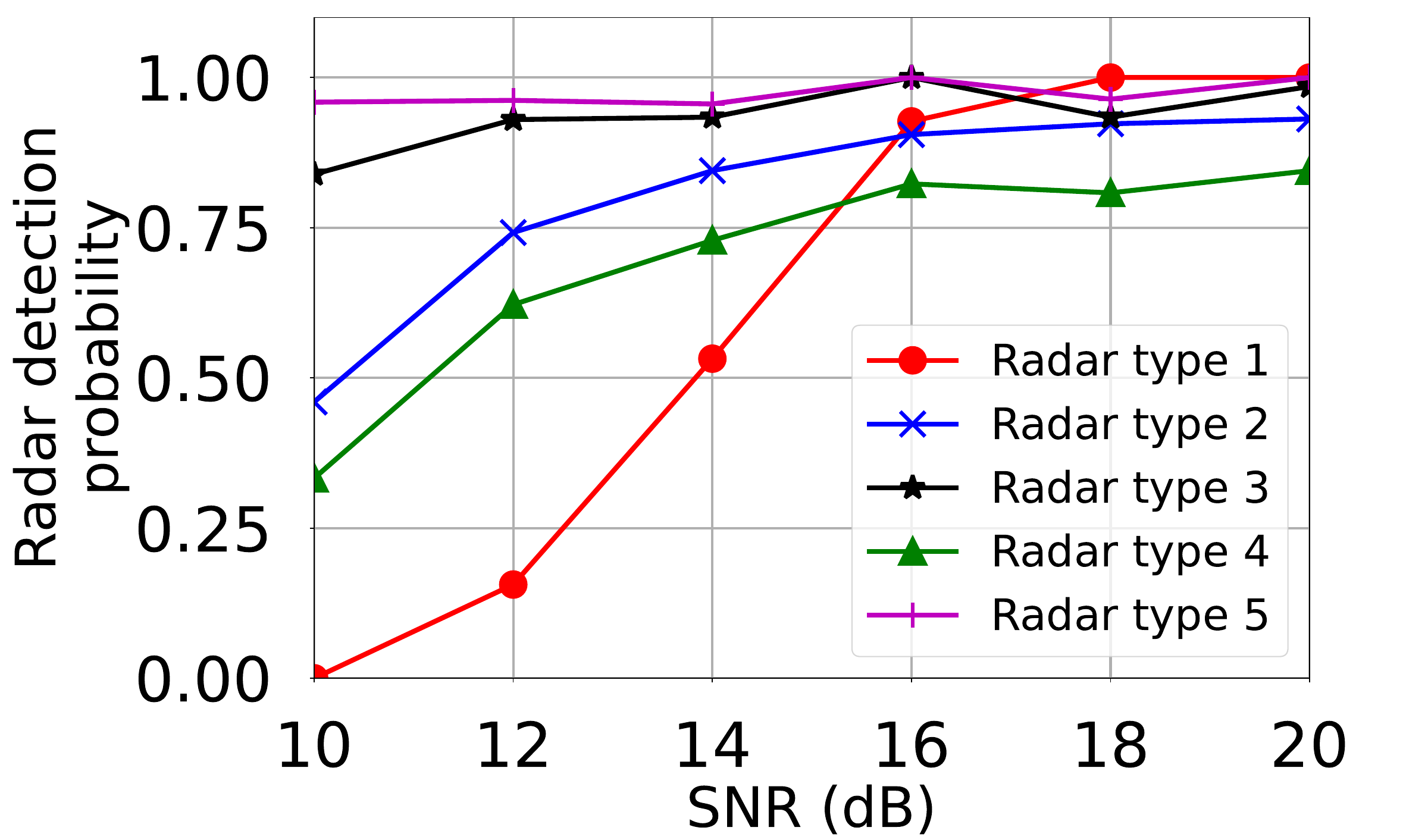}
    \caption{Radar detection accuracy of \YOLOCNN for different SNR. No interference is present, and false positive rate is $\approx$ 0\%.} 
    \label{fig:yolo_cnn_detection_rate}
\end{figure}

\textit{Capabilities of \YOLOCNN:}
The description of \YOLOCNN explains that it has several capabilities that we aimed for.
Recall from Section~\ref{section:intro} that one of our primary goals is to detect low SNR radar signals. In Section~\ref{subsection:spectro}, we chose the number of frequency bins in the spectrograms carefully to improve the detectability of radar signals. However, our experiments suggest that such a measure may not be sufficient for low SNR radar signals. Fig.~\ref{fig:yolo_cnn_detection_rate} shows the radar detection capability of \YOLOCNN for different radar SNR. The details of our experiments are presented later in Section~\ref{section:evaluations}.  We see from Fig.~\ref{fig:yolo_cnn_detection_rate} that as radar SNR reduces below 20 dB, the detectability of radar types 1, 2, and 4 is significantly degraded. To tackle this limitation of \YOLOCNN, we develop another strategy (the second flow in Fig.~\ref{fig:flow_diagran}) described in the following two sections.

\subsection{Wavelet Preprocessing} \label{subsection:wavelet_peprocess}
\begin{figure}[t]
    \centering
    \includegraphics[scale=0.37]{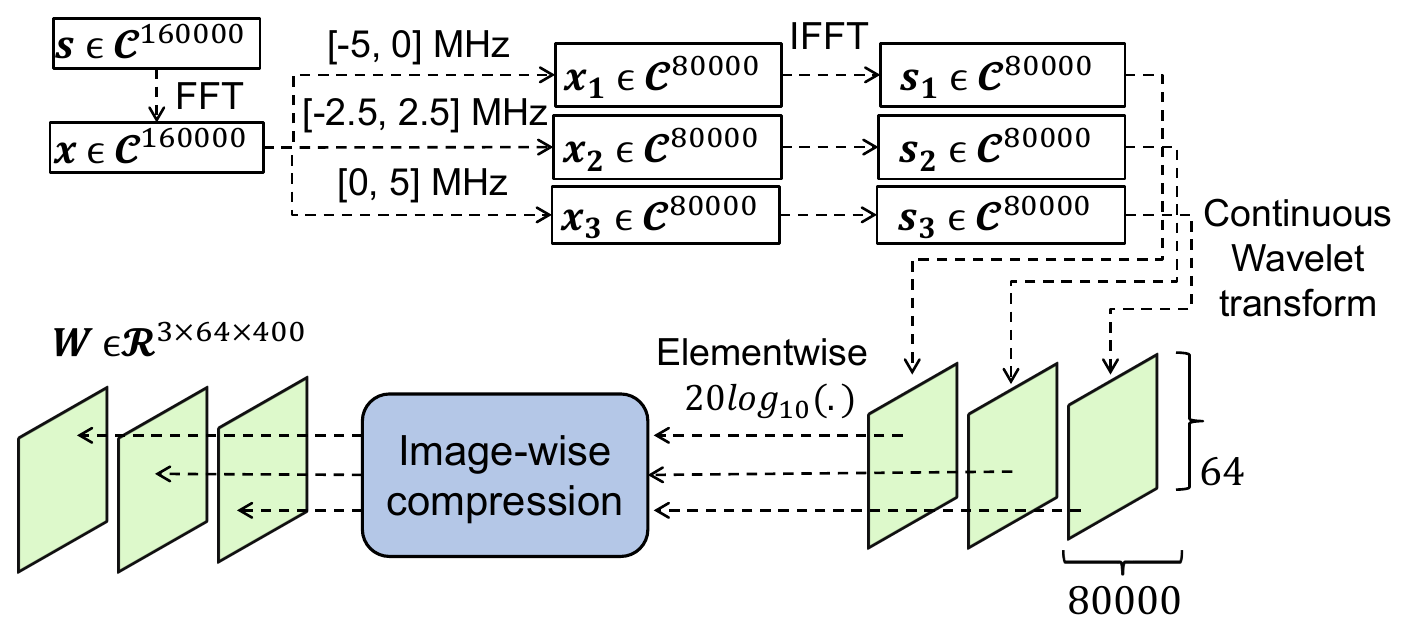}
    \caption{Operations inside the wavelet preprocessing block.} 
\label{fig:wavelet_preprocess}
\end{figure}

\textit{Intuition}: The idea of this block is to mimic the operation of matched filtering for improving radar SNR and, thus, its detectability. However, since the detector has no priori knowledge of the radar signal parameters, we design the steps in this block to overcome this problem. The input to this block is the same as the spectrogram preprocessing block, specifically the complex vector $\mathbf{s}$ that comprises of the I, Q values. The operations in this block are shown in Fig.~\ref{fig:wavelet_preprocess}.

Methods based on matched filtering do not take care of interference. Hence, we take three measures to reduce the impact of interference in the second flow of \sys. Out of the three, one is explained in this section, and the remaining two are in the following section.

\textit{Filtering}: The purpose of this step is the following. From Fig.~\ref{fig:yolo_cnn_detection_rate}, we see that \YOLOCNN's main limitation is with radar types 1, 2, and 4. Both radar types 1 and 2 have a fixed bandwidth of 1.6 MHz, and most of \YOLOCNN's misdetections for radar type 4 are for lower chirp width.
Hence, we use the filtering step to look at smaller sub-bands where that radar signal may reside and possibly improve the radar SNR in these sub-bands compared to the whole monitored band. We select the subbands to be overlapped as we do not know the radar center frequency. 

Another objective of the filtering step is to contrast radar signals from interference, which is our first measure to deal with interference in the second flow of \sys. Recall that the interference signals in our considered system model occupy the whole 10 MHz monitoring band. In contrast, radar types 1, 2, and 4 (main focus of \sys's second flow based on Fig.~\ref{fig:yolo_cnn_detection_rate}) occupy smaller bands. Hence, the sub-bands resulting from the filtering step will have dissimilar patterns on different subbands for radar signals but not for interference.

Now, we present the details of the filtering procedure. First, we perform $N$ point FFT on $\mathbf{s}$ to get a complex vector $\mathbf{x}$.
Next, we perform rectangular windowing on $\mathbf{x}$ to get three different complex vectors $\mathbf{x}_1$, $\mathbf{x}_2$, and $\mathbf{x}_3$. $\mathbf{x}_1$, $\mathbf{x}_2$, and $\mathbf{x}_3$ are the frequency domain representation of $\mathbf{s}$ over -5 to 0 MHz, -2.5 to 2.5 MHz, and 0 to 5 MHz, respectively, assuming that the sensor's monitoring band is -5 to 5 MHz. Then, we perform IFFT on $\mathbf{x}_1$, $\mathbf{x}_2$, and $\mathbf{x}_3$ to obtain $\mathbf{s}_1$, $\mathbf{s}_2$, and $\mathbf{s}_3$, respectively. Essentially, $\mathbf{s}_1$, $\mathbf{s}_2$, and $\mathbf{s}_3$ represent the bandpass filtered time domain signals of the original time domain signal $\mathbf{s}$. From the above description, we can see that the complexity of our filtering step increases with the monitoring bandwidth, $B$. For this reason, using $B = 10$ MHz is convenient for \sys as pointed out in Section~\ref{section:system_model}.

\textit{Wavelet transform}: Next, for each bandpass-filtered signal, we perform a CWT, which can be thought of as an equivalent of time-lagged correlation (convolution) of the signal and the filter impulse response.
Since we do not know the ideal filter impulse response, we need to find a suitable mother Wavelet function that can approximate the filter impulse response. Additionally, since we do not know the radar signal frequency, the transform must be computed for different frequencies, i.e., for different scale parameters in the Wavelet transform. Now, we explain our Wavelet transform procedure for $\mathbf{s}_i$, which is repeated for $i=1, 2, 3$.

In CWT, first, we select a mother wavelet function, $\Psi(t)$. Then, we correlate $\mathbf{s}_i(t)$ with $\Psi(\frac{t - \delta}{s})$ for different values of $\delta$ and $s$, represented by the following equation~\cite{mallat1999wavelet}:
\begin{equation} \label{eq:wavelet_transform}
    \mathbf{W}_i^{u} (\delta, s) = \frac{1}{|\sqrt{s}|} \int_{-\infty}^{\infty} \mathbf{s}_i(t) \Psi^* (\frac{t - \delta}{s}) dt
\end{equation}
where $\delta$ is the time lag parameter and $s$ is the scale factor. Lower values of $s$ correspond to compressed versions of the mother wavelet and extract high-frequency information. Higher values of $s$ correspond to expanded versions of the mother wavelet and extract low-frequency information.
Since $\mathbf{W}_i^{u}$ has two parameters, the output of the CWT can be viewed as a matrix of size $L \times S$, where the rows correspond to different lag parameters and the columns correspond to different scale parameters. We leverage this structure of $\mathbf{W}_i^{u}$ in the Wavelet-CNN block of \sys. We investigate the suitability of various mother Wavelet functions and choose the complex Morlet function, given below~\cite{ball2008low}, based on its similarity with radar pulse shapes.
\begin{equation} \label{eq:cmorlet}
    \Psi(t) = \frac{1}{\sigma \sqrt{2 \pi}} \exp\Big[-\frac{1}{2} \Big(\frac{t}{\sigma}\Big)^2 \Big] \exp(j2\pi f_0 t)
\end{equation}
Here $\sigma$ controls the bandwidth of $\Psi(t)$ and $f_0$ controls its frequency.~\eqref{eq:cmorlet} represents a complex exponential with a Gaussian envelope. In \sys we use $f_0 = 10$ MHz and $\sigma$ such that the bandwidth of $\Psi(t)$ is 1.5 MHz. Our choices of $f_0$ and $\sigma$ are based on the sensor's monitoring bandwidth, $B$, and
radar bandwidth (primarily type 1 and 2). 

\textit{Dimensions of $\mathbf{W}_i^{u}$}: The size of $\mathbf{W}_i^{u}$ is $L \times S$. Since the size of $\mathbf{s}_i$ is 80000
we use 80000 different lag parameters in CWT. Hence, we have $L= 80000$. For the scale parameter, we use 64 different values that are uniformly spaced in logarithmic scale in the range $[\log_{10} 0.5,\log_{10} 64]$, which covers the frequencies relevant to our sensor. Thus, we have $S = 64$.

\textit{Compression of $\mathbf{W}_i^{u}$}: As mentioned before, the matrices $\mathbf{W}_i^{u}$; $i = 1, 2, 3$ are fed to a CNN. However, with $\mathbf{W}_i^{u}$ having a dimension of $80000 \times 64$ complicates the design of the CNN architecture. Hence, we apply a compression technique on $\mathbf{W}_i^{u}$; $i = 1, 2, 3$. This compression strategy, denoted as `Imagewise compression' in Fig.~\ref{fig:wavelet_preprocess}, is similar to the compression technique described in Section~\ref{subsection:spectro}. Specifically, we reshape $\mathbf{W}_i^{u}$ of size $80000 \times 64$ to $400 \times 200 \times 64$. Then for each of the $200 \times 64$ matrices, we retain the column-wise maximum value, resulting in a compressed version of $\mathbf{W}_i^{u}$. Let us denote the compressed version of $\mathbf{W}_i^{u}$ as $\mathbf{W}_i \in \mathcal{R}^{400 \times 64}$. For a particular scale, collapsing 200 consecutive values along the time lag dimension in $\mathbf{W}_i^{u}$ can be justified in a similar manner as done in the context of $\mathbf{X}_u$. 200 consecutive values along the time lag dimension correspond to $200 \times 0.1 \mu$sec (inter-sample duration) 
$= 20 \mu$sec, which is much smaller than the radar inter-pulse intervals (refer to Table~\ref{tab:radar_types}). 

\begin{figure}[t]
    \centering
    \includegraphics[scale=0.25]{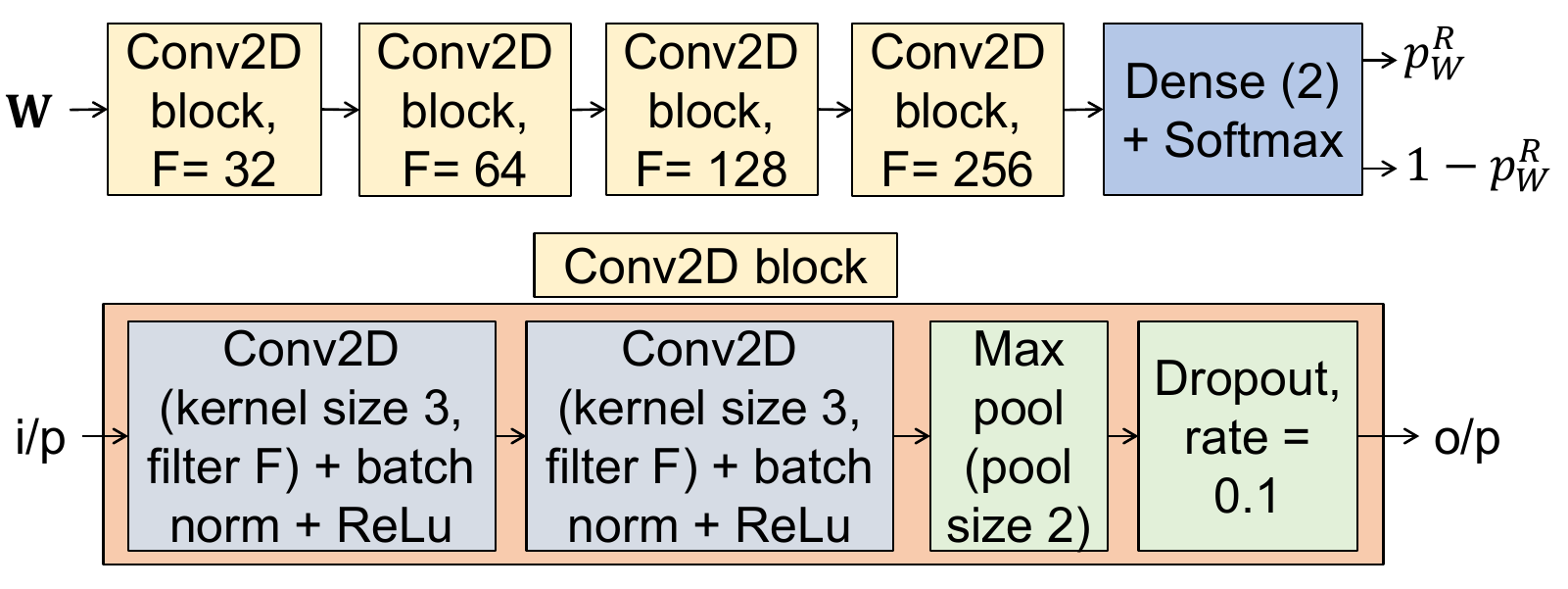}
    \caption{Neural network architecture of \WaveletCNN.} 
\label{fig:wavelet_cnn}
\end{figure}
\subsection{Wavelet-CNN}\label{subsection:wavelet_cnn}
% ...
As discussed in the previous section, the Wavelet preprocessing block tries to mimic the operation of matched filtering. Since the filter impulse response is unknown, we perform the computation in~\eqref{eq:cmorlet} for different values of $s$. However, we still need to make the detection decision after the approximated matched filtering step. Based on the computed Wavelet transforms, we decide whether a radar signal is present or not. For that, we use a CNN as described next.

The input to \WaveletCNN is the tensor $\mathbf{W} = [\mathbf{W}_1, \mathbf{W}_2, \mathbf{W}_3]$, whose dimension is $3 \times 400 \times 64$, as shown in Fig.~\ref{fig:wavelet_cnn}. The matrices in $\mathbf{W}$ are similar to the spectrograms discussed in Section~\ref{subsection:spectro}. However, the primary difference is that STFTs can represent high resolution in either time or frequency, whereas the matrices in $\mathbf{W}$ (the CWTs) represent high-resolution information in both time and frequency domains. Hence we use $\mathbf{W}$ as the input features to \WaveletCNN. The neural network acts as a function that performs the following mapping: $\mathcal{F}: \mathcal{R}^{3 \times 400 \times 64} \rightarrow \mathcal{R}^2$, where the input to $\mathcal{F}$ is $\mathbf{W}$ and output is the tuple $(\hat{p}_W^R, 1 - \hat{p}_W^R)$. Here $\hat{p}_W^R$ is the predicted probability of the presence of radar signal. Clearly, $\mathcal{F}$ acts as a binary classifier.

\textit{Training procedure:} During training, we learn the function $\mathcal{F}$ by minimizing the binary cross-entropy loss function, $\mathcal{L}_W = \sum_{b \in \mathcal{B}} p_{W,b}^R \times \log_2(\hat{p}_{W,b}^{R}) + (1 - p_{W,b}^R) \log_2(1 - \hat{p}_{W,b}^R)$, using the Adam optimizer.
$p_{W,b}^R$ and $\hat{p}_{W,b}^{R}$ are the true and predicted probabilities of the presence of radar in training example $b$. After training, we extract the following parameter for the prediction phase.
    
$p_{W}^{R,true}$: We pass all the training examples through the trained CNN. We note the predicted radar probability $\hat{p}_{W,b}^R$ for all the examples where radar is present (based on ground truth) and form the set $\mathcal{P}_{W}^{R, true}$. Finally, we find the $g^{th}$ percentile of $\mathcal{P}_{W}^{R, true}$ and denote it as $p_{W}^{R,true}$.

\textit{Prediction procedure:} During prediction, we pass the input tensor $\mathbf{W}_b$ through the trained CNN, $\mathcal{F}$ and note the predicted radar probability, $p_{W,b}^{R}$. Then, we compare $p_{W,b}^{R}$ to a threshold, $t_w$, and declare the presence of radar only if $p_{W,b}^{R} \geq t_w$. We select the threshold as $p_{W}^{R,true}$. Instead of just declaring the presence of radar if $p_{W,b}^{R} \geq 0.5$, we use the thresholding operation because of the following reason. Since we do not train \WaveletCNN to differentiate between radar and interference, the network may predict $p_{W,b}^{R}$ to be greater than 0.5, but not necessarily very high, when only the interference is present. In such cases, we will have false alarms. Hence, we must choose a non-zero but small value of $g$ in the definition of $p_{W}^{R,true}$. This is our second measure for tackling interference in \sys's second flow. The third measure is the following.

We assume that the sensor is aware of the interference signal's center frequency and introduce a slight frequency offset $\Delta_{CF}$ at the sensor with respect to the interference signal. The center frequency offset will introduce distortions in the digitally modulated interference signals and appear as noise to the subsequent signal processing steps. 

\textit{Parameter estimation:} \WaveletCNN cannot estimate signal parameters. We develop a strategy to partially address this limitation. Our strategy is to reuse the neural network output of \YOLOCNN, $\mathbf{P}$, but apply a different post-processing technique.
Recall from Section~\ref{subsection:yolo_cnn} that during prediction, \YOLOCNN first decides whether an object is present or not. Then, it performs object localization ($\hat{x}_i, \hat{y}_i, \hat{w}_i, \hat{h}_i$) only if an object has been detected. Once we are into the second flow of \sys, it is evident that no radar object was detected by \YOLOCNN. However, if \WaveletCNN predicts the presence of a radar signal, we can override the object detection decision of \YOLOCNN.  Overriding the object detection decision of \YOLOCNN implies using an object detection threshold, say $t_o^w$, that is different from $t_o$ (refer to Fig.~\ref{fig:yolo_cnn}). $t_o^w$ must be lower than $t_o$; otherwise, no radar object would be detected, as was the case with \YOLOCNN in the first place. Note that by using a lower object detection threshold, we are not affecting \sys's radar false alarm rate as the decision regarding the presence of radar has already been made by \WaveletCNN. However, a very low value for $t_o^w$ may cause many false object detections and adversely affect the radar parameter estimation quality. Based on these factors, we choose $t_o^w$ to be $c_{R,O}^{min}$. Recall from Section~\ref{fig:yolo_cnn} that $c_{R,O}^{min}$ considers the minimum confidence of all the radar objects on a spectrogram, whereas $c_{R,O}^{max}$ (used in $t_o$) considers the maximum. 
Using $t_o^w$, we perform the object detection and localization on $\mathbf{P}$ as shown in Fig.~\ref{fig:yolo_cnn} with the only difference that the procedure is applied only to radar class as \WaveletCNN only impacts radar detection.
\begin{table*}[t] 
\vspace{-5mm}
\caption{Evaluations metrics} 
\label{tab:eval_metrics}
\resizebox{6.5in}{!}{
    {
    \begin{tabular}
    {|c|c|}
        \hline
        Metric & Definition \\
        \hline
        $p_c^{R}$ & Binary classification accuracy between radar and non-radar (both with and without interference)  \\ 
        \hline
        $p_d^{R}$ & Fraction of examples with radar that is correctly detected; radar true positive rate \\
        \hline
        $p_f^{R}$ & Fraction of examples with no radar but falsely detected; radar false positive rate \\
        \hline
        $p_c^{I}$ & Binary classification accuracy between interference and non-interference (with and without radar) \\
        \hline
        $p_d^{I}$ & Fraction of examples with interference that is correctly detected; interference true positive rate\\
        \hline
        $p_f^{I}$ & Fraction of examples without interference but falsely detected; interference false positive rate \\
        \hline
        $b_M^R$ & Average radar missed bandwidth (for examples where radar is correctly detected and pulse parameters estimated) \\
        \hline
        $b_E^R$ & Average radar bandwidth estimated in excess (for examples where radar is correctly detected and pulse parameters estimated) \\
        \hline
        $n_P^R$ & Average percentage of detected radar pulses (for examples where radar is correctly detected and pulse parameters estimated) \\
        \hline
        $e_{PW}^R$ & Average pulse width absolute error (for examples where radar is correctly detected and pulse parameters estimated) \\
        \hline
        $e_{PI}^R$ & Average pulse interval estimation absolute error (for examples where radar is correctly detected and pulse parameters estimated) \\
        \hline
        $t_M^I$ & Average interference missed ON time (for examples where interference is correctly detected and parameters estimated)\\
        \hline
        $t_E^I$ & Average interference ON time estimated in excess (for examples where interference is correctly detected and parameters estimated) \\
        \hline
    \end{tabular}
    }
    }  
\end{table*}

\section{Evaluations} \label{section:evaluations}
In this section, first, we describe the datasets and experiments. Next, we present the evaluation metrics and the baseline methods. Finally, we present the evaluation results.

\subsection{Datasets}
\textbf{Radar}: For radar signals, we rely on a dataset generated synthetically by NIST~\cite{caromi2019rf}. This dataset provides several captures, each of 80 msec, in the form of I, Q values. The captures correspond to a 10 MHz band. Almost half of the captures have no radar signal (receiver noise only), and the remaining ones with radar. Each of the captures containing a radar has at most one radar signal, chosen randomly from the five radar types listed in Table~\ref{tab:radar_types}. The radar parameters are randomly chosen from the ranges specified in Table~\ref{tab:radar_types}. The SNR of the radar signals is chosen randomly from $[10, 12, 14, 16, 18, 20]$ dB, and no interference signal is present in the captures. 
For our evaluations, we use 9000 captures with 50:50 split between radar and noise. 

\textbf{Interference}: For evaluating \sys in interference, we generate several interference datasets. As discussed in Section~\ref{section:system_model}, the interference signals are assumed to be downlink signals from a BS. For the following datasets, INR is defined as the `average interference-plus-noise to average noise ratio' over a band of 1 MHz around the interference signal's center frequency. This is done so that the SINR values can be easily computed and to make the SINR values meaningful. (Recall from Section~\ref{section:intro} that radar SNR values are also defined similarly).

\textit{QPSK ON dataset}: Using MATLAB, we generate 2000 captures of QPSK signals with a bandwidth of 9.1 MHz and a center frequency offset of $\Delta_{CF}$ = 0.35 MHz with respect to the sensor's center frequency. Each capture is 80 msec, and the QPSK signal is always ON within one capture. The QPSK signal changes at the symbol rate.
The INR is randomly chosen from [2, 4, 6, 8, 10] dB across captures but is kept constant within a capture. 
    
\textit{QPSK ON-OFF dataset}: Similar to QPSK ON dataset but the QPSK signal turns ON for 3 msec and then off for 2 msec. One such ON-OFF pattern is shown in Fig.~\ref{fig:yololet_if_obj}.
    
\textit{LTE FDD dataset}: Using MATLAB LTE toolbox~\cite{matlabLTEToolbox}, we generate 2000 captures LTE downlink captures that occupy 50 resource blocks (9 MHz). This dataset's frequency offset, capture duration, and INR values are similar to other datasets. For this dataset, we use LTE frequency division duplexing (FDD) mode~\cite{matlabTDDFDD}, where BSs and UEs use different frequencies.
    
\textit{LTE TDD dataset}: This dataset is similar to the above dataset, but we use LTE time division duplexing (TDD)~\cite{matlabTDDFDD}, where BSs and UEs use the same band for their transmissions but take turns in multiples of LTE slot duration (1 msec) defined via the uplink/downlink (UL/DL) configurations. For each capture, we randomly choose one of seven possible UL/DL configurations~\cite{matlabTDDFDD}.

\subsection{Experiments}\label{subsection:experiments}
Using the above datasets, we conduct three experiments.

\textbf{Experiment 1}: We create a training dataset of 2500 radar, 2500 AWGN (receiver noise), 2500 interference, and 2500 radar plus interference captures. We randomly select between QPSK ON and QPSK ON-OFF for the interference captures. For the radar plus interference captures, we simply add the I,Q values of radar and interference (randomly chosen between QPSK ON and QPSK ON-OFF) while ensuring that receiver noise is not added twice. For each of the captures, we select a 10 msec long capture from the 80 msec captures in the datasets. The radar and interference start time within the 10 msec capture are randomized. We train different methods (described later in this section), including \sys, using these 10,000 examples. For all the deep learning methods, we use the Keras~\cite{chollet2015keras} framework. 
% ...
We evaluate the trained models using various metrics (explained later) on the test set. The test set has 4000 examples, almost half radar and the remaining AWGN. Note that the test set has no interference. 
    
\textbf{Experiments 2A and 2B}: We use the models trained in experiment 1, but
the test sets are different. For both experiments 2A and 2B, the test set contains 4000 examples, with almost half of them radar plus interference and the remaining interference only. The interference signals are QPSK ON and QPSK ON-OFF for experiments 2A and 2B, respectively.
    
\textbf{Experiment 3}: This experiment is similar to experiment 1, but in this case, we use the LTE FDD and TDD interference signals for the training instead of QPSK interference.
    
\textbf{Experiments 4A and 4B}: These experiments are similar to experiments 2A and 2B, but we use the models trained as part of experiment 3. The interference signals are LTE FDD and LTE TDD for experiments 4A and 4B, respectively.

\begin{figure}[t]
    \centering{
    \hspace{-7mm}
    \begin{subfigure}[Experiment 1, AWGN]{
    \label{fig:tp_fp_all_methods_awgn_tain_qpsk}
    \includegraphics[scale=0.14]{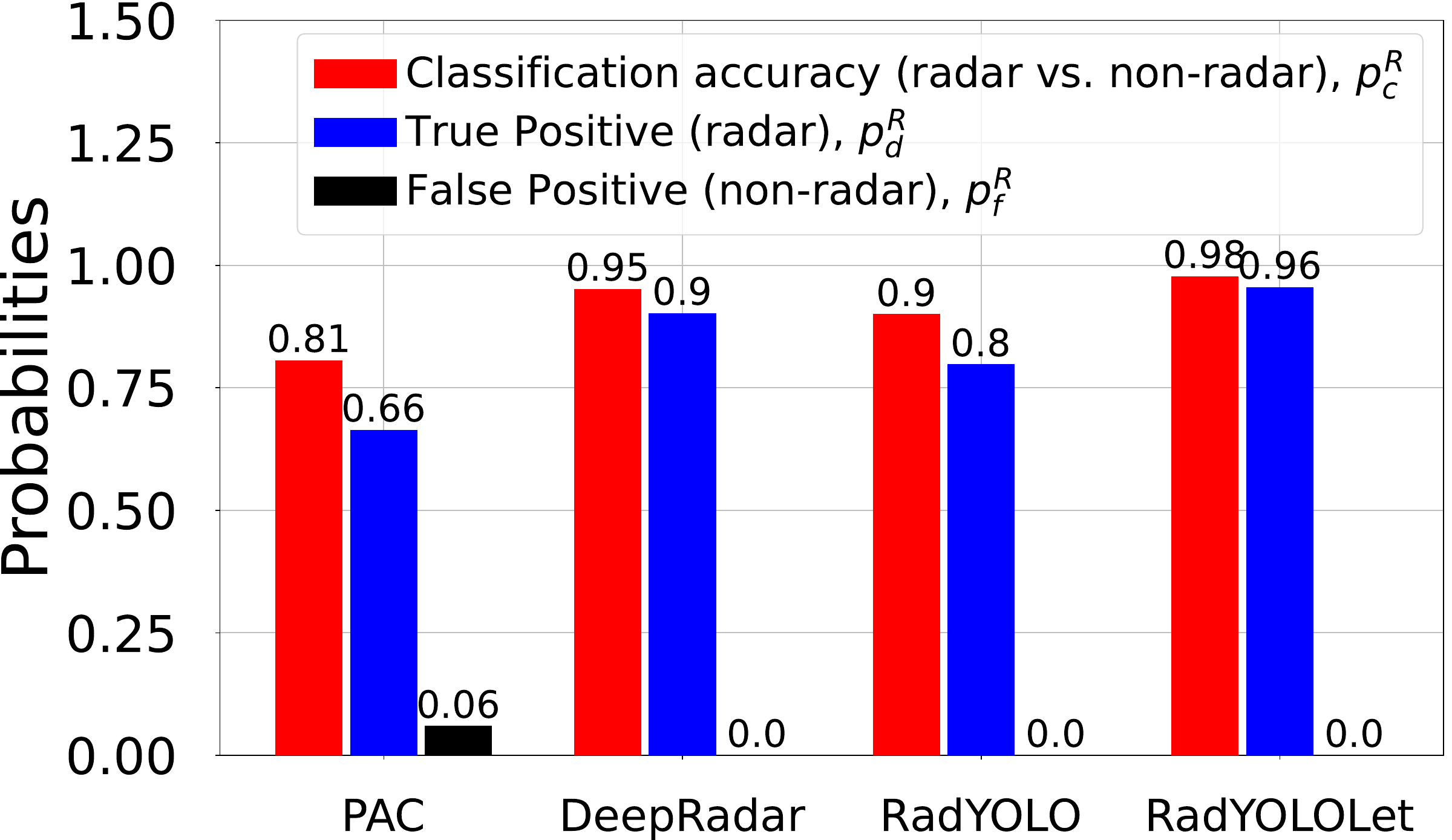} }  
    \end{subfigure}
    \hspace{6mm}
    \begin{subfigure}[Experiment 3, AWGN]{
    \label{fig:tp_fp_all_methods_awgn_tain_lte}
    \includegraphics[scale=0.14]{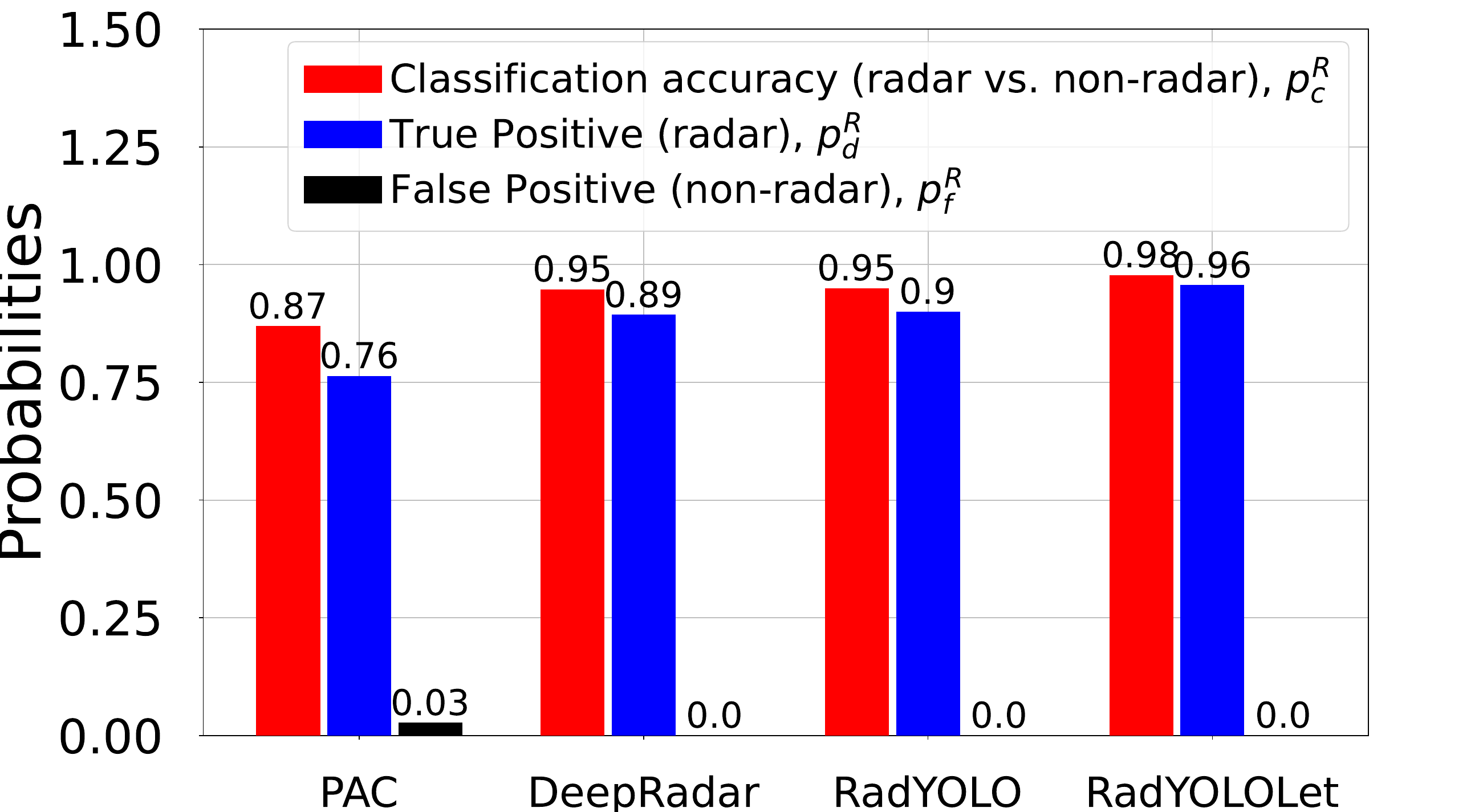}   }
    \end{subfigure}
    }
    \caption{Binary classification between radar and non-radar signals, which for these plots is AWGN, for different methods.}
    \label{fig:tp_fp_all_methods_awgn}
\end{figure}

\subsection{Metrics}
We evaluate \sys and compare it with other methods using the metrics in Table~\ref{tab:eval_metrics}.

\subsection{Methods for comparison}
In this section, we describe the methods that we use to compare \sys's performance.

\textbf{Peak analysis classifier (PAC)}~\cite{caromi2018detection}: This method computes the following features and uses them for training an SVM-based binary classifier. 
The features are mean, variance, maximum of the time intervals between the peaks of the amplitude, and the mean amplitude of the peaks of the captured signal.
This method can only distinguish between radar and non-radar signals. Hence, it cannot detect interference signals, and also cannot estimate the signal parameters. 
    
\textbf{DeepRadar}~\cite{sarkar2021deepradar}: This method treats all the radar pulses on a spectrogram as a single object and applies YOLO for detecting and localizing those objects. As a result, this method can estimate radar center frequency and bandwidth but cannot estimate the temporal parameters. For a fair and meaningful comparison with \sys, we make some modifications in DeepRadar. First, DeepRadar considers 100 MHz monitoring bandwidth and uses multiple grid cells in YOLO along the frequency axis of the spectrograms. However, in this paper, we consider the monitoring bandwidth to be 10 MHz. Accordingly, we use only one grid cell in DeepRadar's YOLO along the frequency axis of the spectrograms. Second, we apply the preprocessing proposed in this paper also to the spectrograms fed to DeepRadar.

\textbf{\YOLOCNN}: This is simply our proposed scheme, but without using the second flow of Fig.~\ref{fig:flow_diagran}.

\subsection{Results}

\subsubsection{Radar vs. non-radar classification in AWGN} \label{subsection:binary_classification_radar_awgn}
Using experiments 1 and 3, we compare the classification (radar versus AWGN) accuracy of different methods in Fig.~\ref{fig:tp_fp_all_methods_awgn}. The results are combined for SNR range $[10-20]$ dB and all radar types. We make the following observations. 

\begin{figure}[t]
    \centering
    \hspace{-2.5mm}
    \begin{subfigure}[PAC]
    {\label{fig:pac_pd}
    \includegraphics[scale=0.122]{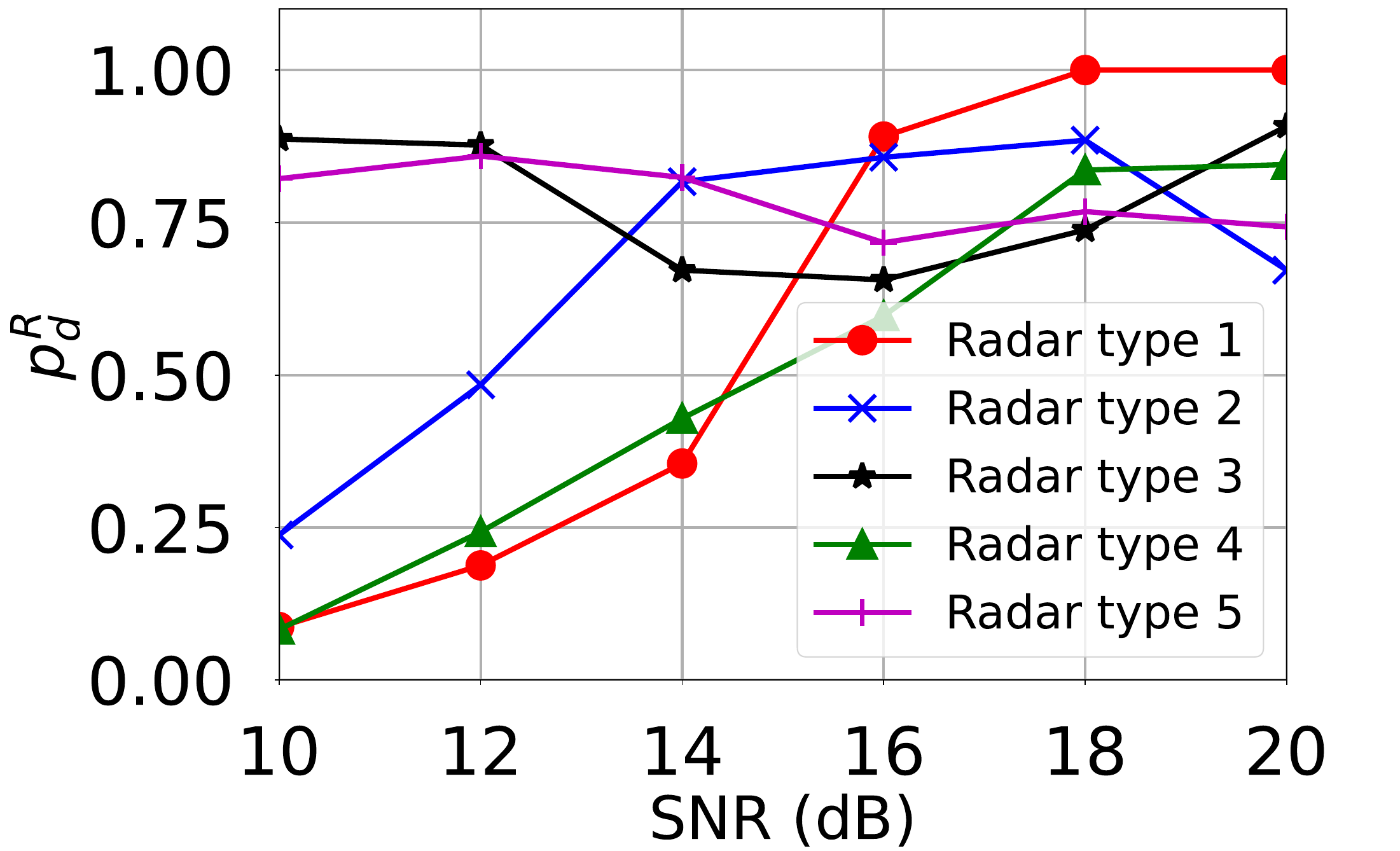}}
    \end{subfigure} 
    \hspace{-7.5mm}
    \begin{subfigure}[DeepRadar]
    {\label{fig:deepradar_pd}
    \includegraphics[scale=0.122]{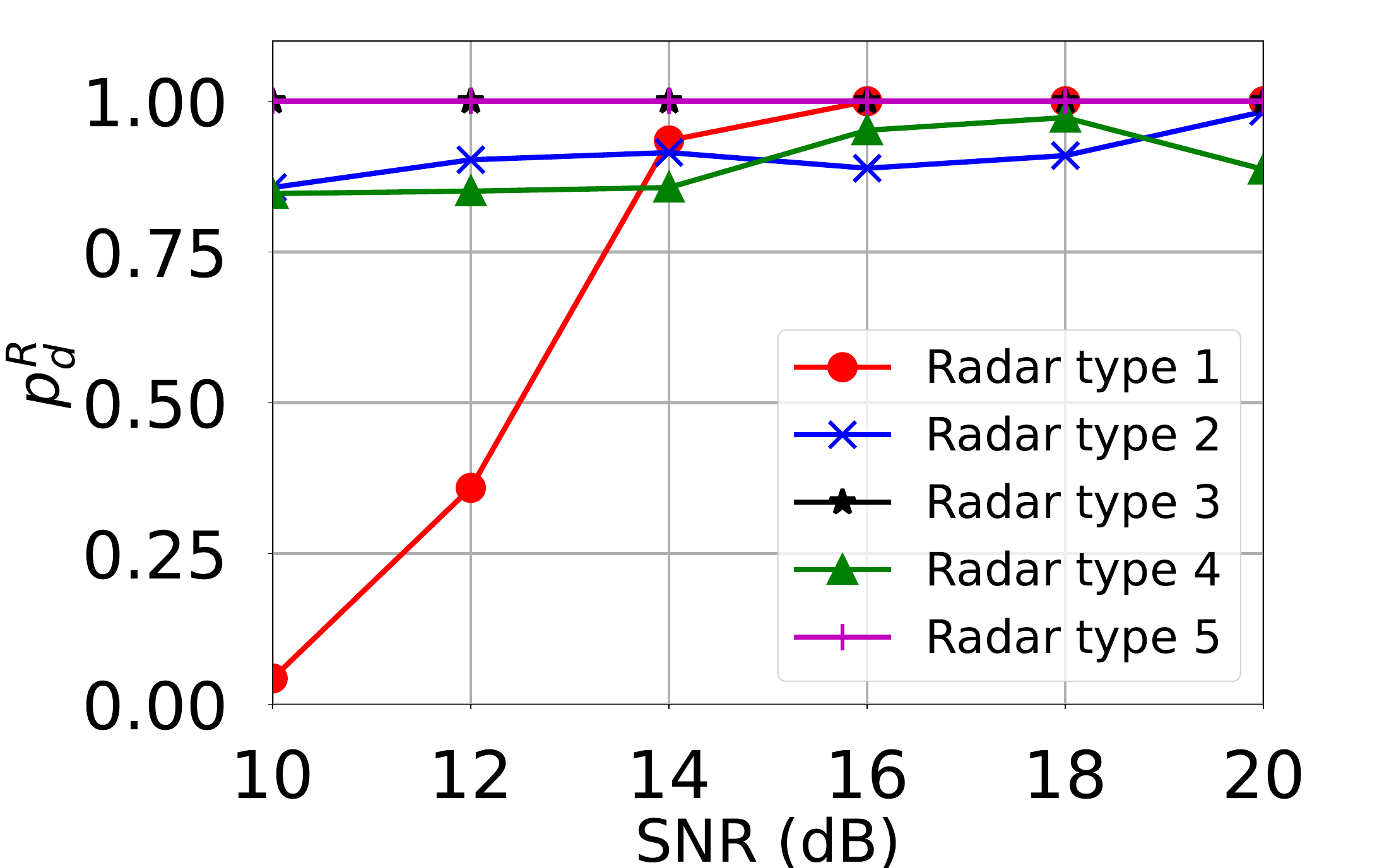}}
    \end{subfigure} 
    \hspace{-2.5mm}
    \begin{subfigure}[\sys]
    {\label{fig:yololet_pd}
    \includegraphics[scale=0.122]{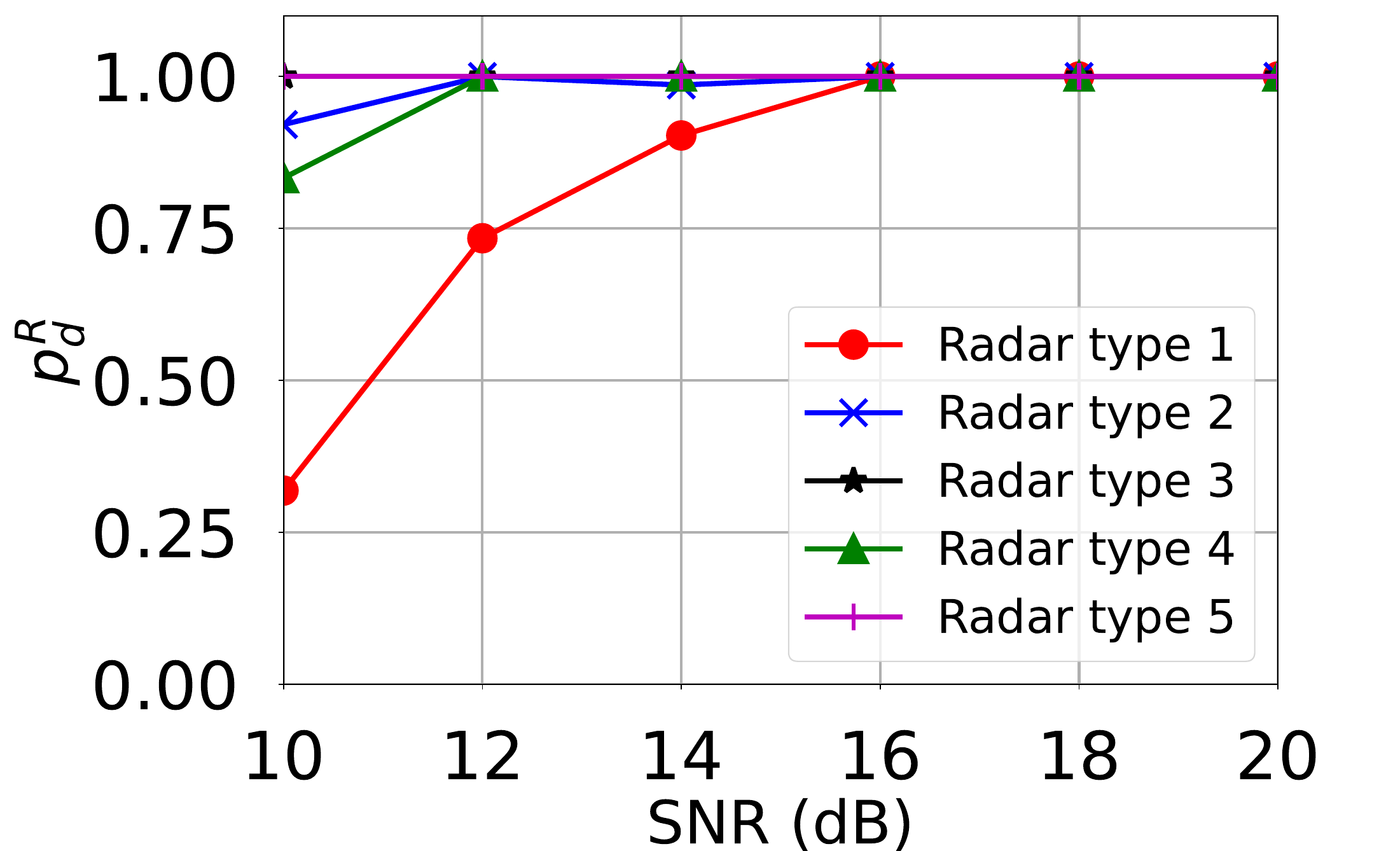}}
    \end{subfigure} 
    \hspace{-7.5mm}
    \begin{subfigure}[\YOLOCNN vs. \sys]{\label{fig:yololet_pd_improve}
    \includegraphics[scale=0.123]{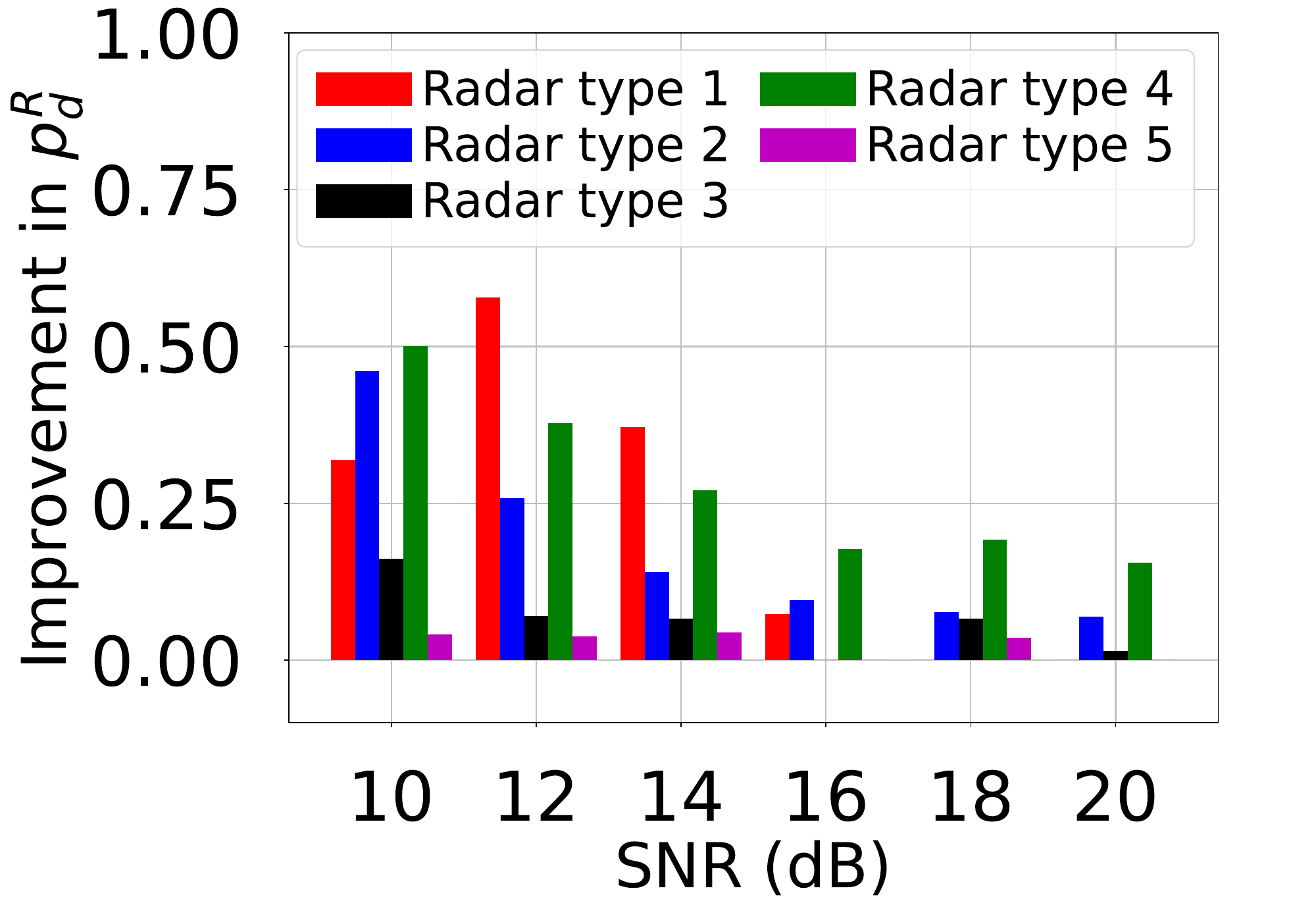}}
    \end{subfigure} 
    \caption{Comparison of different methods in terms of radar detection accuracy, $p_d^{R}$, for fixed AWGN power but varying radar signal power. 
    } 
    \label{fig:radar_detection_per_snr}
\end{figure}
\begin{figure*}[t]
    \centering
    \hspace{-2mm}
    \begin{subfigure}[Expt. 2A, QPSK ON interference]
    {\label{fig:tp_fp_qpsk_on}
    \includegraphics[scale=0.135]{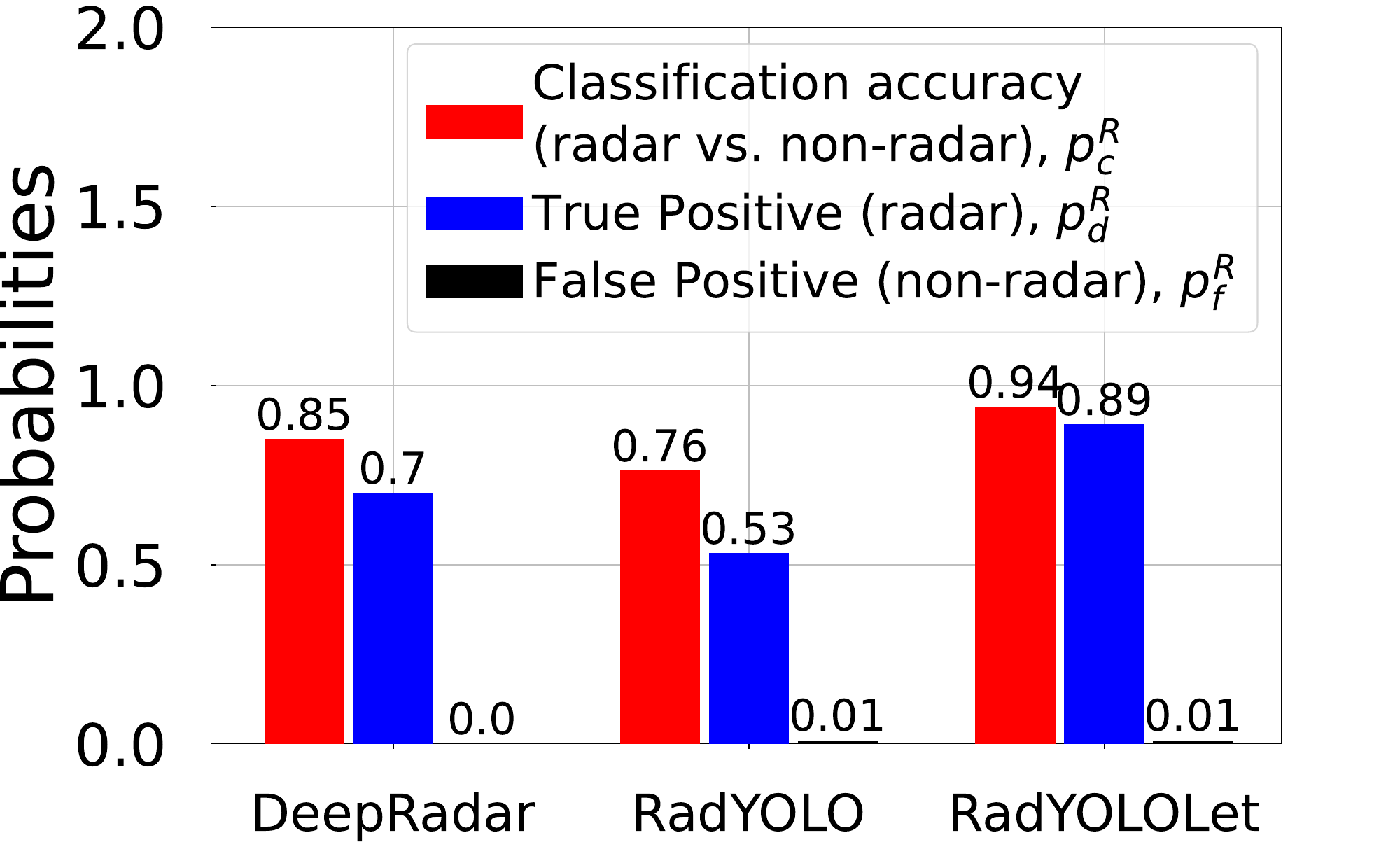}}
    \end{subfigure} 
    \hspace{-4.5mm}
    \begin{subfigure}[Expt. 2B, QPSK ON-OFF interference]
    {\label{fig:tp_fp_qpsk_on_off}
    \includegraphics[scale=0.138]{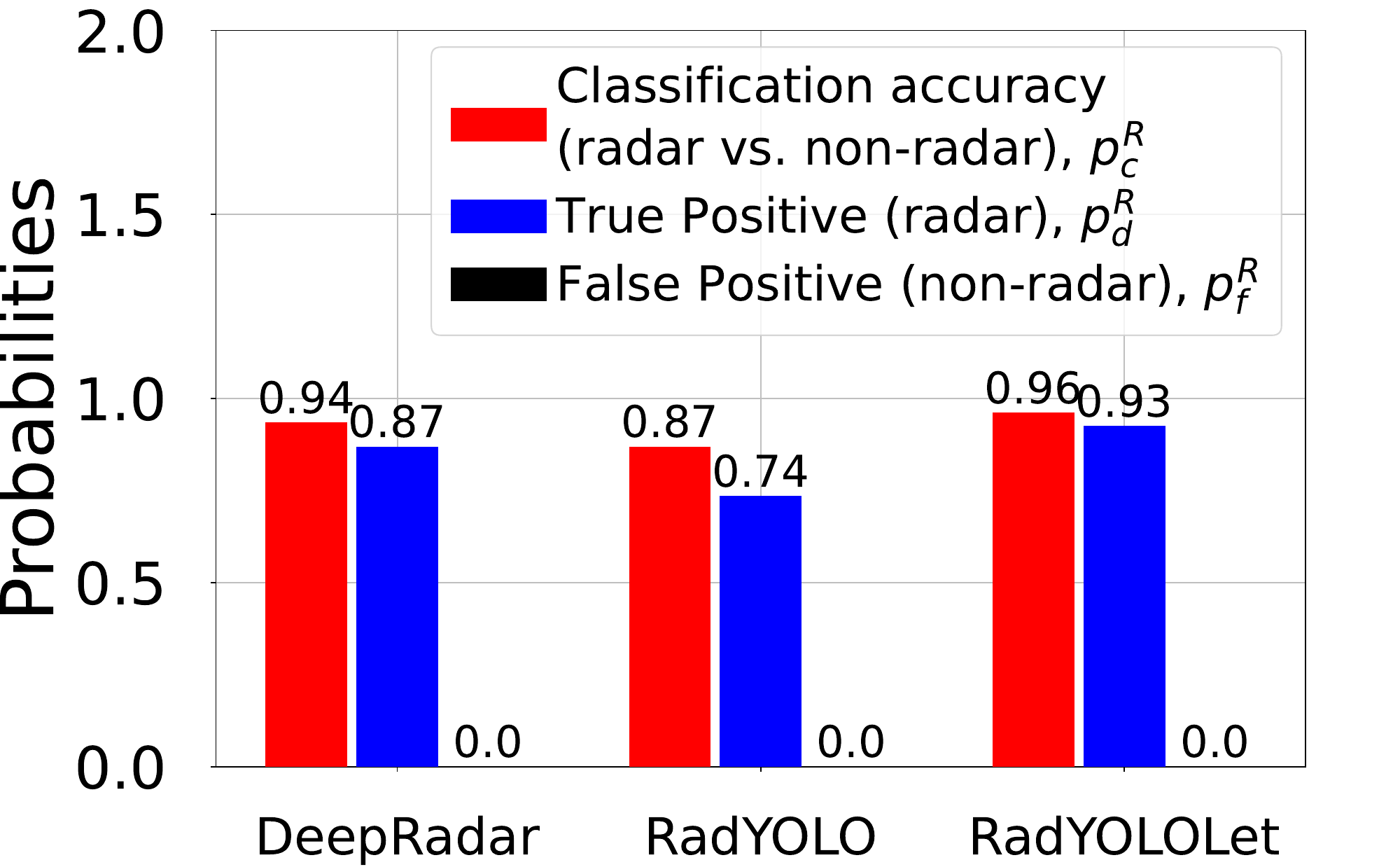}}
    \end{subfigure} 
    \hspace{-5mm}
    \begin{subfigure}[Expt. 4A, LTE FDD interference]
    {\label{fig:tp_fp_lte_on}
    \includegraphics[scale=0.134]{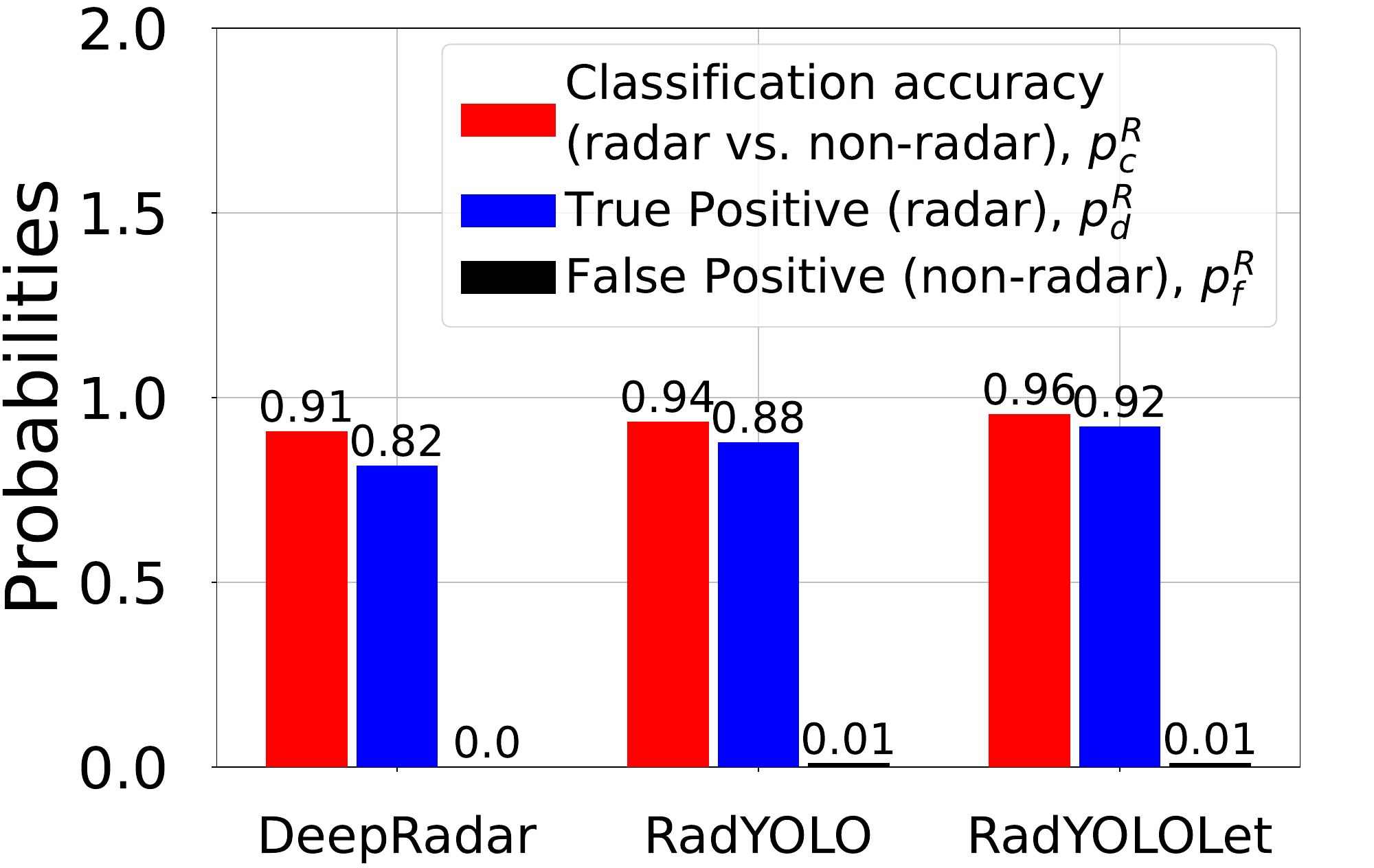}}
    \end{subfigure} 
    \hspace{-7mm}
    \begin{subfigure}[Expt. 4B, LTE TDD interference]
    {\label{fig:tp_fp_lte_on_off}
    \includegraphics[scale=0.134]{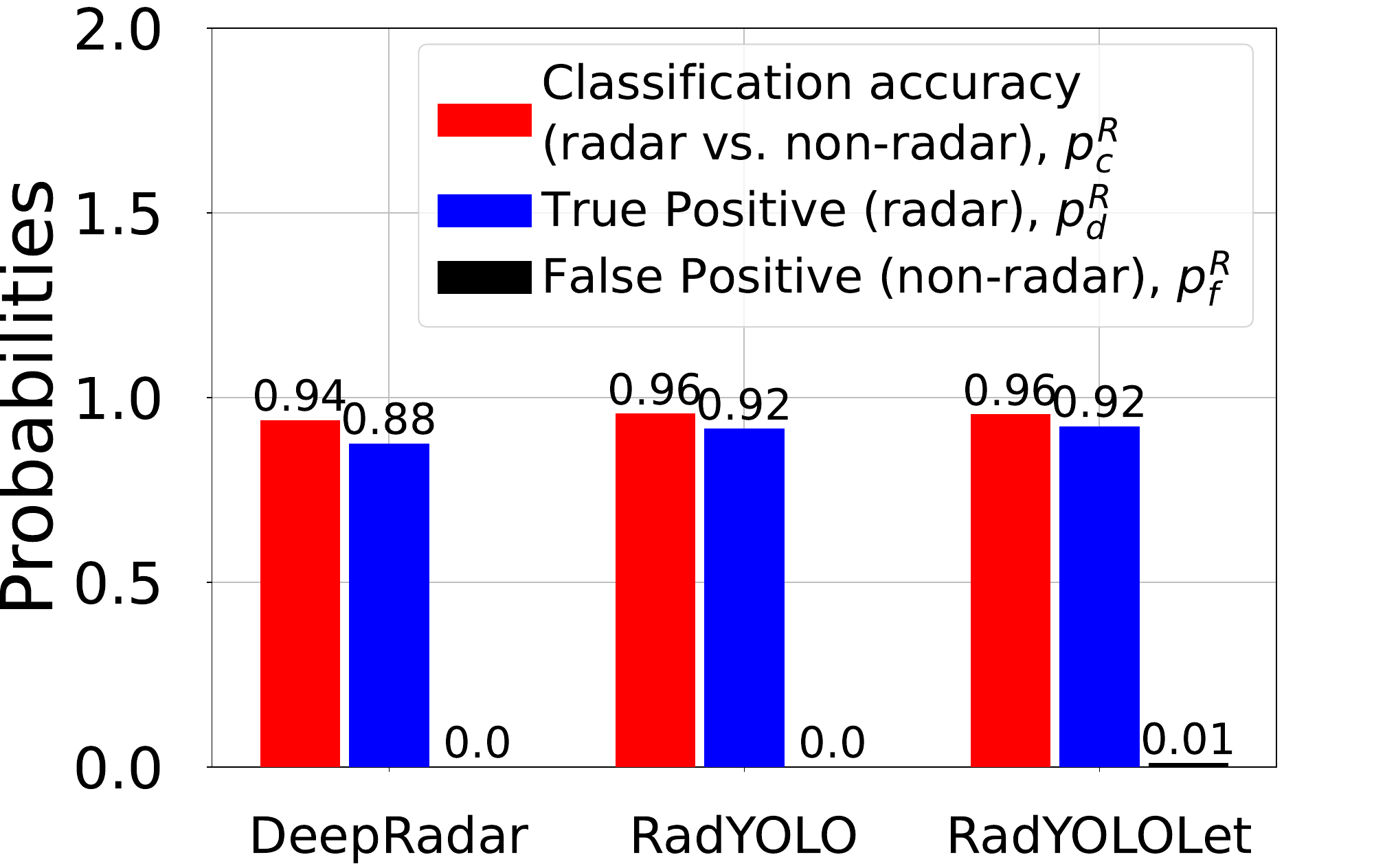}}
    \end{subfigure} 
    \caption{Classification between radar and non-radar for different methods. Both radar and non-radar signals have interference on top of the noise floor. Results are combined for SNR range [10-20] dB and INR range [2-10] dB, i.e., SINR range [0-18] dB.}
    \label{fig:pd_in_interference}
\vspace{3mm}
\end{figure*}
\begin{figure*}[t]
    \centering
    \begin{subfigure}[QPSK interference]
    {\label{fig:tp_fp_qpsk_if}
    \includegraphics[scale=0.134]{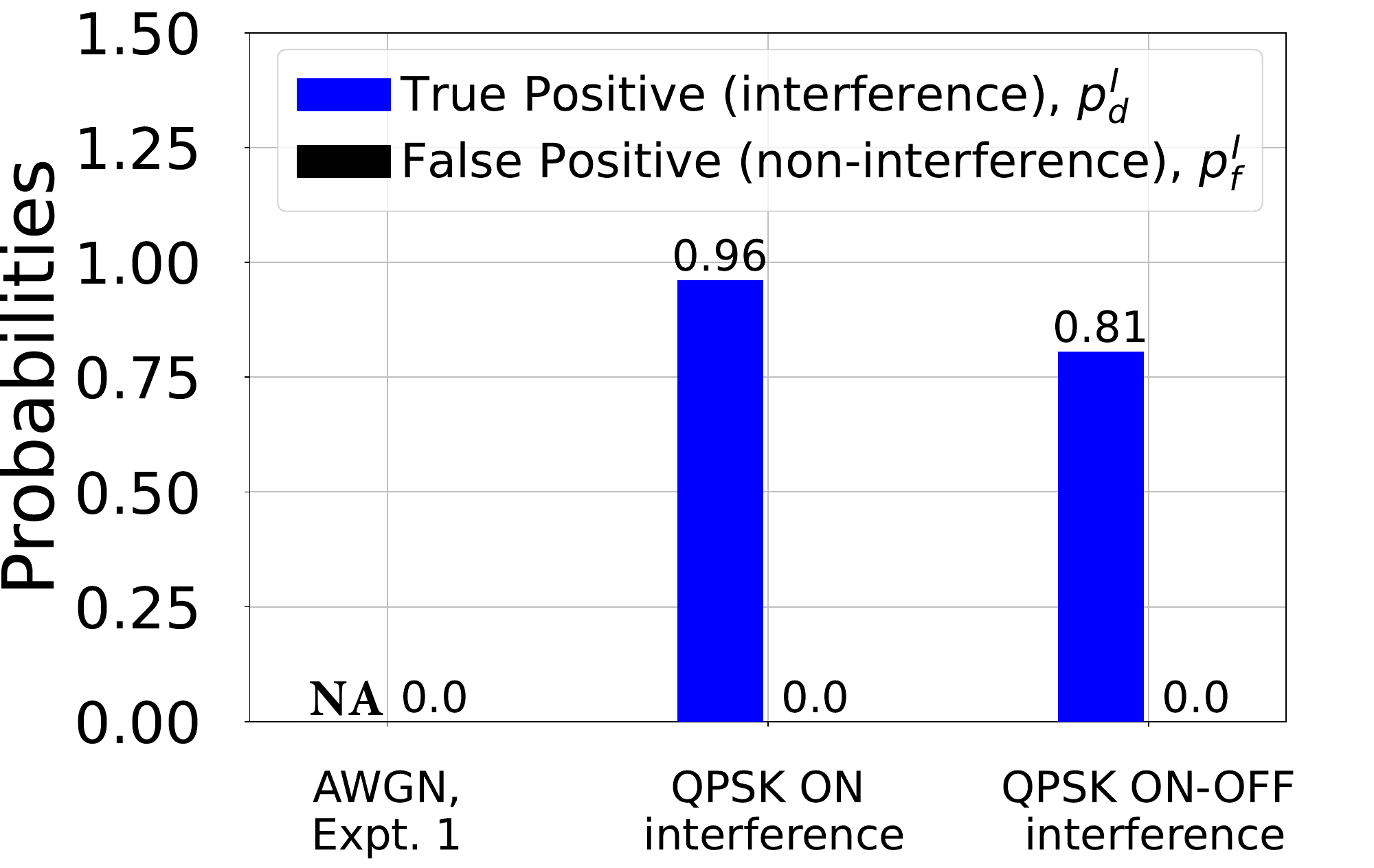}}
    \end{subfigure} 
    \hspace{4mm}
    \begin{subfigure}[LTE interference]
    {\label{fig:tp_fp_lte_if}
    \includegraphics[scale=0.134]{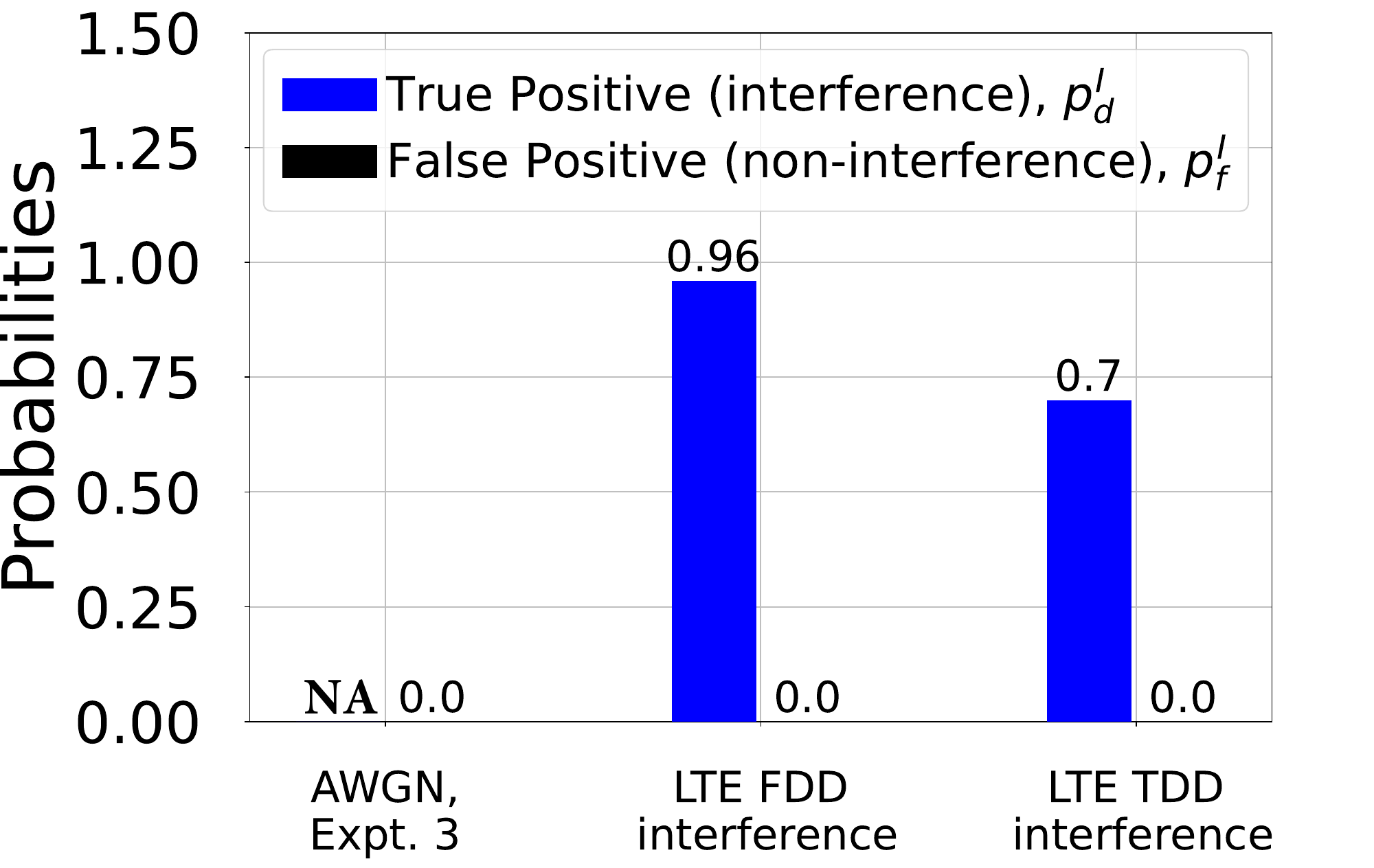}}
    \end{subfigure} 
    \begin{subfigure}[Interference detection rate]
    {\label{fig:if_detection_vs_INR}
    \includegraphics[scale=0.135]{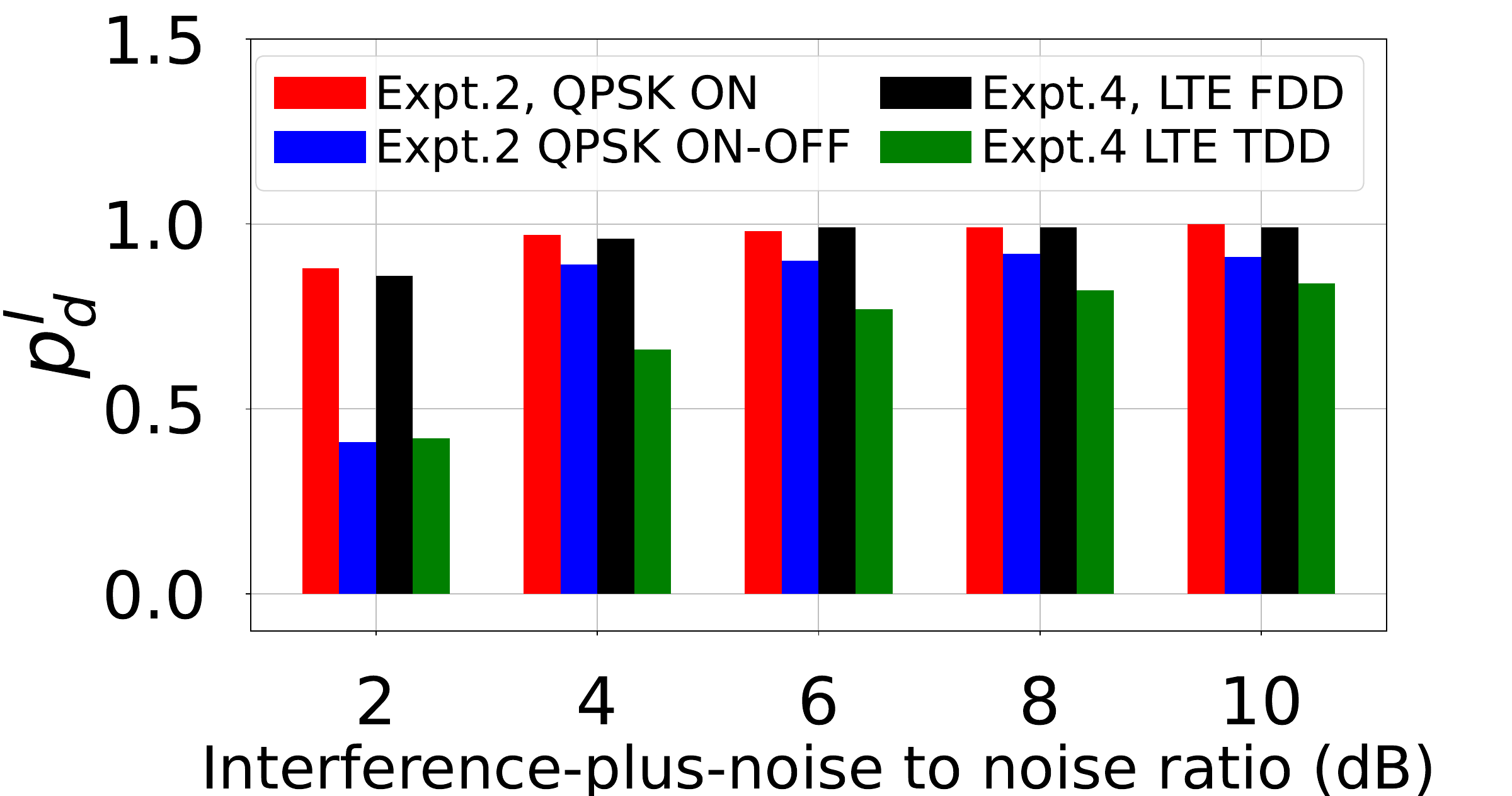}}
    \end{subfigure} 
    \caption{Classification between interference and non-interference using \YOLOCNN. The radar SNR and interference INR ranges are the same as in Fig.~\ref{fig:pd_in_interference}. Fig.~\ref{fig:if_detection_vs_INR} shows the interference detection rate of \YOLOCNN for different INR in different experiments.
    } 
    \label{fig:tp_fp_inr_yolomc}
\end{figure*}
First, differences in training data lead to different models for the same method. Thus, the performances of PAC and \YOLOCNN are different for experiments 1 and 3, which have similar test data but different training data. The trained models for experiment 3 perform better than those for experiment 1 because the interference patterns in experiment 3's training data are less obscuring. Specifically, the LTE TDD interference has a lower average ON time (6 msec) due to the different possible UL/DL configurations than the average ON
time of QPSK ON-OFF signals (9 msec). \YOLOCNN is sensitive to interference as it treats interference as a different class, unlike \WaveletCNN and DeepRadar. The reason for PAC's sensitivity to interference is different. PAC relies on the amplitude peak statistics, which are affected by interference. This figure also shows the robustness of our overall scheme \sys as it is unaffected by the differences in the training data. It appears that DeepRadar is also unaffected by the differences in the training data, but subsequent results will show that this is not the case in all experiments. 

Second, for all the metrics considered in Fig.~\ref{fig:tp_fp_all_methods_awgn}, PAC does not perform as well as the other methods, justifying the need for deep learning-based classification in our problem. The handcrafted feature extraction also limits PAC's pattern recognition capability.

Third, DeepRadar performs better than \YOLOCNN in Fig.~\ref{fig:tp_fp_all_methods_awgn_tain_qpsk}. This is because our object detection formulation, described in Section~\ref{subsection:yolo_cnn}, is more complex than that of DeepRadar. We use a complex framework to estimate more signal parameters than DeepRadar. However, the added complexity hurts \YOLOCNN's radar detection performance to some extent. 

Fourth, although DeepRadar performs better than \YOLOCNN in experiment 1, our overall method, \sys, outperforms all other methods due to the combined use of \YOLOCNN and \WaveletCNN. Thus, by combining the benefits of the two flows, \sys can achieve both superior classification accuracy and diverse parameter estimation capability. 

Fifth, \YOLOCNN and \sys have radar false positive rate below 1\% due to careful selection of thresholds, $t_o$ and $p_W^{R, true}$, described in Section~\ref{subsection:yolo_cnn} and~\ref{subsection:wavelet_cnn}. DeepRadar has low $p_f^R$ as it deals with larger objects, unlike \YOLOCNN, and is less prone to false object detections. 

\subsubsection{Radar detection accuracy versus SNR} Fig.~\ref{fig:tp_fp_all_methods_awgn} showed the radar true positive rate, $p_d^R$, combined across all SNR values and all radar types. To gain insights about performance across different SNRs, in Fig.~\ref{fig:radar_detection_per_snr}, we show different methods' $p_d^R$ for individual radar types and different values of SNR using the results of experiment 1. Fig.~\ref{fig:radar_detection_per_snr} does not show the results for \YOLOCNN as it was presented before in Section~\ref{subsection:yolo_cnn}. Hence in Fig.~\ref{fig:yololet_pd_improve}, we show \sys's improvement in $p_d^R$ over that of \YOLOCNN. Our primary observations are the following. 

First, for all the methods detecting radar type 1 for low SNR becomes difficult. This can be explained using Table~\ref{tab:radar_types}, which shows that the pulse width of radar type 1 is much smaller than others. A lower value of radar pulse width implies that the sensor must detect the radar using less signal energy. We also see that the performance trends corresponding to different radar types are similar for all the methods. This implies some radar types are inherently more challenging to detect than others. 
For example, different radar types have different bandwidth. Hence, even for the same per MHz SNR (the way we defined SNR as per CBRS rules), the total SNR can be different for different radar types.

Second, although the $p_d^R$ improvement in \sys over DeepRadar is 6\% in Fig.~\ref{fig:tp_fp_all_methods_awgn}, we see a more important difference from Fig~\ref{fig:deepradar_pd}, (c).
\sys can achieve 100\% $p_d^R$ for all radar types up to 16 dB SNR. DeepRadar can achieve high $p_d^R$ for all radar types up to 14 dB, but it cannot guarantee 100\% $p_d^R$ while ensuring a false positive rate below 1\%. 
\sys's superior performance in low SNR is due to the robustness of \WaveletCNN and its preprocessing.

Third, Fig.~\ref{fig:yololet_pd_improve} shows that the main $p_d^R$ improvement of \sys over \YOLOCNN is in the low SNR regime, especially for radar types 1, 2, and 4, which is precisely what we aimed for in the second flow of \sys.

\subsubsection{Radar vs. non-radar classification in interference} In Fig.~\ref{fig:pd_in_interference}, we compare the radar versus non-radar classification accuracy for different methods in the presence of different types of interference. The results are combined for SNR range [10-20] dB and INR range [2-10] dB, i.e., SINR range [0-18] dB. Results in Fig.~\ref{fig:tp_fp_qpsk_on},~\ref{fig:tp_fp_qpsk_on_off} are obtained using the model trained in experiment 1. Results in Fig.~\ref{fig:tp_fp_lte_on},~\ref{fig:tp_fp_lte_on_off} are obtained using the model trained in experiment 3.

First, we see that for all four experiments, \sys has the best performance in terms of $p_c^R$ and $p_d^R$. This shows that our proposed method can detect radar accurately, even in interference. Importantly, Fig.~\ref{fig:pd_in_interference} also shows that \sys achieves this high radar detection accuracy while ensuring that the radar false positive rate, $p_f^R$, is less or equal to 1\%. This demonstrates that the measures we took for dealing with interference in Section~\ref{subsection:wavelet_peprocess} and~\ref{subsection:wavelet_cnn} are effective. 

Second, comparing Fig.~\ref{fig:tp_fp_qpsk_on},~\ref{fig:tp_fp_qpsk_on_off}, we see that, for all the methods, the classification accuracy, $p_c^R$, is higher with QPSK ON-OFF interference than QPSK ON interference. The reason is that QPSK ON interference is always ON, making it more difficult to detect the radar signals. On the other hand, QPSK ON-OFF interference turns ON intermittently. A similar observation can be made by comparing Fig.~\ref{fig:tp_fp_lte_on},~\ref{fig:tp_fp_lte_on_off}. Fig.~\ref{fig:tp_fp_lte_on} corresponds to LTE FDD interference, which is always ON, and Fig.~\ref{fig:tp_fp_lte_on_off} corresponds to LTE TDD, which is intermittently ON. However, the performance gap between Fig.~\ref{fig:tp_fp_lte_on},~\ref{fig:tp_fp_lte_on_off} is less than that of Fig.~\ref{fig:tp_fp_qpsk_on},~\ref{fig:tp_fp_qpsk_on_off}. The explanation for the following observation also answers this.
\begin{figure}[t]
    \centering
    \begin{subfigure}
    {
    \includegraphics[scale=0.185]{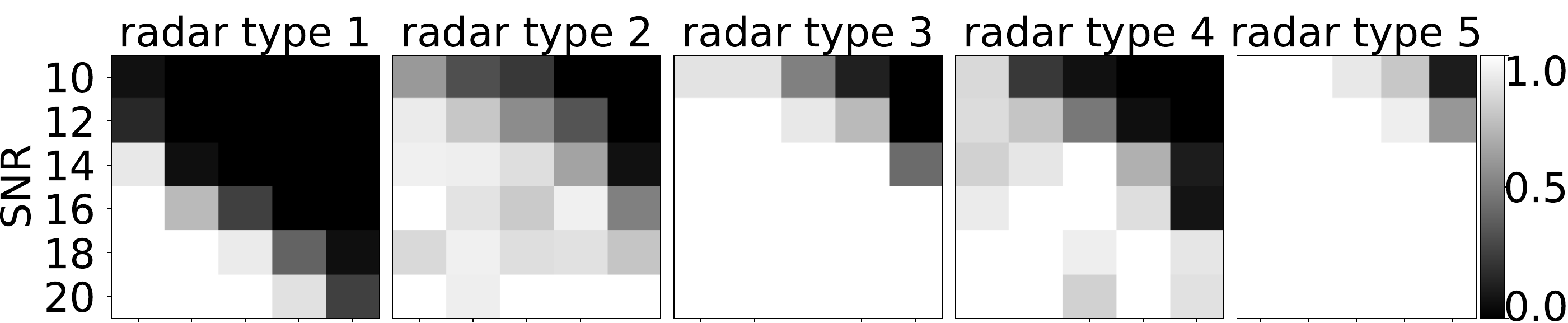}}
    \end{subfigure} 
    \vspace{-1mm}
    \vspace{-3mm}
    \begin{subfigure}
    {
    \includegraphics[scale=0.185]{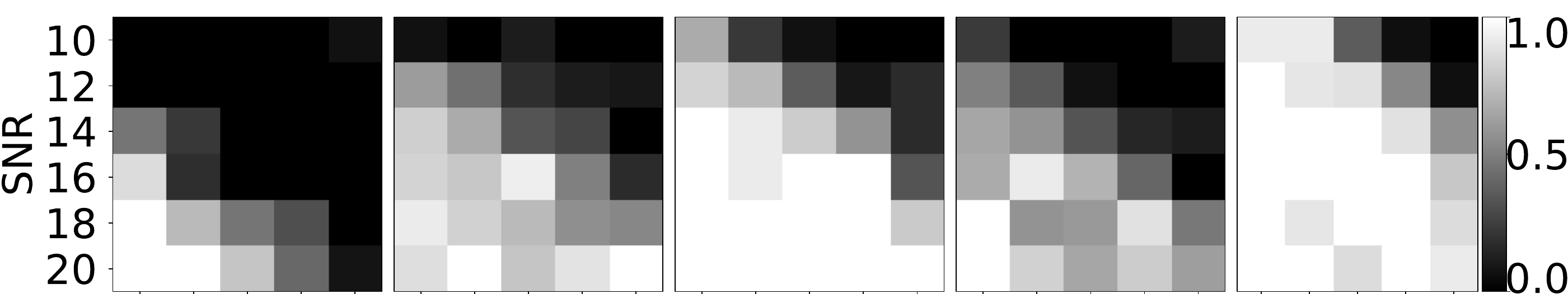}}
    \end{subfigure} 
    \vspace{-1mm}
    \vspace{-3mm}
    \begin{subfigure}
    {
    \includegraphics[scale=0.185]{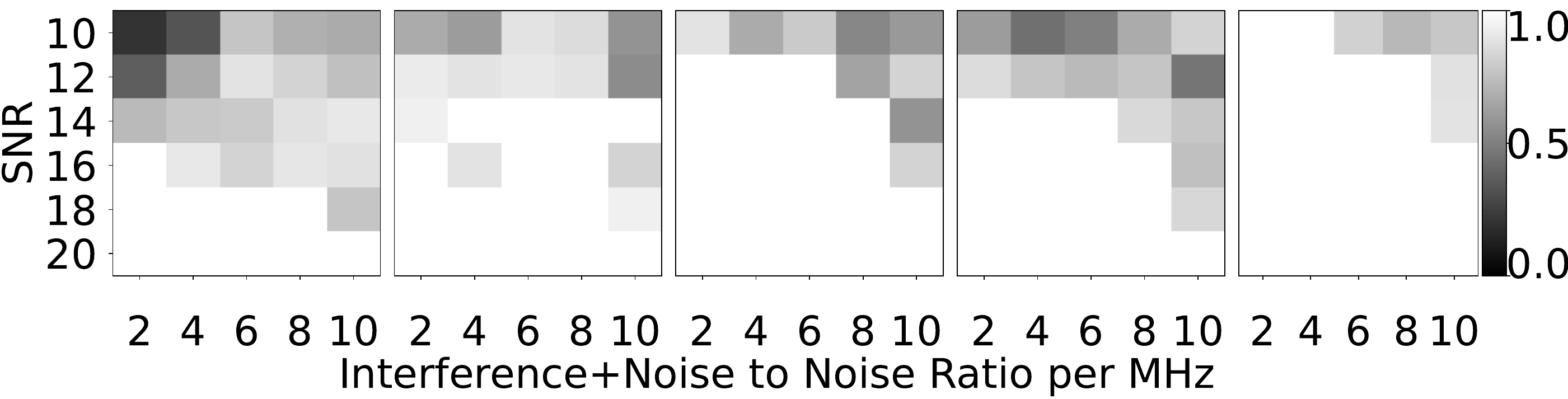}}
    \end{subfigure} 
    \caption{Radar detection probability, $p_d^R$, for DeepRadar (top), \YOLOCNN (middle), and \sys (bottom) for experiment 2A.
    } 
    \label{fig:qpsk_on_acc_img}
\end{figure}
\begin{figure}[t]
    \centering
    \begin{subfigure}
    {\label{fig:yololet_qpsk_on}
    \includegraphics[scale=0.185]{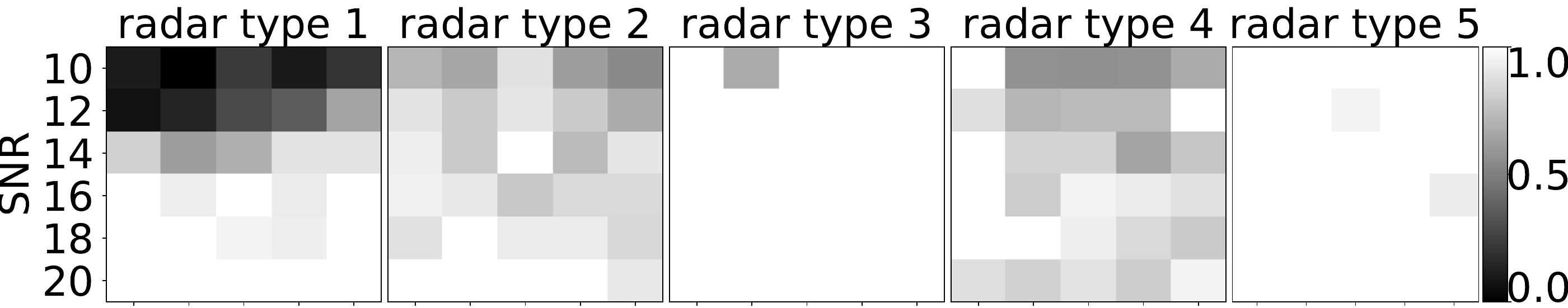}}
    \end{subfigure} 
    \vspace{-1mm}
    \vspace{-3mm}
    \begin{subfigure}
    {\label{fig:yolo_qpsk_on_off}
    \includegraphics[scale=0.185]{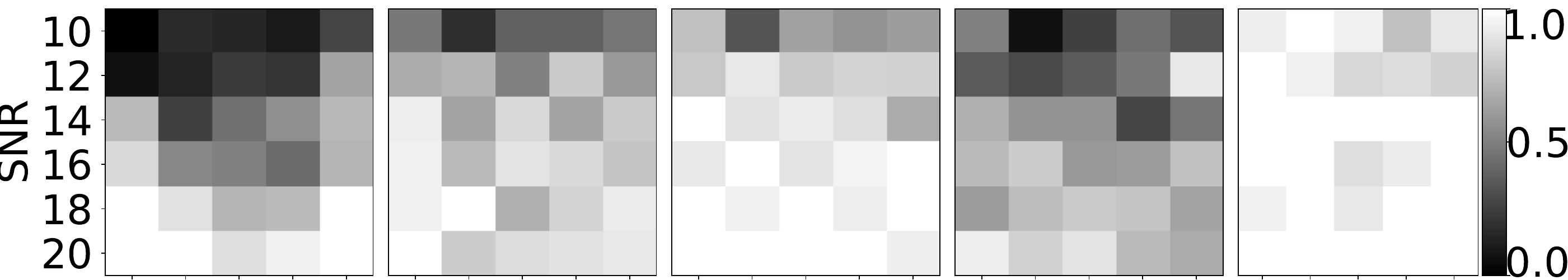}}
    \end{subfigure} 
    \vspace{-1mm}
    \vspace{-3mm}
    \begin{subfigure}
    {\label{fig:yololet_qpsk_on_off}
    \includegraphics[scale=0.185]{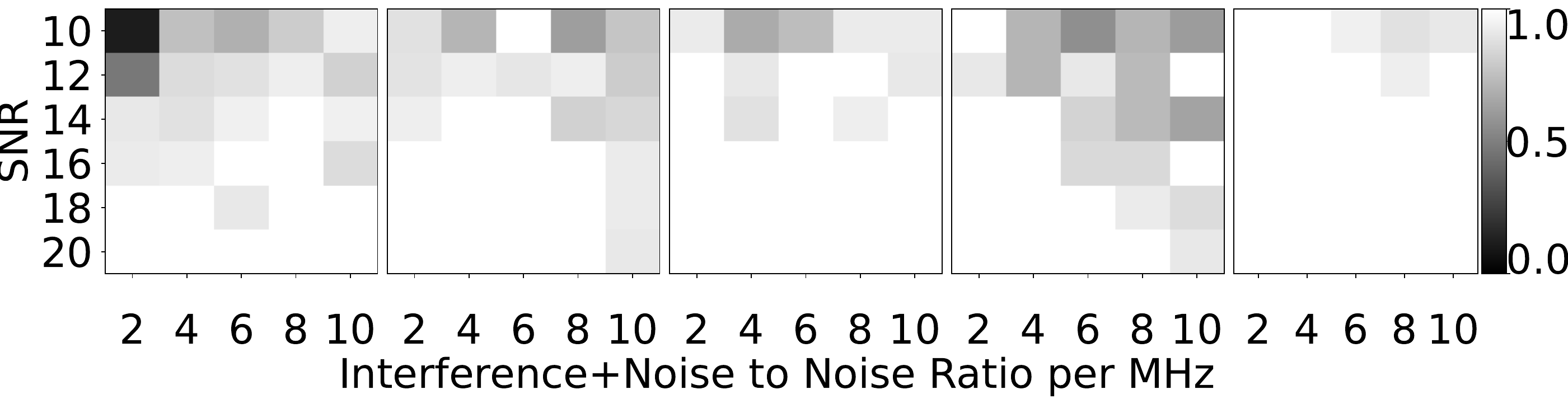}}
    \end{subfigure} 
    \caption{Radar detection probability, $p_d^R$, for DeepRadar (top), \YOLOCNN (middle), and \sys (bottom) for experiment 2B.
    } 
    \label{fig:qpsk_on_off_acc_img}
\end{figure}
\begin{figure}[t]
    \centering
    \begin{subfigure}
    {\label{fig:yololet_qpsk_on}
    \includegraphics[scale=0.185]{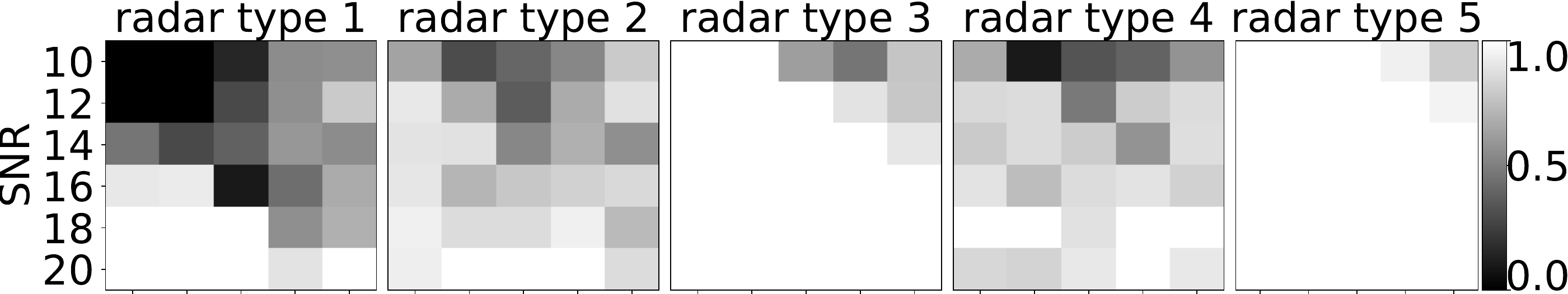}}
    \end{subfigure} 
    \vspace{-1mm}
    \vspace{-3mm}
    \begin{subfigure}
    {\label{fig:yolo_qpsk_on}
    \includegraphics[scale=0.185]{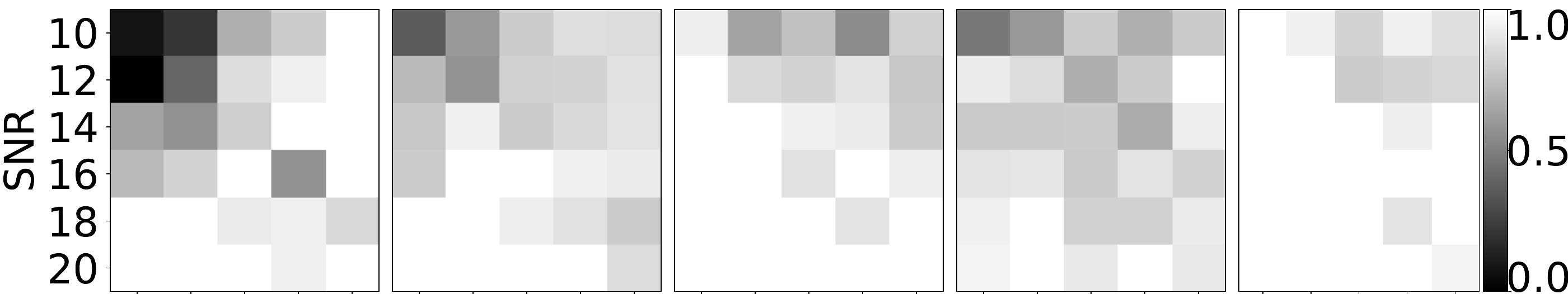}}
    \end{subfigure} 
    \vspace{-1mm}
    \vspace{-3mm}
    \begin{subfigure}
    {\label{fig:yololet_lte_on}
    \includegraphics[scale=0.185]{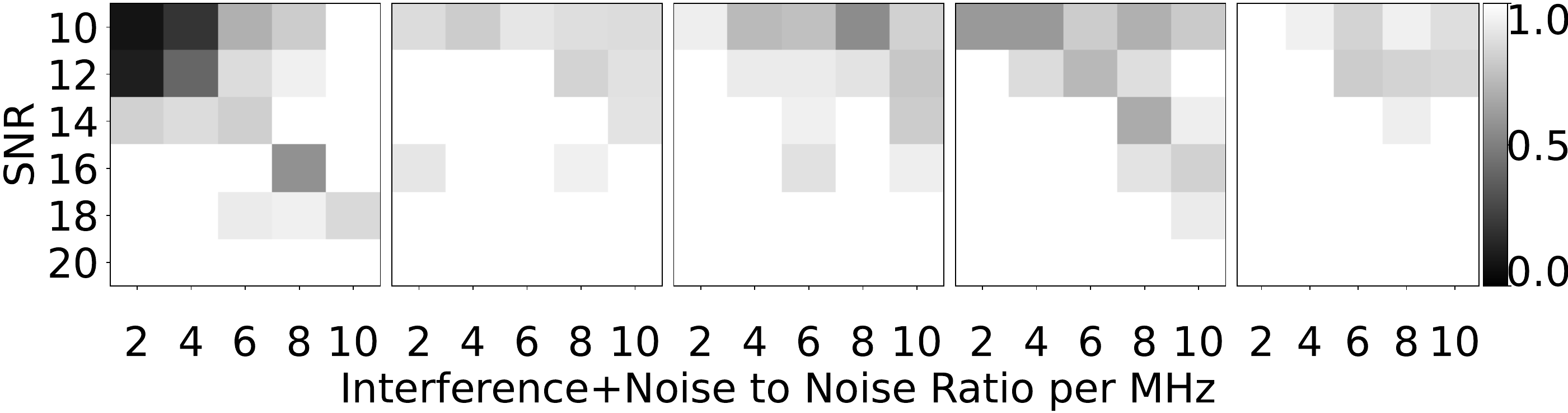}}
    \end{subfigure} 
    \caption{Radar detection probability, $p_d^R$, for DeepRadar (top), \YOLOCNN (middle), and \sys (bottom) for experiment 4A.
    } 
    \label{fig:lte_on_acc_img}
\end{figure}
\begin{figure}[t]
    \centering
    \begin{subfigure}
    {\label{fig:yololet_qpsk_on}
    \includegraphics[scale=0.185]{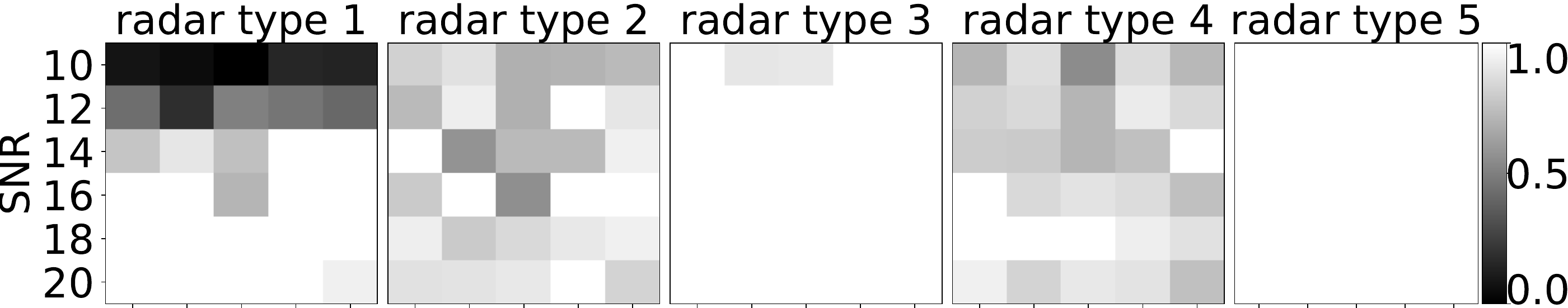}}
    \end{subfigure} 
    \vspace{-1mm}
    \vspace{-3mm}
    \begin{subfigure}
    {\label{fig:yolo_qpsk_on}
    \includegraphics[scale=0.185]{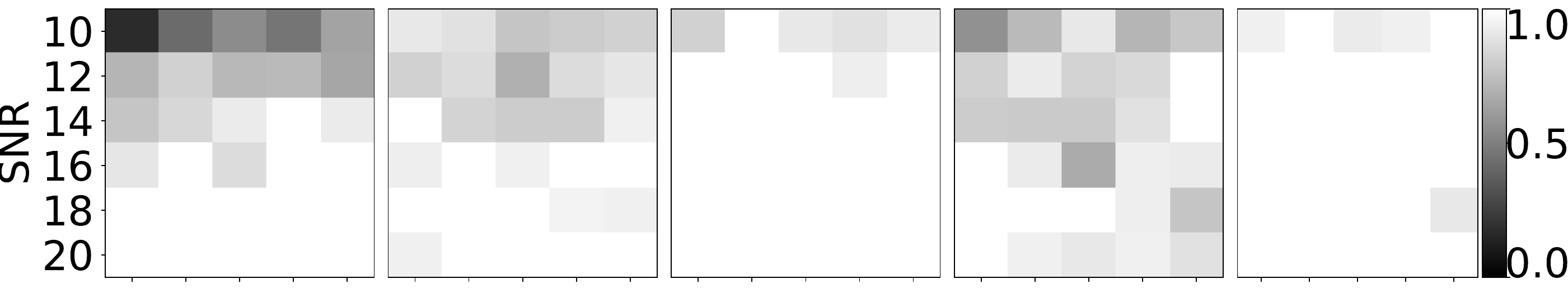}}
    \end{subfigure} 
    \vspace{-1mm}
    \vspace{-3mm}
    \begin{subfigure}
    {\label{fig:yololet_lte_on}
    \includegraphics[scale=0.185]{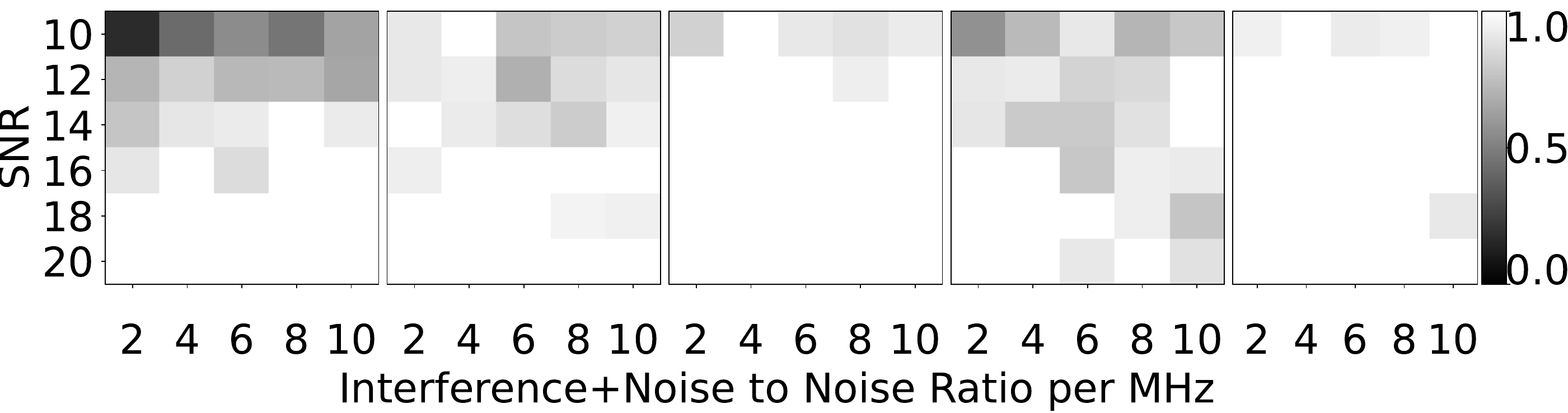}}
    \end{subfigure} 
    \caption{Radar detection probability, $p_d^R$, for DeepRadar (top), \YOLOCNN (middle), and \sys (bottom) for experiment 4B.
    } 
    \label{fig:lte_on_off_acc_img}
\end{figure}

Third, by comparing Fig.~\ref{fig:tp_fp_qpsk_on},~\ref{fig:tp_fp_qpsk_on_off} with Fig.~\ref{fig:tp_fp_lte_on},~\ref{fig:tp_fp_lte_on_off} we observe that \YOLOCNN and DeepRadar perform better with LTE interference than QPSK interference. 
The reason for that is the following.
When \YOLOCNN and DeepRadar are trained with QPSK interference (refer to experiment 1), their object detection models have some overfitting because of interference.
I.e., the models produce high confidence for the detected radar objects based on training data (the confidences influence the object detection thresholds)
but the confidence for test radar objects is lower. This primarily affects the test data in experiment 2A, where the test signals have QPSK ON interference. Also, this problem is more prominent for \YOLOCNN because of the small radar pulse objects. On the other hand, for the \YOLOCNN and DeepRadar models trained with LTE interference, this issue is less severe because the interference patterns in the training data are less obscuring, as discussed in Section~\ref{subsection:binary_classification_radar_awgn}. However, Fig.~\ref{fig:pd_in_interference} also reveals that our overall scheme \sys is less affected by the above issue because of the use of \WaveletCNN to aid \YOLOCNN. \WaveletCNN does not use an object detection and is less interference-sensitive. Hence, \sys has consistent performance across all four cases in Fig.~\ref{fig:pd_in_interference}.


 \begin{figure*}[t]
    \centering
    \hspace{-8mm}
    \begin{subfigure}[Example parameter estimation]
    {\label{fig:example_params_estimation}
    \includegraphics[scale=0.21]{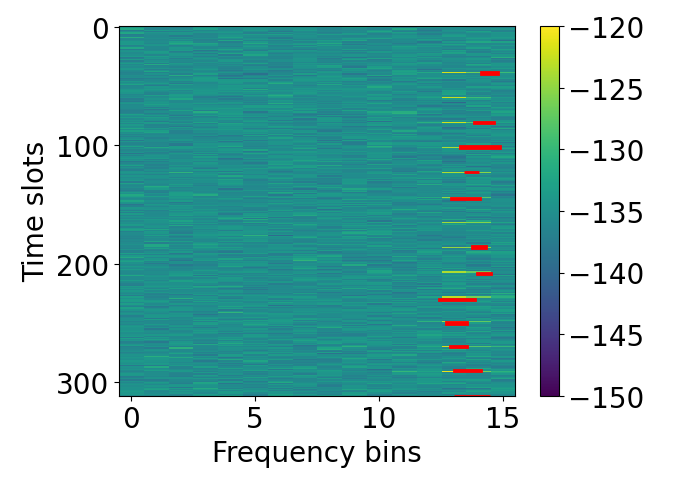}}
    \end{subfigure} 
    \hspace{-6mm}
    \begin{subfigure}[Missed bandwidth]
    {\label{fig:missed_bw_all}
    \includegraphics[scale=0.117]{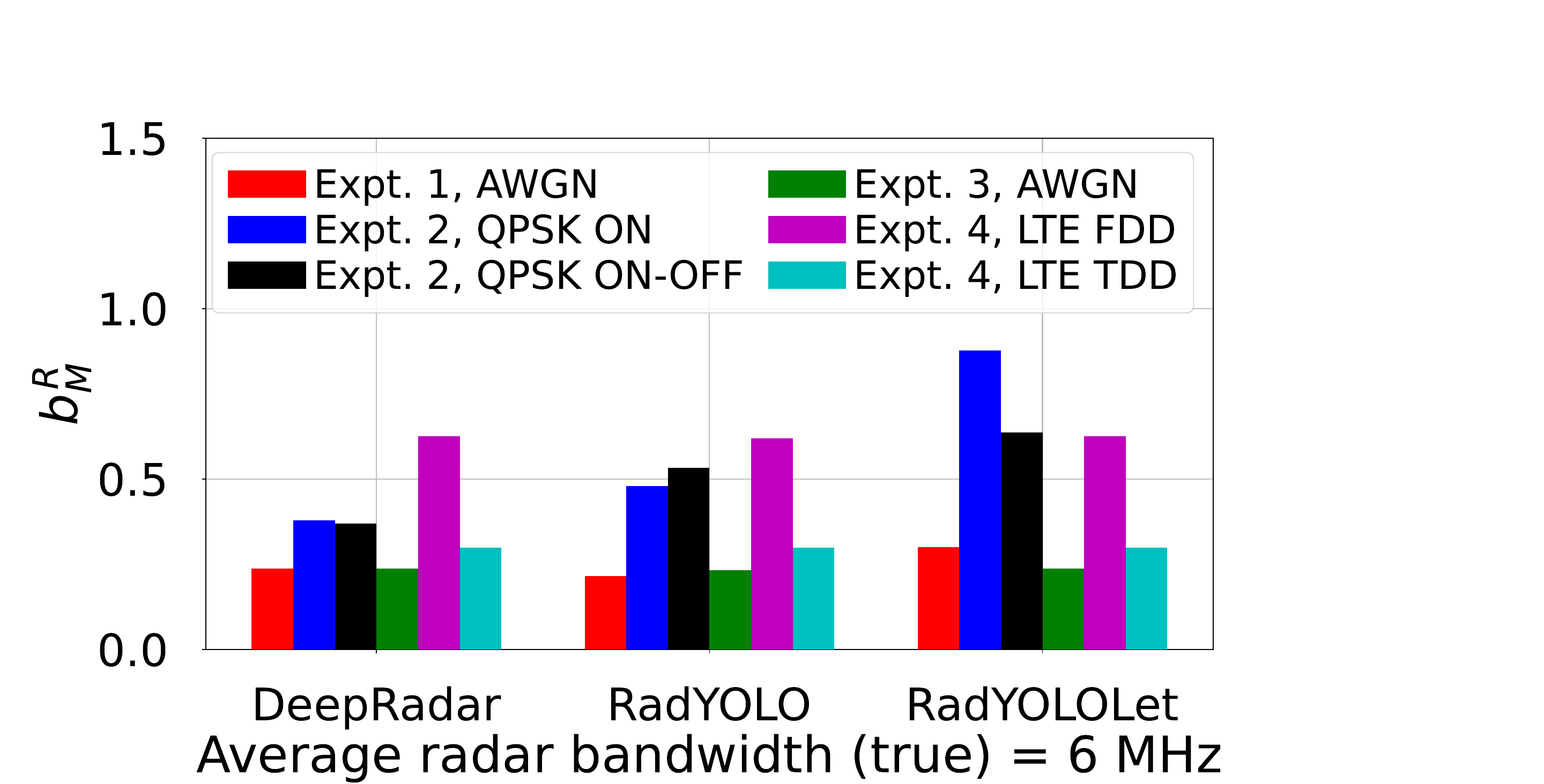}}
    \end{subfigure} 
    \hspace{-6mm}
    \begin{subfigure}[Excess bandwidth]
    {\label{fig:excess_bw_all}
    \includegraphics[scale=0.117]{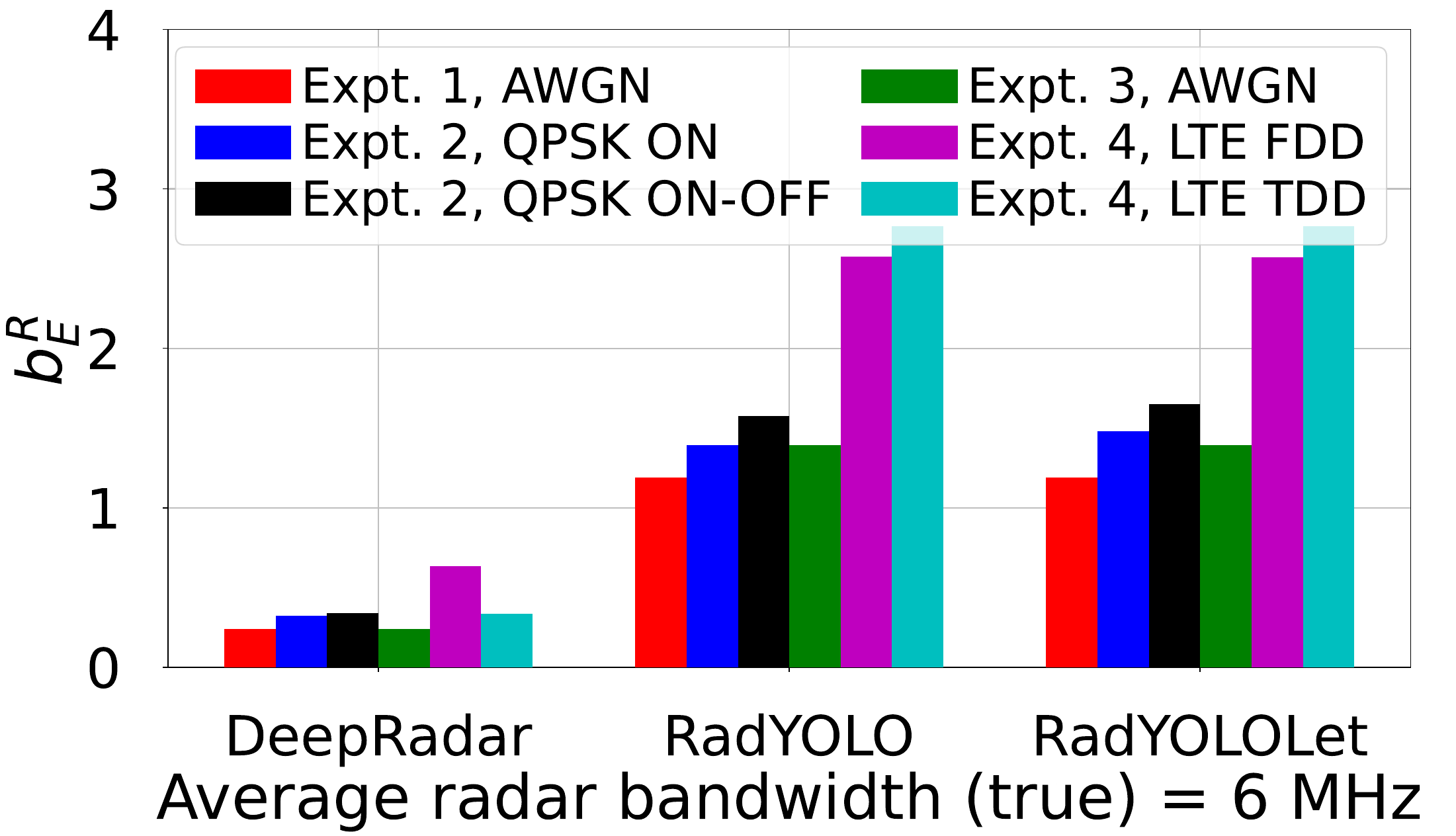}}
    \end{subfigure}
    \hspace{-5mm}
    \begin{subfigure}[Fraction of detected pulses]
    {\label{fig:num_pulses_all}
    \includegraphics[scale=0.117]{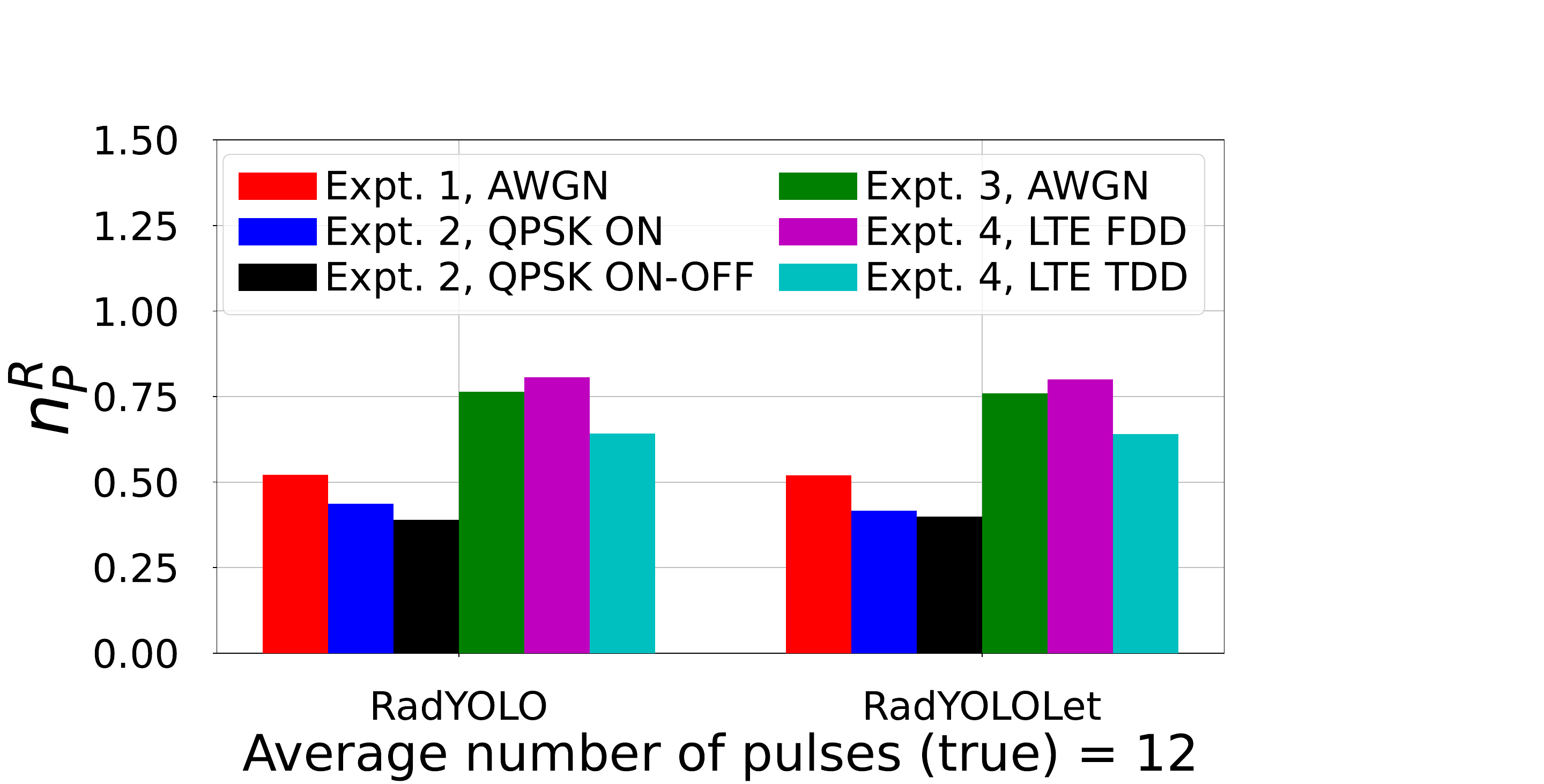}}
    \end{subfigure} \\
    \hspace{-7mm}
    \begin{subfigure}[Pulse width estimation error]
    {\label{fig:pw_all}
    \includegraphics[scale=0.115]{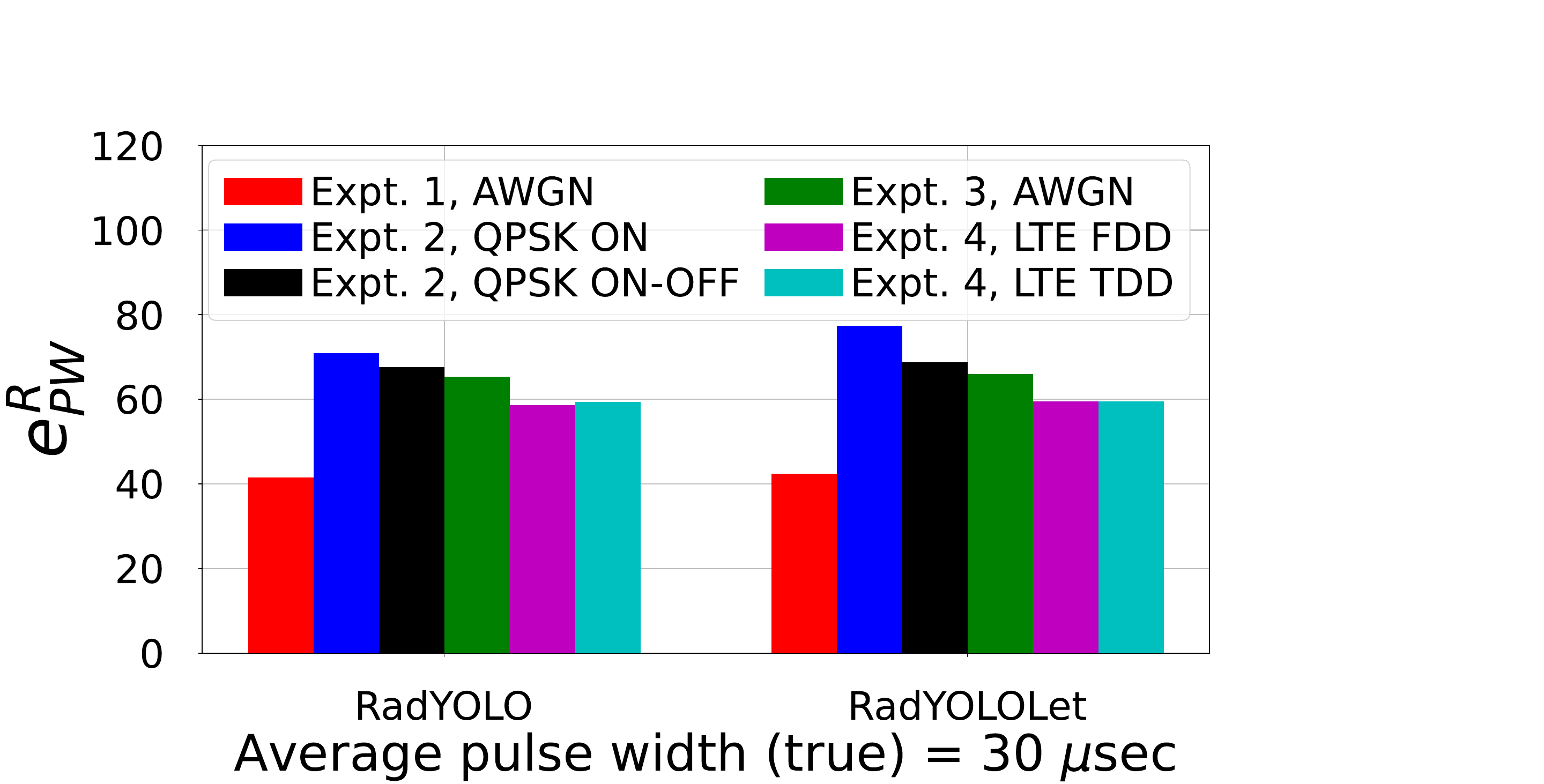}}
    \end{subfigure} 
    \hspace{-7.5mm}
    \begin{subfigure}[Pulse interval estimation error]
    {\label{fig:pri_all}
    \includegraphics[scale=0.115]{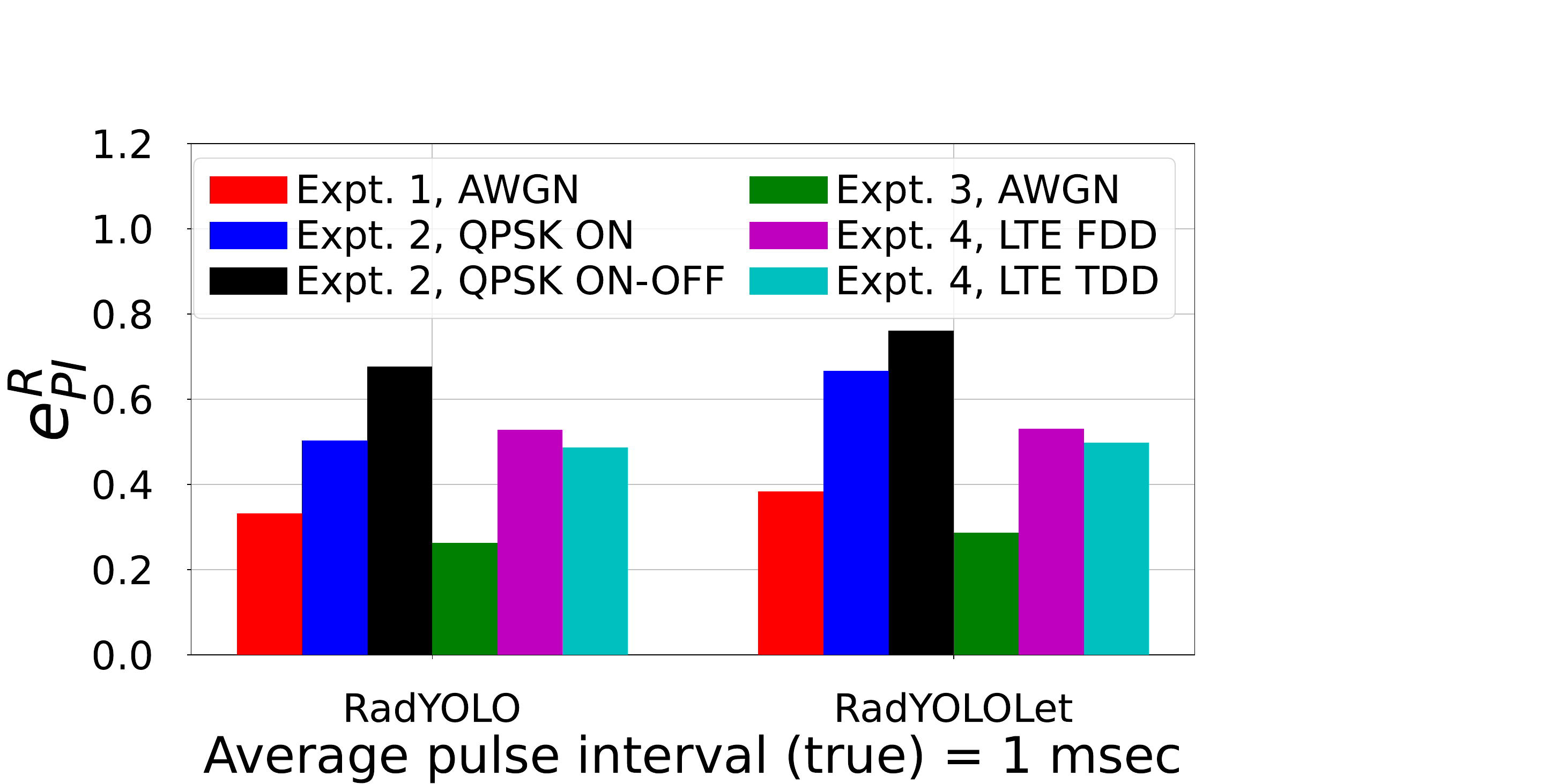}}
    \end{subfigure} 
    \hspace{-7mm}
    \begin{subfigure}[Number of estimations]
    {\label{fig:num_estimations}
    \includegraphics[scale=0.105]{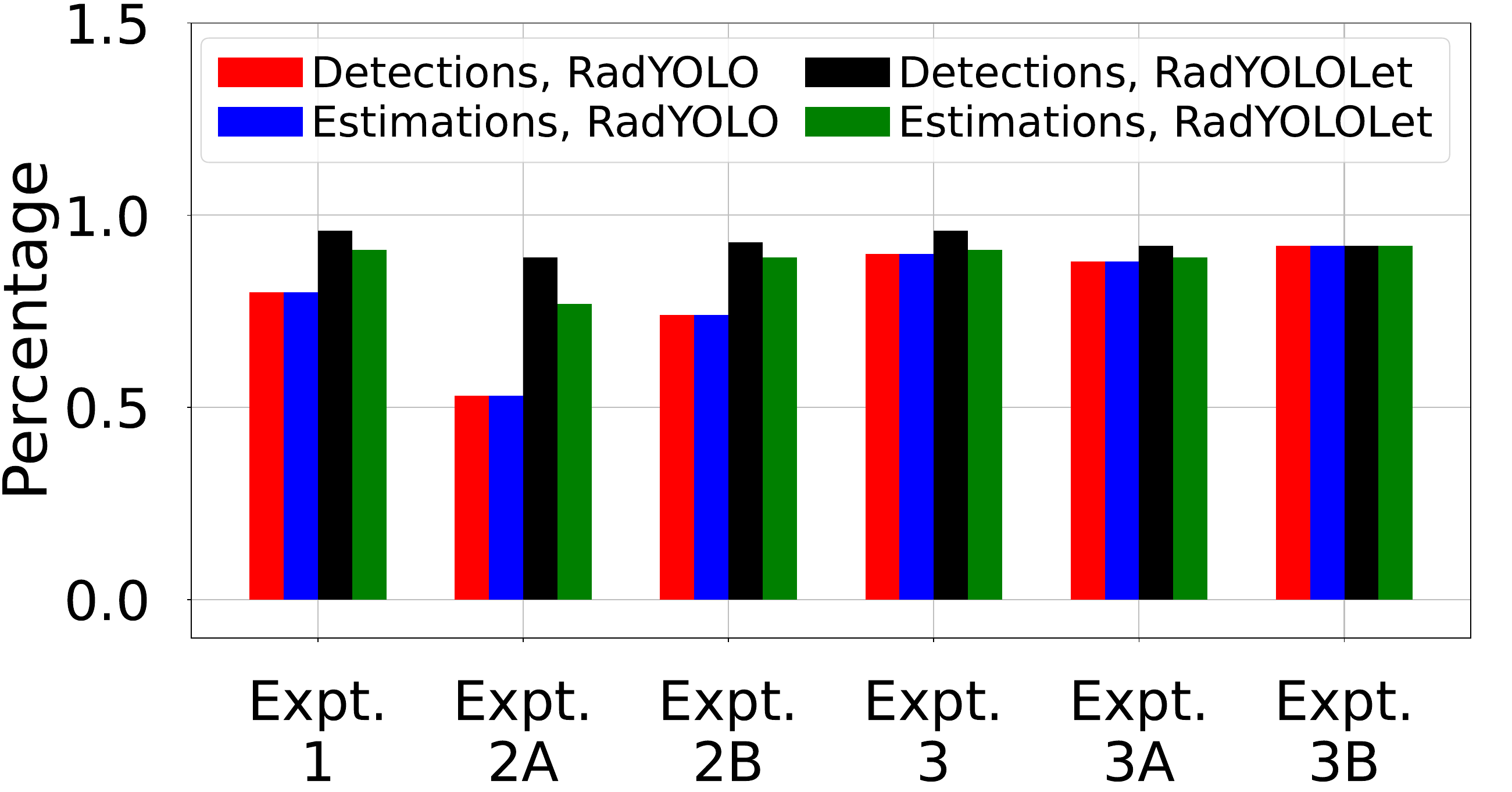}}
    \end{subfigure}
    \hspace{-4mm}
    \begin{subfigure}[Interference ON time estimation]
    {\label{fig:if_on_time_est}\includegraphics[scale=0.115]{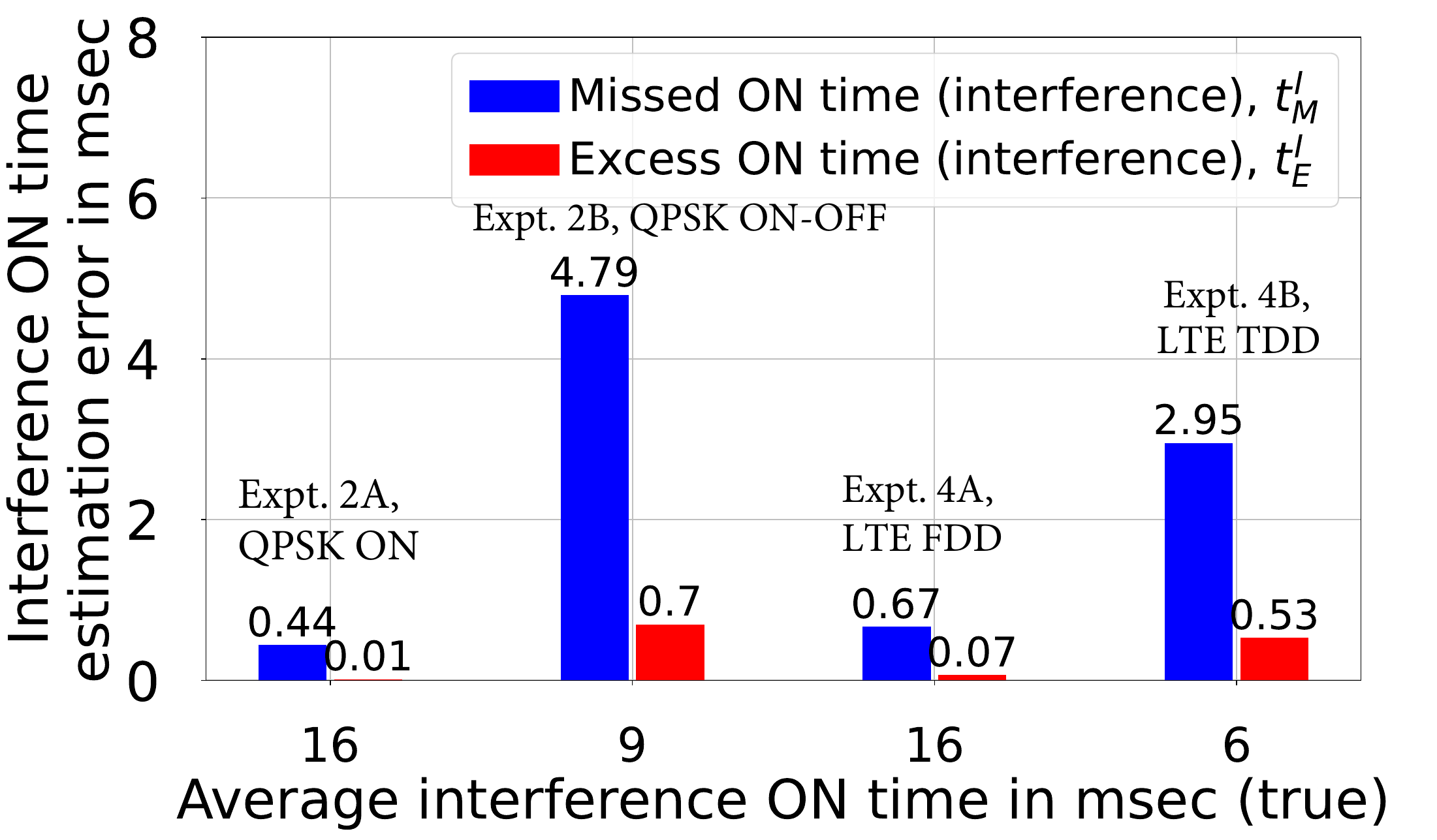}}
    \end{subfigure} 
    \caption{Signal parameter estimation of different methods. Except for the last one, all other figures are for radar signal parameters. 
    } 
    \label{fig:signal_params_est}
\end{figure*}
\subsubsection{Interference vs. non-interference classification} In Fig.~\ref{fig:pd_in_interference}, we argued that the undetected radar signals in \YOLOCNN are miss detections, not miss classifications. This is demonstrated in Fig.~\ref{fig:tp_fp_qpsk_if} and~\ref{fig:tp_fp_lte_if}, showing that the interference false positive rate is always 0\%.
Note that the results in Fig.~\ref{fig:tp_fp_inr_yolomc} are only for \YOLOCNN as the other methods cannot classify interference signals. In experiments 1 and 3, we do not have any interference signal in the test set. Hence for these two experiments, we use `NA' for interference true positive rate, $p_d^I$. Fig.~\ref{fig:tp_fp_qpsk_if} and~\ref{fig:tp_fp_lte_if} also show that the $p_d^I$ with QPSK ON and LTE FDD interference is very high. However, that is not the case for QPSK ON-OFF and LTE TDD. The reason is interference objects corresponding to QPSK ON and LTE FDD signals are much bigger than that of QPSK ON-OFF and LTE TDD signals. Hence, they are easier to detect with high confidence. Due to similar reasoning, $p_d^I$ for QPSK ON-OFF is higher than that of LTE TDD. 

Fig.~\ref{fig:if_detection_vs_INR} shows the interference detection rate, $p_d^I$, for different values of INR and different experiments. This figure shows that the interference signals are missed more in the low INR region. This is expected because, at low INR regions, the interference objects are less detectable. Importantly, the missed interference signals are not miss classified as radar as demonstrated by the $\leq 1\%$ radar false positive rate in Fig.~\ref{fig:pd_in_interference}.

\subsubsection{Radar detection performance at different SINR} 
Next, we analyze $p_d^R$ for experiments 2A, 2B, 4A, and 4B in Fig.~\ref{fig:qpsk_on_acc_img}, ~\ref{fig:qpsk_on_off_acc_img}, ~\ref{fig:lte_on_acc_img}, and ~\ref{fig:lte_on_off_acc_img}, respectively, for different values SINR and the different radar types. Since we vary both the radar SNR and the interference INR, the results are shown as images where the pixel values are $p_d^R$, the INR increases along the x-axis, and the SNR increases along the y-axis. 
The primary observation is that irrespective of the interference type, \sys can tolerate up to 4 dB INR and still achieve 100\% radar detection accuracy when the radar SNR is fixed at 20 dB. On the other hand, if we fix the INR to be 2 dB, then \sys can achieve 100\% $p_d^R$ for SNR up to 18 dB. The above two observations suggest that \sys can accurately function up to 16 dB radar SINR. This is not achievable by the other two methods. These figures also show that the second flow of \sys assists \YOLOCNN not only in AWGN but also in different types of interference. 

\subsubsection{Signal parameter estimation} 
In Fig.~\ref{fig:signal_params_est}, we show the parameter estimation of different methods.
Fig.~\ref{fig:example_params_estimation} is an example estimation of \YOLOCNN that will help us explain some observations. 
In Fig.~\ref{fig:signal_params_est}, along with \YOLOCNN and \sys, we evaluate DeepRadar for bandwidth estimation but not for the other parameters as DeepRadar cannot estimate them.

 \begin{figure*}[t]
    \centering
    \hspace{-2mm}
    \begin{subfigure}[Missed bandwidth]
    {\label{fig:miss_bw_binwise}
    \includegraphics[scale=0.112]{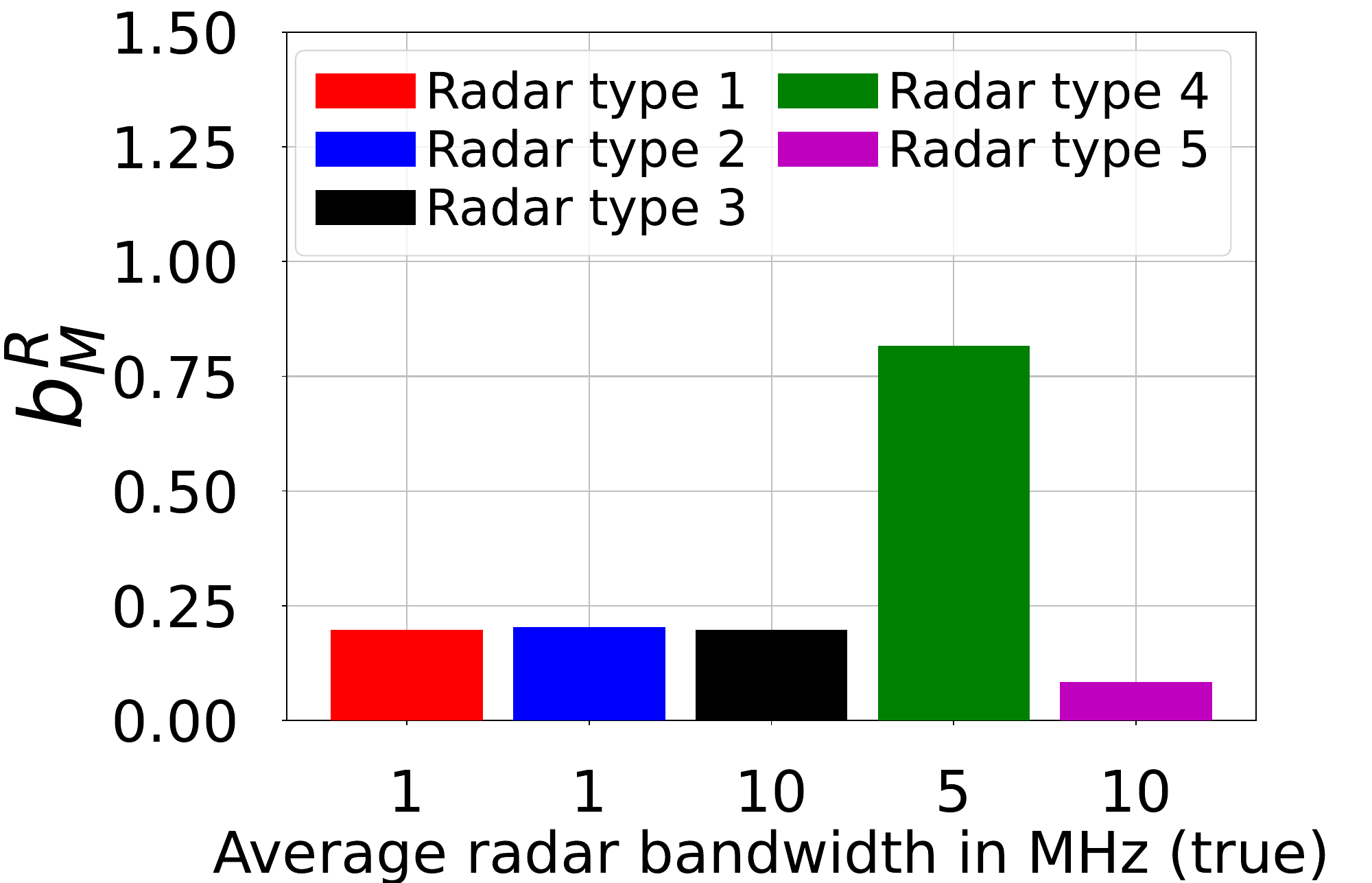}}
    \end{subfigure} 
    \hspace{-6mm}
    \begin{subfigure}[Excess bandwidth]
    {\label{fig:excess_bw_binwise}
    \includegraphics[scale=0.112]{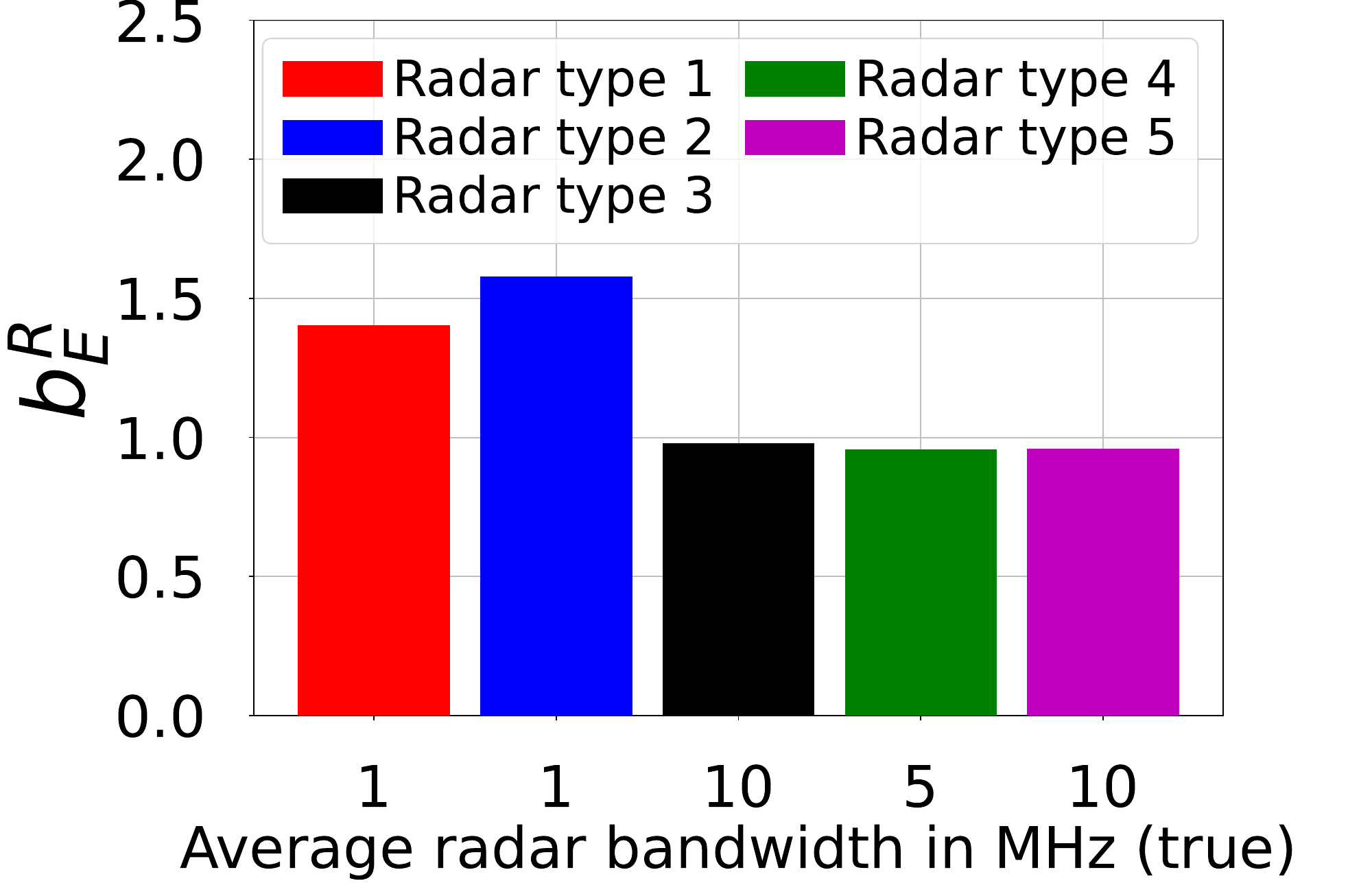}}
    \end{subfigure} 
    \hspace{-7mm}
    \begin{subfigure}[Fraction of detected pulses]
    {\label{fig:num_pulses_binwise}
    \includegraphics[scale=0.114]{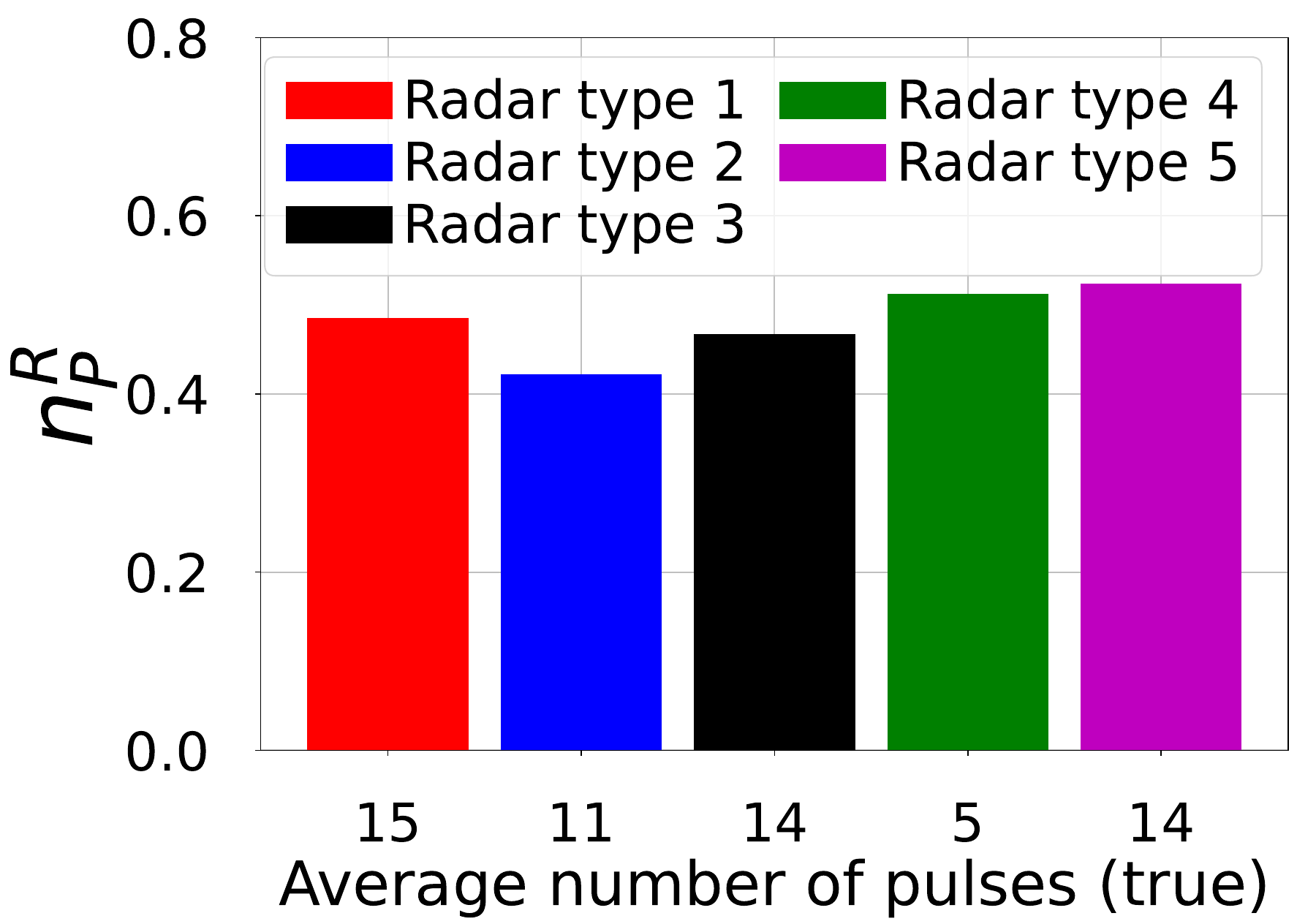}}
    \end{subfigure} 
    \hspace{-5.5mm}
    \begin{subfigure}[Pulse width estimation]
    {\label{fig:pw_binwise}
    \includegraphics[scale=0.115]{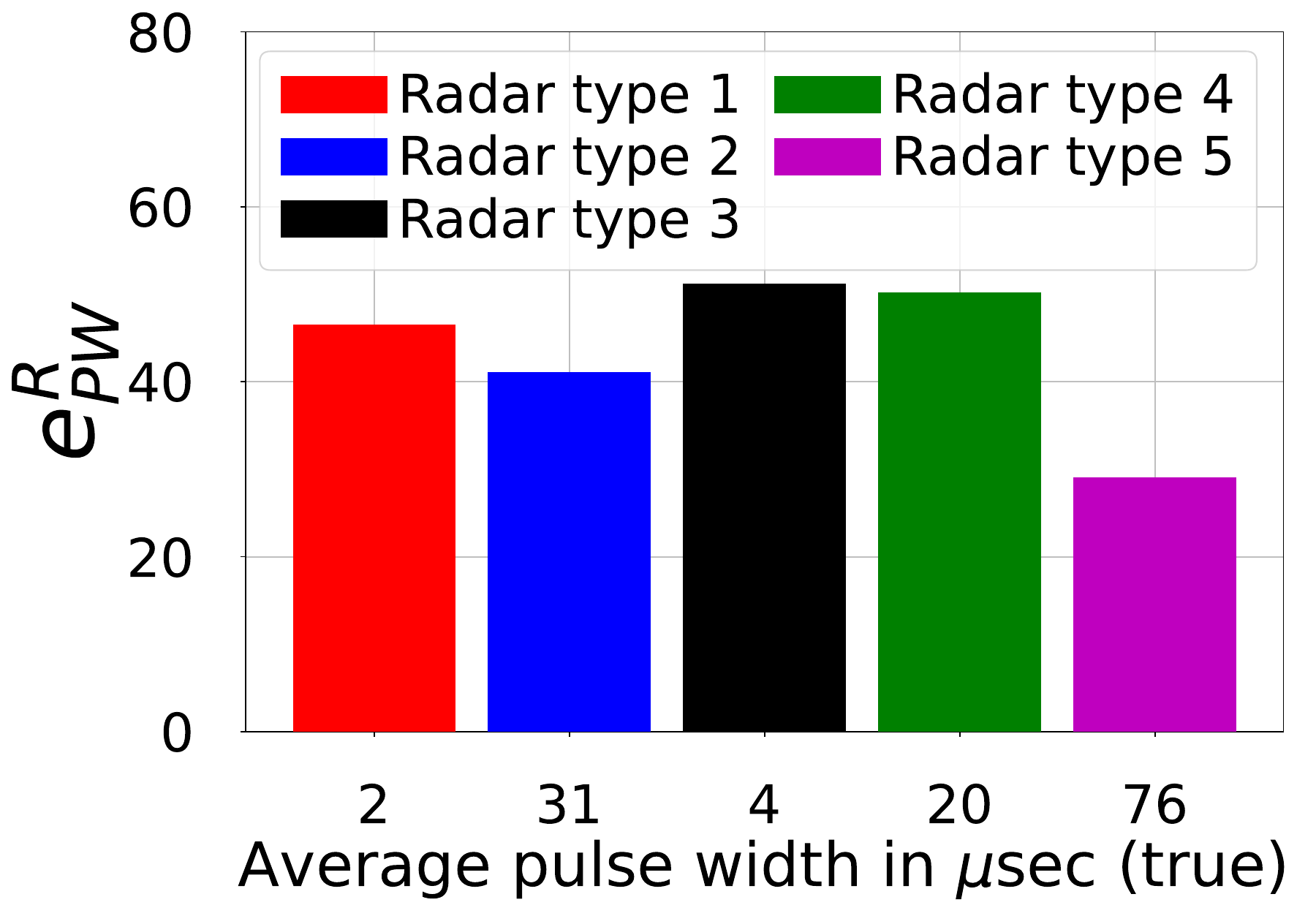}}
    \end{subfigure} 
    \hspace{-5mm}
    \begin{subfigure}[Pulse interval estimation]
    {\label{fig:pri_binwise}
    \includegraphics[scale=0.114]{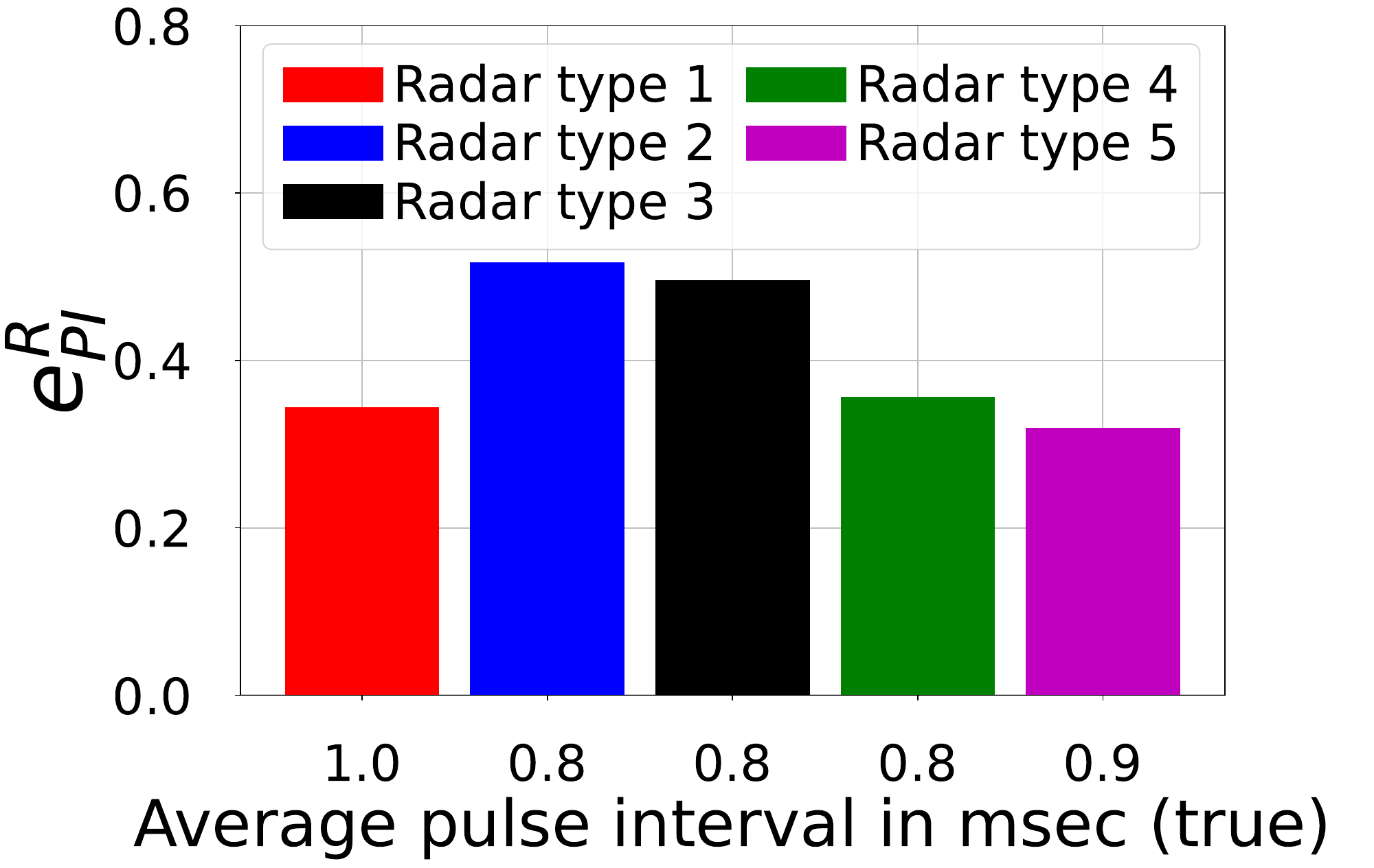}}
    \end{subfigure} 
    \caption{Radar signal parameter estimation for different types using experiment 1.} 
    \label{fig:signal_params_binwise}
\end{figure*}
First, we see from Fig.~\ref{fig:missed_bw_all},~\ref{fig:excess_bw_all} that both in terms of missed and excess bandwidth, DeepRadar performs better than \YOLOCNN and \sys. The difference is marginal in terms of $b_M^R$, and all the methods have a low missed bandwidth error.
However, $b_E^R$ for \YOLOCNN and \sys is at least 1 MHz higher than that of DeepRadar.
This happens because \YOLOCNN attempts to detect each radar pulse individually 
and decides the signal bandwidth to be the union of the bandwidth of the individual pulses. As we can see from Fig.~\ref{fig:example_params_estimation}, the bandwidth estimation of individual pulses can be different, leading to a wider estimated bandwidth than the actual. Since \sys relies on \YOLOCNN's bandwidth estimation, it also has the same problem. DeepRadar does not have this problem because it treats all the pulses as a single object. However, this also is a limitation of DeepRadar as it cannot estimate the temporal parameters.

Second, Fig.~\ref{fig:num_pulses_all} shows that both \YOLOCNN and \sys detect 40-75\% of the radar pulses. This shows the difficulty of detecting small radar pulse objects. Even after prioritizing the small objects in the loss function in~\eqref{eq:yolo_cnn_loss}, \YOLOCNN cannot detect all pulses. However, this is not a major problem as \sys does not require detecting all the radar pulses for estimating pulse width and interval as explained in Section~\ref{subsection:yolo_cnn}. Fig.~\ref{fig:num_pulses_all} also shows that \YOLOCNN and \sys detect more pulses for the model trained in experiment 3. The reason is the difference in training data in experiments 1 and 3, as explained in the context of Fig.~\ref{fig:pd_in_interference}. However, a higher $n_P^R$ also increases the possibility of excess bandwidth estimation. Thus, $b_E^R$ for \YOLOCNN and \sys is higher in experiments 4A, 4B as shown in Fig.~\ref{fig:excess_bw_all}.

Third, Fig.~\ref{fig:pw_all} shows that the pulse width estimation error of \YOLOCNN is high with respect to the true average radar pulse width. However, Fig.~\ref{fig:example_params_estimation} shows that the estimated width of the pulses (height of objects on spectrogram) is well aligned with the actual pulses. This incongruence arises from the fact that the radar pulse width is very low, on average 30 $\mu$sec, and the errors are also shown in $\mu$sec, which is difficult to interpret visually from Fig.~\ref{fig:example_params_estimation}. The error in pulse width can be attributed to the difficulty in estimating the extremely small radar pulse width. 

Fourth, we see from Fig.~\ref{fig:pri_all} that the pulse interval errors are lower when the test set contains no interference signal. This happens because in the presence of interference, \YOLOCNN misses a higher number of pulses, which is reflected by Fig.~\ref{fig:num_pulses_all}. This, in turn, affects the pulse interval estimation, which is the minimum gap between any pair of detected pulses.

Fifth, for almost all the plots in Fig.~\ref{fig:signal_params_est} \sys's performance is comparable to that of \YOLOCNN. While this is expected as \sys relies on \YOLOCNN's estimations, it is important to note that \sys's results are based on more test examples than \YOLOCNN. This is demonstrated via Fig.~\ref{fig:num_estimations}, which shows the percentages of test data on which the methods perform successful radar detection and parameter estimation. \YOLOCNN's detection and estimation percentage is always same as they are done jointly. \sys's detection percentage is strictly better than that of \YOLOCNN as discussed in Fig.~\ref{fig:tp_fp_all_methods_awgn},~\ref{fig:pd_in_interference}. \sys's estimation percentage is better than that of \YOLOCNN due to \WaveletCNN's parameter estimation approach presented in Section~\ref{subsection:wavelet_cnn}. However, as the detections and estimations are decoupled in \WaveletCNN, its estimation percentage is always lower than its detection percentage.


Fig.~\ref{fig:if_on_time_est} shows the parameter estimation for interference signals. Recall that \WaveletCNN cannot improve interference detection and estimation performance compared to \YOLOCNN. Hence, the results in Fig.~\ref{fig:if_on_time_est} are for \YOLOCNN. We see that the ON time estimation errors are very small for experiments 2A, 4A. This is because the interference objects in these experiments have fixed sizes,
i.e., lesser uncertainty. On the other hand, the estimation errors are higher for experiments 2B and 4B because the interference objects in these experiments can have unknown locations and sizes. The missed ON time is higher in experiment 2A because the interference objects for QPSK ON-OFF interference are larger than the interference objects for LTE FDD.

\subsection{Radar signal parameter estimation for different types} Fig.~\ref{fig:signal_params_est} shows the results for all radar types combined. To provide more insights, we show \YOLOCNN's radar parameter estimation for different radar types in Fig.~\ref{fig:signal_params_binwise}. The results in this figure are for experiment 1. The observations from this figure are the following. First, the missed bandwidth is higher for radar type 4 as its bandwidth variability is higher than that of other radar types. Second, the excess bandwidth is higher for radar types 1 and 2 because they are narrower than the other radar types. Third, the pulse width estimation error is lower for radar type 5 as its pulse width is higher than the remaining ones. Fourth, the pulse interval estimation error is relatively higher for radar types 2 and 3. This happens because the number of detected pulses affects the pulse interval estimation. We can see from Fig.~\ref{fig:num_pulses_binwise} that the fraction of detected pulses is lower for these two radar types.

\section{Conclusions and Future Work} \label{section:conclusions}
We presented \sys, a novel deep-learning-based versatile spectrum sensing method for detecting radar and estimating their parameters. We developed two different CNNs, \YOLOCNN and \WaveletCNN, that are the workhorse for \sys. Both the CNNs and their inputs and outputs were carefully designed. We thoroughly evaluate \sys using a diverse set of experiments. Our evaluations demonstrate the efficacy of \sys both in low SNR and low SINR. Specifically, \sys can achieve 100\% radar detection accuracy to 16 dB SNR, as well as 16 dB SINR, which cannot be guaranteed by other comparable methods.


\bibliographystyle{IEEEtran}
\bibliography{Main}

\end{document}